\documentclass[pdflatex,sn-mathphys-num]{sn-jnl}%

\usepackage{here}
\usepackage{csquotes}
\usepackage{comment}

\usepackage{graphicx}%
\usepackage{multirow}%
\usepackage{amsmath,amssymb,amsfonts}%
\usepackage{amsthm}%
\usepackage{mathrsfs}%
\usepackage[title]{appendix}%
\usepackage{xcolor}%
\usepackage{textcomp}%
\usepackage{manyfoot}%
\usepackage{booktabs}%
\usepackage{algorithm}%
\usepackage{algorithmicx}%
\usepackage{algpseudocode}%
\usepackage{listings}%

\theoremstyle{thmstyleone}%

\theoremstyle{thmstyletwo}%

\theoremstyle{thmstylethree}%

\begin{document}

\title[Evolution and Prospects of Higgs Boson Searches \\ Beyond the Standard Model]{Evolution and Prospects of Higgs Boson Searches \\ Beyond the Standard Model}

\author*[]{\fnm{Andr\'e} \sur{Sopczak}}\email{andre.sopczak@cern.ch}

\affil[]{\orgdiv{Institute of Experimental and Applied Physics}, \orgname{Czech Technical University in Prague}, \orgaddress{\street{Husova 240/5}, \city{Prague}, \postcode{11000}, \state{Prague}, \country{Czechia}}}

\abstract{The search for Higgs bosons Beyond the Standard Model (BSM) began intensively at the Large Electron--Positron (LEP) collider (1989--2000), continued at the Tevatron (2001--2011), and entered a new era with the start of the Large Hadron Collider (LHC) in 2010. While the discovery of a Higgs boson in 2012 established the Standard Model mechanism of electroweak symmetry breaking, it also opened a wide landscape of open questions that motivate ongoing BSM Higgs searches.

This review traces the evolution of these searches across experiments, highlighting both experimental techniques and underlying theoretical ideas in an accessible manner. Emphasis is placed on how different collider environments shaped search strategies and how seemingly subtle experimental effects---such as small excesses, non-standard decay topologies, or suppressed couplings---can provide insight into the possible structure of extended Higgs sectors. 

By connecting historical developments with current analyses and future prospects, this article aims not only to summarise results but also to provide students and newcomers with intuitive insight into how Higgs searches are designed, interpreted, and continuously refined. It illustrates how experimental challenges translate into physics opportunities, thereby inviting the next generation of researchers to contribute to one of the central questions in particle physics: whether the Higgs sector is minimal or part of a more complex structure.}


\maketitle

\section{Introduction and Outline}

Over the past decades, the experimental search for Higgs bosons Beyond the Standard Model (BSM) has undergone a profound evolution, spanning the Large Electron--Positron collider (LEP), the Tevatron, and the Large Hadron Collider (LHC). This progression is often described in terms of increasing centre-of-mass energy and integrated luminosity. However, such a view captures only part of the underlying physics. The results of Higgs boson searches across these experimental programmes demonstrate that sensitivity to BSM Higgs sectors is not determined solely by kinematic reach, but by the structure of Higgs couplings and the interplay of production and decay mechanisms.

Such a view provides a unifying perspective for interpreting both past results and future prospects. At LEP, the complementarity between Higgsstrahlung ($e^+e^- \to HZ$) and pair production ($e^+e^- \to H_1H_2$) enabled robust exclusions across a wide range of models, even at relatively modest energies. At the same time, scenarios in which this complementarity is weakened---such as CP-violating or highly suppressed coupling configurations---led to unexcluded regions that remain phenomenologically relevant. At the Tevatron, sensitivity to Higgs bosons arose not from extended mass reach, but from strong couplings in specific decay channels, particularly in associated production with vector bosons and decay into bottom quarks. At the LHC, the availability of multiple production modes, including gluon--gluon fusion, vector boson fusion, associated production, and top-associated processes, has enabled a comprehensive mapping of Higgs boson interactions with Standard Model particles.

Taken together, these results highlight a central principle: no single production or decay channel suffices to probe the structure of the Higgs sector. Instead, the experimental programme relies on the complementarity of multiple processes, each sensitive to different couplings and kinematic regimes. This complementarity explains both the remarkable success of Higgs boson searches in constraining large classes of BSM models and the persistence of unexcluded scenarios that evade detection through suppressed couplings or non-standard decay topologies.

The discovery of a Higgs boson with a mass of approximately 125\,GeV at the LHC in 2012 established the mechanism of electroweak symmetry breaking within the Standard Model. At the same time, it shifted the focus of Higgs physics from discovery to precision measurement and to the search for deviations that could signal the presence of an extended scalar sector. Despite increasingly precise measurements of Higgs boson properties, no conclusive evidence for BSM physics has been observed in the Higgs sector to date. However, this absence of deviations must be interpreted in the context of experimental sensitivity: in many well-motivated theoretical frameworks, including effective field theory descriptions of new physics at the TeV scale, deviations in Higgs couplings are expected at the percent level, below the current precision of LHC measurements.

This situation raises a number of important and timely questions. To what extent can extended Higgs sectors remain hidden through coupling suppression or unconventional decay modes? Is there room for additional scalar states below or above 125\,GeV, as suggested by longstanding experimental hints? How do the historical results from LEP and the Tevatron inform the interpretation of current LHC data? And what level of precision is required to probe the natural parameter space of weakly coupled BSM scenarios?

The purpose of this review is to address these questions by tracing the development of BSM Higgs boson searches across collider experiments, while emphasising the physical principles that underlie their sensitivity. Rather than presenting results as an isolated collection of exclusions and limits, the discussion focuses on how different experimental environments and analysis strategies probe complementary aspects of the Higgs sector. In particular, attention is given to the role of coupling structures, to the interpretation of apparent gaps or excesses in the data, and to the connections between historical results and current searches, including the possibility of light Higgs bosons with non-standard decays motivated by excesses observed at LEP and at the LHC.

The structure of the paper follows the historical progression of experimental programmes. Section~2 reviews the searches performed at LEP, including neutral and charged Higgs bosons and their interpretation in various benchmark scenarios. Section~3 discusses the Tevatron results, with particular emphasis on associated production and top-quark decays. Section~4 presents the current status of Higgs boson measurements and searches at the LHC and the prospects for the High-Luminosity LHC, including precision measurements, rare processes, and searches for additional scalar states. Throughout, the emphasis is placed on drawing connections between experimental results and the underlying physical mechanisms that determine their sensitivity.

By combining historical perspective with current developments, this review aims to provide not only a summary of results but also an interpretative framework for understanding how Higgs boson searches constrain, and continue to probe BSM physics.

Basic characteristics of these colliders are summarised in Table~\ref{tab:colliders}.%

\begin{figure}[h!]
\vspace{-8mm}
\centering
\includegraphics[width=0.49\columnwidth]%
{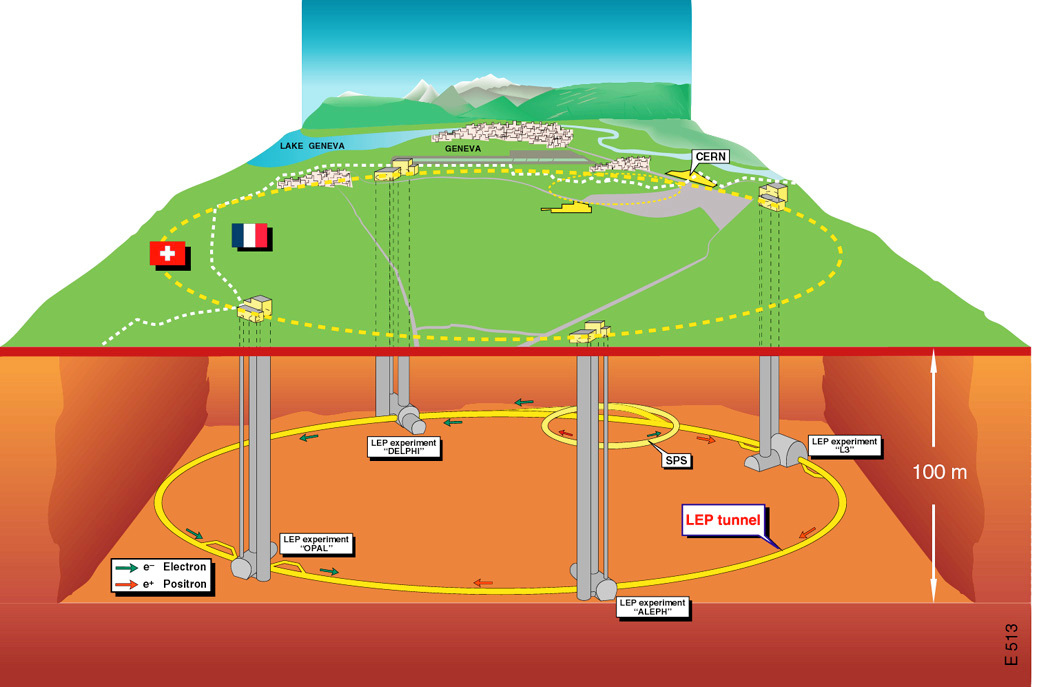}
\includegraphics[width=0.49\columnwidth]{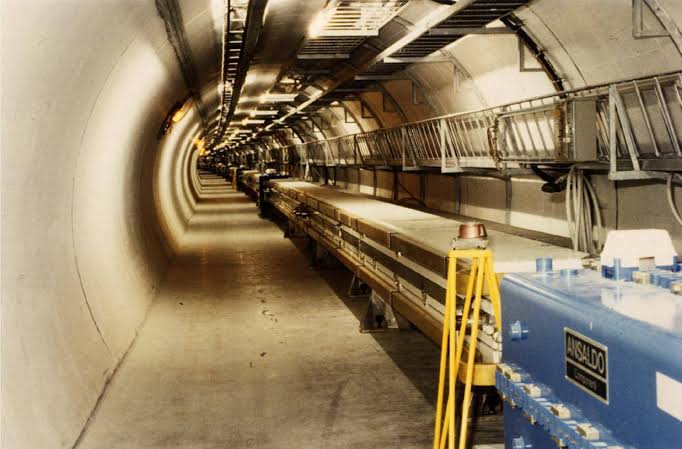}
\caption{LEP area view and magnets in the tunnel.}
\label{fig:lep}
\end{figure}

The Large Electron--Positron {\bf LEP} collider was approved on
22 May 1981 at the CERN
Council. On 
14 July 1989,  electron and positron beams were first injected.
On 2 November 2000, 
the final shutdown of the LEP collider took place. Figure~\ref{fig:lep} shows the LEP area view and magnets in the tunnel.

\begin{table}[htbp]
\vspace{-3mm}
\centering
\caption{Basic characteristics of past (LEP, Tevatron, LHC Run-1,2,3) and future (HL-LHC) colliders, including the approximate integrated luminosity per experiment.}
\label{tab:colliders}
\begin{tabular}{cccc}
\toprule
 & \multicolumn{3}{c}{Characteristics} \\
Collider & Particles & $\sqrt{s}$\,[TeV] & $\mathcal{L}_{\rm int}$\,[fb$^{-1}$] \\
\midrule
LEP        & $e^+e^-$   & 0.09--0.209 & $ 1$ \\
Tevatron   & $p\bar p$  & 1.96 & $ 10$ \\
LHC Run-1  & $pp$ & 7--8 & $ 25$ \\
LHC Run-2  & $pp$ & 13 & $ 140$ \\
LHC Run-3  & $pp$ & 13.6 & $ 330$ \\
HL-LHC     & $pp$       & 14 & 3000  \\
\bottomrule
\end{tabular}
\vspace{-9mm}
\end{table}

\begin{figure}[h!]
\vspace{-2mm}
\centering
\includegraphics[width=0.49\columnwidth]{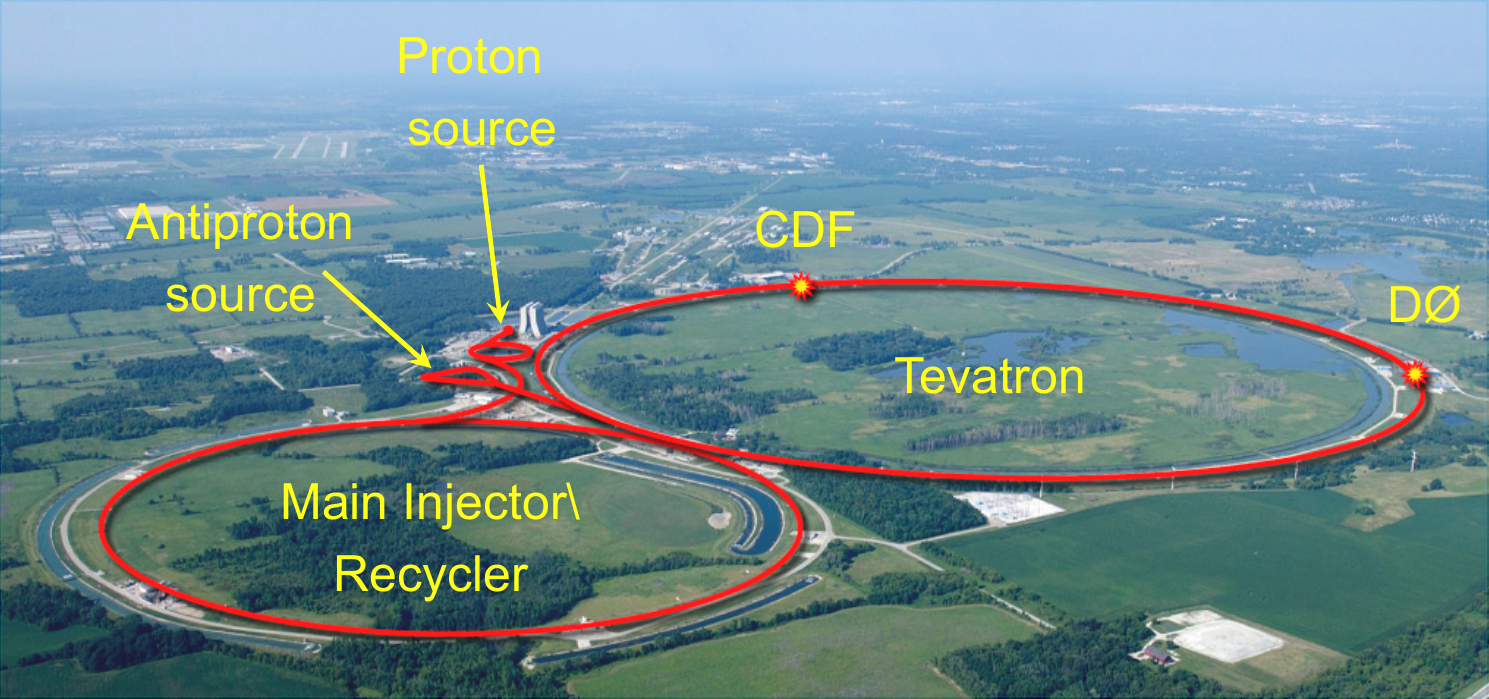}
\includegraphics[width=0.49\columnwidth]{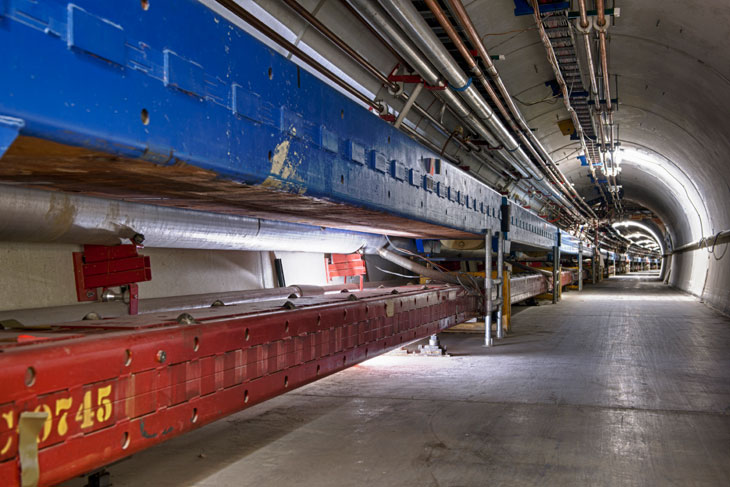}
\caption{Tevatron area view and magnets in the tunnel.}
\label{fig:tevatron}
\end{figure}

At Fermilab on 1 March 2001, the {\bf Tevatron} collider Run II began operation after successful completion of experiment, computing and facility upgrades.
The Tevatron, at that time the world's highest-energy proton-antiproton collider, shut down on 30 September 2011.
Figure~\ref{fig:tevatron} shows the area view and magnets in the tunnel.

\begin{figure}[h!]
\centering
\vspace{-3mm}
\includegraphics[width=0.49\columnwidth]{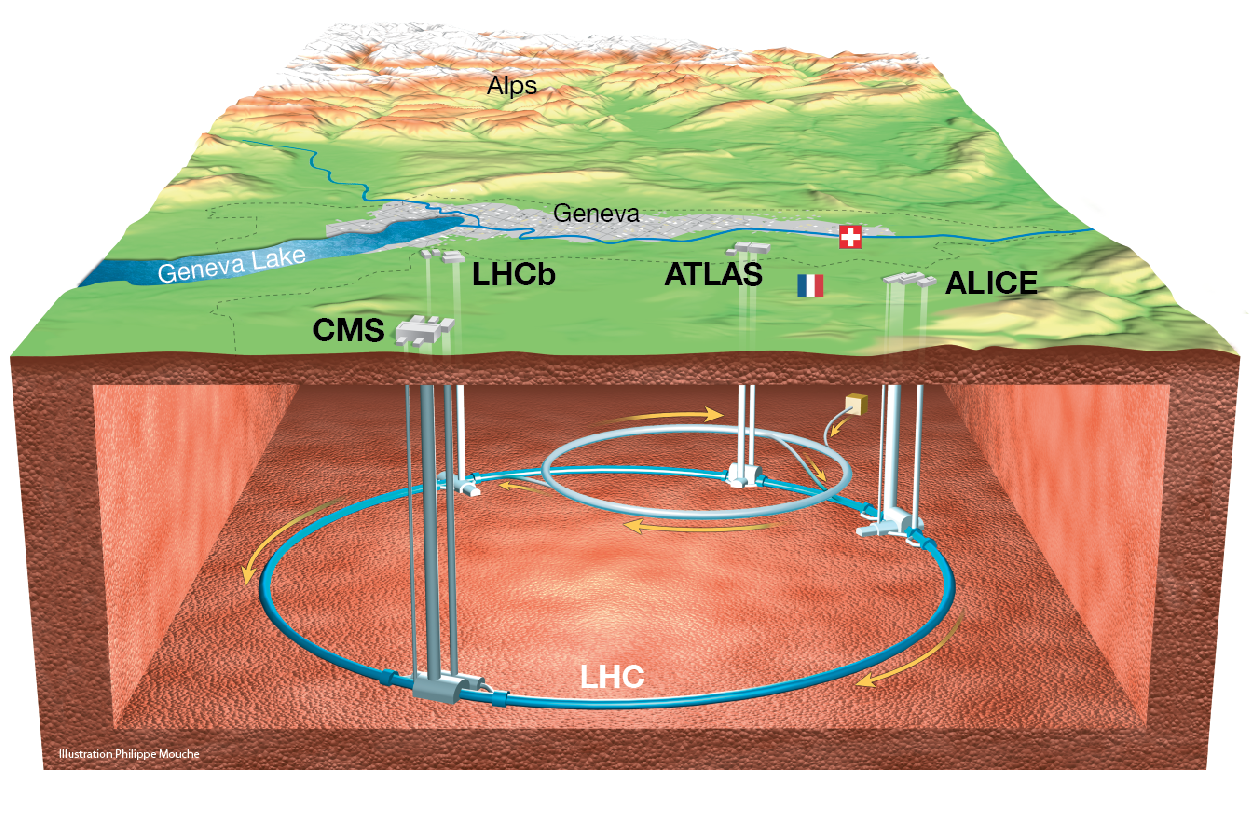}
\includegraphics[width=0.49\columnwidth]{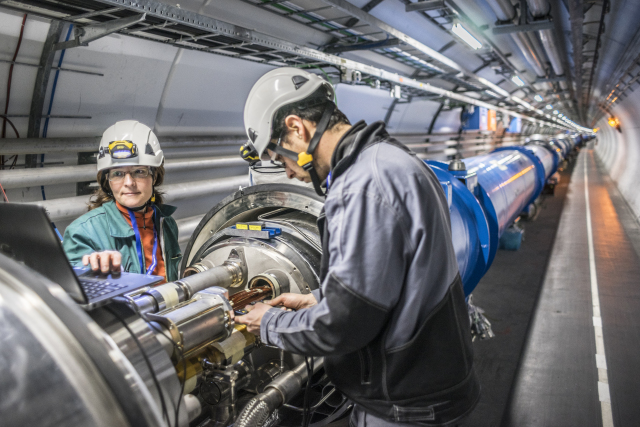}
\caption{LHC area view and magnets in the tunnel.}
\label{fig:lhc}
\end{figure}

On 24 June 1994, 
CERN advanced towards approval of {\bf LHC}, and on 
5 September 2008, CERN reiterated the safety of LHC on the eve of first beam.
The LEP tunnel was reused to host the LHC.
LHC Run-3 data-taking ended in June 2026.
Figure~\ref{fig:lhc} shows the area view and magnets in the tunnel.

\begin{figure}[h!]
\vspace{-1mm}
\centering
\includegraphics[width=0.49\columnwidth]{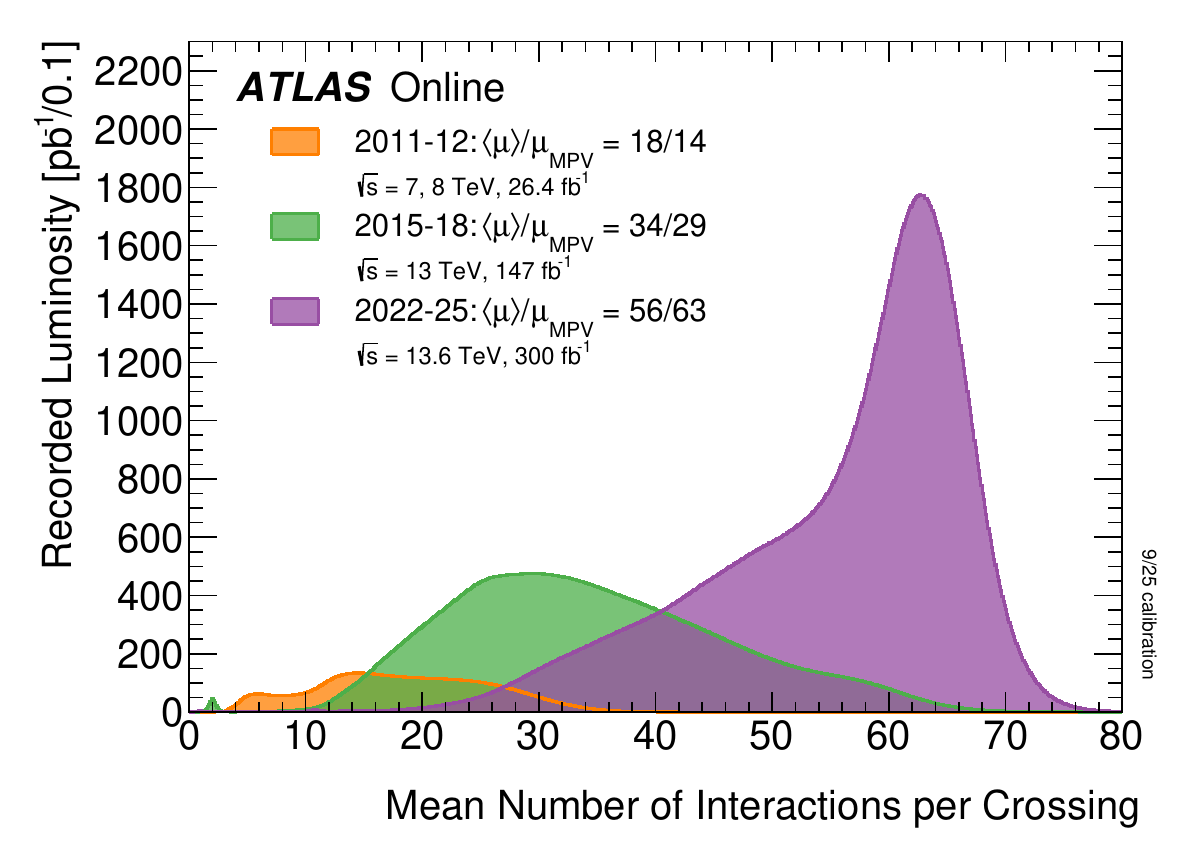}
\includegraphics[width=0.49\columnwidth]{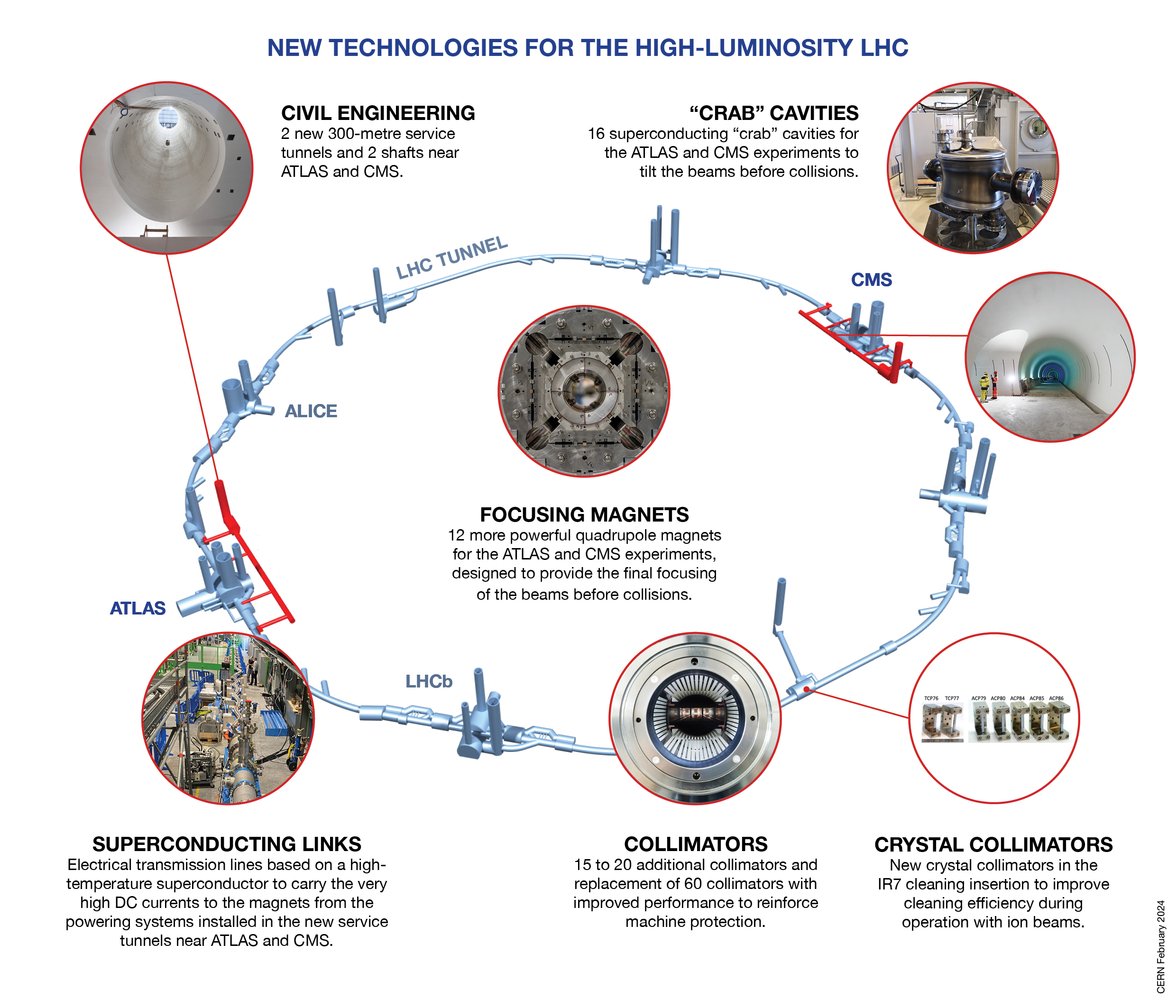}
\vspace{-1mm}
\caption{Current accumulated luminosity at the LHC as a function of the interactions per bunch crossing. At the HL-LHC, 250\,fb$^{-1}$/year and 
140 up to 200 interactions per
bunch crossing are anticipated. 
The figure also summarises new technologies for the HL-LHC.
From~\cite{lumi} and~\cite{HL-LHC}.}
\label{fig:hl-lhc}
\end{figure}

For the {\bf HL-LHC}, in April 2018
civil engineering work began at 
LHC Point 1 (ATLAS side)
During 2026-2029
equipment will be upgraded. 
In 2030 the 
HL-LHC (Run-4) will start.
Figure~\ref{fig:hl-lhc} shows the current accumulated luminosity at the LHC as a function of the interactions per bunch crossing. At the HL-LHC, 250\,fb$^{-1}$/year and 
140 up to 200 interactions per
bunch crossing are anticipated. 
The figure also summarises new technologies for the HL-LHC.

\section{LEP Experimental Searches}

\subsection{LEP Neutral Higgs Boson Results}

\begin{figure}[h!]
\centering
\includegraphics[width=0.32\columnwidth]{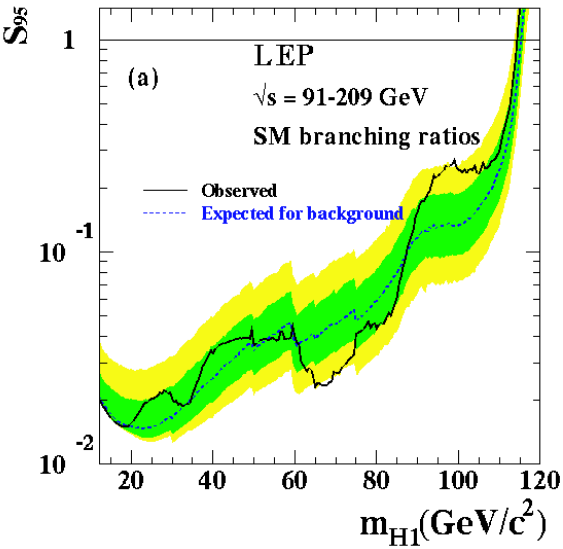}
\includegraphics[width=0.32\columnwidth]{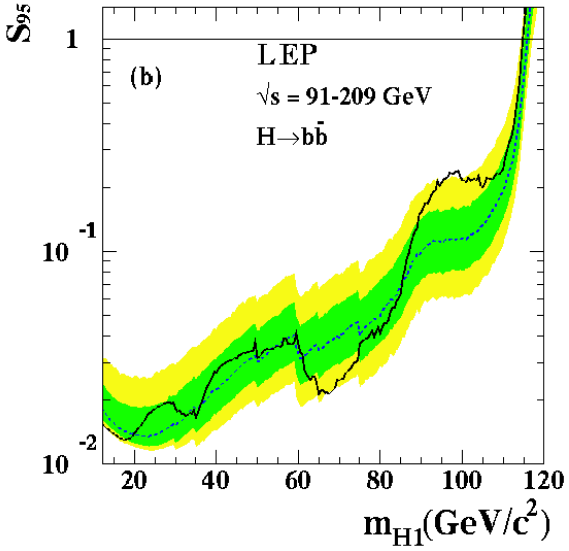}
\includegraphics[width=0.32\columnwidth]{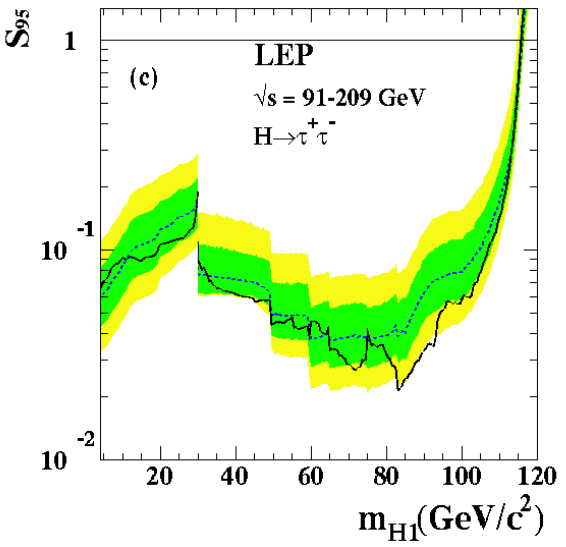}
\caption{95\% CL upper limits $S_{95}$ on $\xi^{2} = \left({g_{HZZ}}/{g^{\mathrm{SM}}_{HZZ}}\right)^{2}$ times the assumed branching fraction (see text).
Left: SM Higgs boson decay.
Centre: $H \to b\bar{b}$.
Right: $H \to \tau^{+}\tau^{-}$.
From~\cite{sopczak2006mssmhiggsbosonsearches}.}
\label{fig:lepsearcheslimits}
\end{figure}

\begin{figure}[h!]
\centering
\includegraphics[width=0.32\columnwidth]{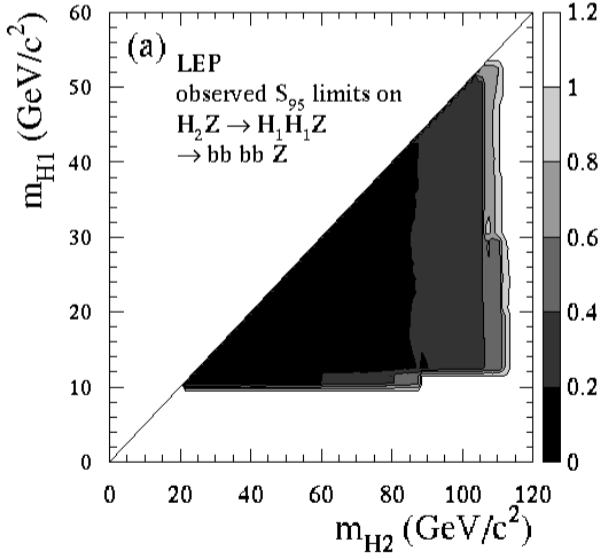}
\includegraphics[width=0.32\columnwidth]{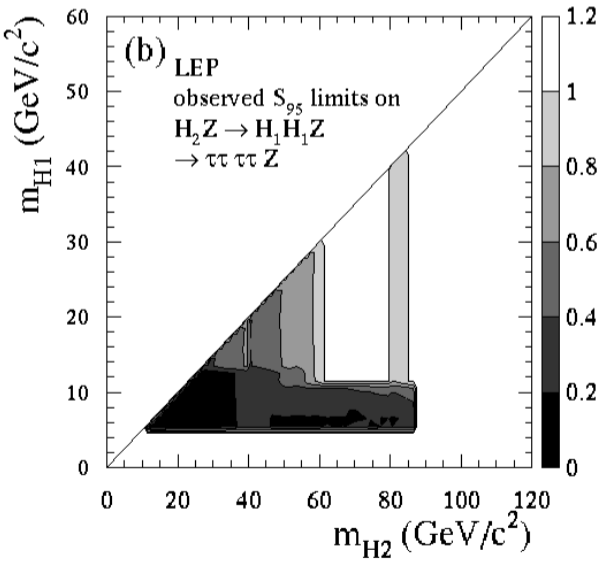}
\includegraphics[width=0.32\columnwidth]{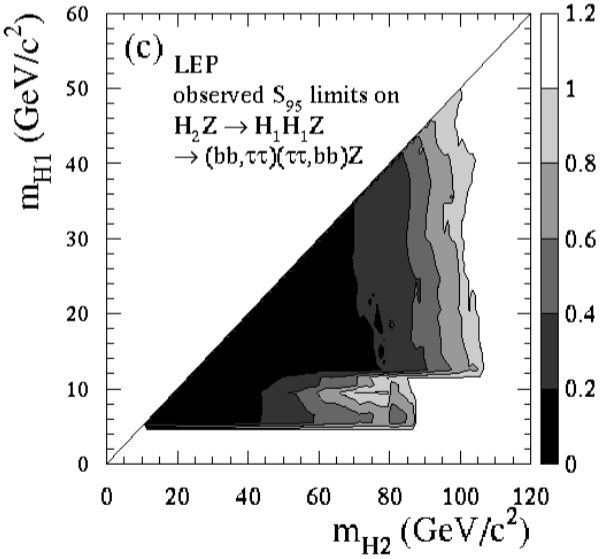}
\caption{Limits on $S_{95}$ for $H_2 Z \to H_1 H_1 Z$ in various decay modes.
Left: $b\bar{b}b\bar{b}$.
Centre: $\tau^{+}\tau^{-}\tau^{+}\tau^{-}$.
Right: $b\bar{b}\tau^{+}\tau^{-}$ and $\tau^{+}\tau^{-}b\bar{b}$.
From~\cite{sopczak2006mssmhiggsbosonsearches}.}
\label{fig:lepsearcheslimits2}
\end{figure}

The experimental searches at LEP for Higgsstrahlung and Higgs boson pair production are summarised in Table~\ref{tab:lepsearches}. 
The results are expressed as limits on the topological production cross-section relative to the Standard Model (SM) expectation. 
Figures~\ref{fig:lepsearcheslimits} and~\ref{fig:lepsearcheslimits2} summarise the limits for Higgsstrahlung. 

\begin{figure}[h!]
\vspace{-5mm}
\centering
\includegraphics[width=0.34\columnwidth]{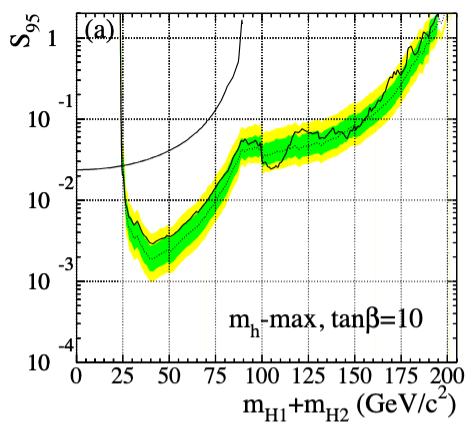}
\includegraphics[width=0.34\columnwidth]{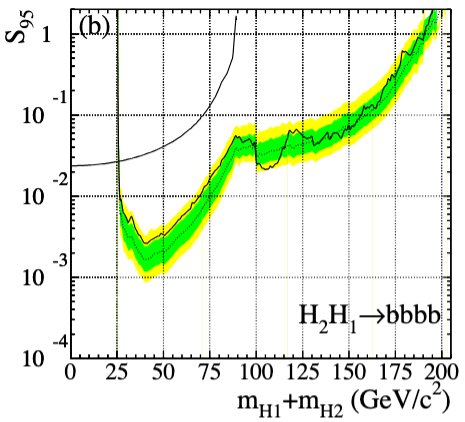}
\includegraphics[width=0.34\columnwidth]{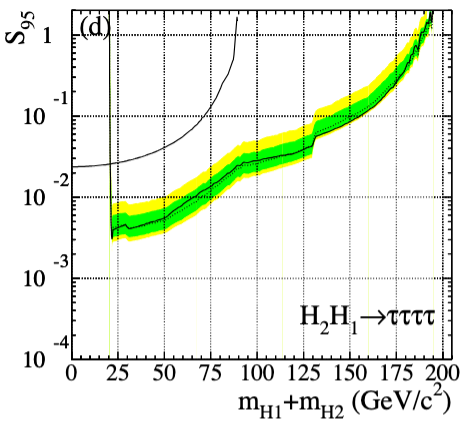}
\includegraphics[width=0.34\columnwidth]{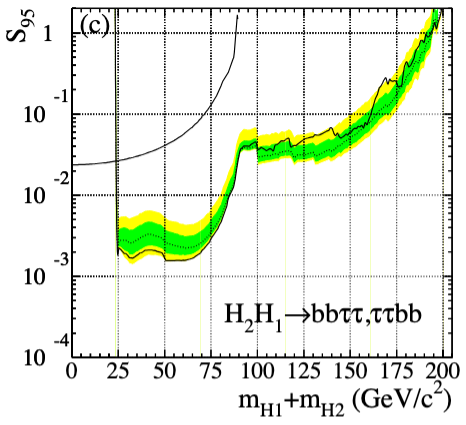}
\caption{
Limits on $S_{95}$ for the $H_1 H_2$ process with different decay modes.
Regions of low $m_{H_1}+m_{H_2}$ are excluded by a 
$\Delta \Gamma_Z < 2\,\mathrm{MeV}$ constraint.
Upper-left: assuming branching-ratio decays of the $h$-max scenario.
Upper-right: $b\bar{b}b\bar{b}$. 
Lower-left: $\tau^{+}\tau^{-}\,\tau^{+}\tau^{-}$.
Lower-right: 
$b\bar{b}\tau^{+}\tau^{-}$ and 
$\tau^{+}\tau^{-}b\bar{b}$.
From~\cite{sopczak2006mssmhiggsbosonsearches}.
}
\label{fig:lepsearchespair}
\end{figure}

\begin{figure}[h!]
\vspace{-2mm}
\centering
\includegraphics[width=0.34\columnwidth]{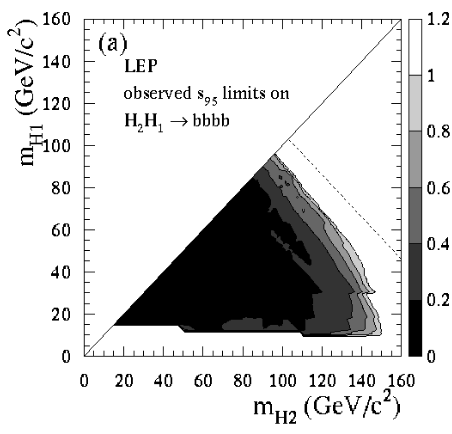}
\includegraphics[width=0.34\columnwidth]{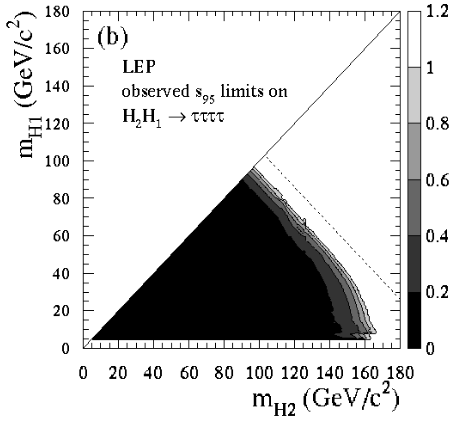}
\includegraphics[width=0.34\columnwidth]{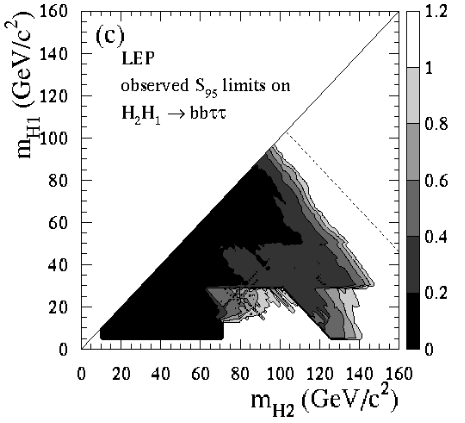}
\includegraphics[width=0.34\columnwidth]{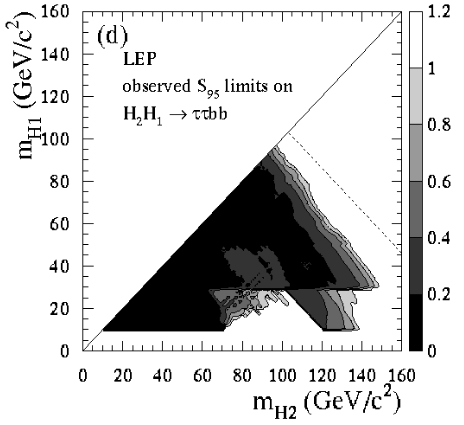}
\caption{Limits on $S_{95}$ for the $H_1 H_2$ pair-production cascade process $H_2 H_1 \to (H_1 H_1) H_1$ with different final-state combinations.
Regions of low $m_{H_1}+m_{H_2}$ are excluded by a
$\Delta \Gamma_Z < 2\,\mathrm{MeV}$ constraint.
Upper-left: $b\bar{b}b\bar{b}$.
Upper-right: $\tau^{+}\tau^{-}\tau^{+}\tau^{-}$.
Lower-left: $b\bar{b}\tau^{+}\tau^{-}$.
Lower-right: $\tau^{+}\tau^{-}b\bar{b}$.
From~\cite{sopczak2006mssmhiggsbosonsearches}.}
\label{fig:lepsearchespair2}
\vspace{-2mm}
\end{figure}

The quantity $S_{95}$ shown in Figs.~\ref{fig:lepsearcheslimits}--\ref{fig:lepsearchespair2} is the 95\% CL upper bound on the signal production rate normalised to a reference cross-section. For Higgsstrahlung, the reference is the SM cross-section for $e^+e^-\to HZ$ at the same Higgs boson mass, so that the limit applies to $\xi^2 = (g_{HZZ}/g^{\rm SM}_{HZZ})^2$ multiplied by the branching fraction assumed for the final state considered (SM-like decays, or 100\% decays into $b\bar b$ or $\tau^+\tau^-$). Since the Higgsstrahlung cross-section scales with the square of the $HZZ$ coupling, any model that predicts a value of $\xi^2\times{\rm BR}$ above the observed $S_{95}$ curve at a given Higgs boson mass is excluded at 95\% CL, while models below the curve remain allowed. For pair production and the cascade topologies, $S_{95}$ is defined analogously, relative to a reference cross-section with maximal $H_1H_2Z$ coupling. The observed limits are shown as solid lines, the median expected limits in the absence of a signal as dashed lines, and the $\pm1\sigma$ and $\pm2\sigma$ ranges of the expectation as bands.

\begin{table}[htbp]
\centering
\caption{Search channels for Higgsstrahlung and pair production.}
\label{tab:lepsearches}
\begin{tabular}{p{0.95\linewidth}}
\toprule
\textbf{Higgsstrahlung:} $e^+e^- \rightarrow H_1 Z$ \\[2mm]

Four-jet: $(H_1 \rightarrow b\bar{b})(Z \rightarrow q\bar{q})$ \\

Missing energy: $(H_1 \rightarrow b\bar{b},\, \tau^+\tau^-)(Z \rightarrow \nu\bar{\nu})$ \\

Leptonic: $(H_1 \rightarrow b\bar{b},\, q\bar{q})(Z \rightarrow e^+e^-,\, \mu^+\mu^-)$ \\

Tau-leptons: $(H_1 \rightarrow \tau^+\tau^-)(Z \rightarrow q\bar{q})$ and 
$(H_1 \rightarrow b\bar{b},\, \tau^+\tau^-)(Z \rightarrow \tau^+\tau^-)$ \\

Cascade: $H_2 Z \rightarrow (H_1 H_1) Z$ \\[3mm]

\textbf{Pair production:} $e^+e^- \rightarrow H_2 H_1$ \\[2mm]

Four-b: $(H_2 \rightarrow b\bar{b})(H_1 \rightarrow b\bar{b})$ \\

Mixed: $(H_2 \rightarrow \tau^+\tau^-)(H_1 \rightarrow b\bar{b})$ and 
$(H_2 \rightarrow b\bar{b})(H_1 \rightarrow \tau^+\tau^-)$ \\

Four-tau: $(H_2 \rightarrow \tau^+\tau^-)(H_1 \rightarrow \tau^+\tau^-)$ \\

Cascade: $H_2 H_1 \rightarrow (H_1 H_1) H_1$ \\
\bottomrule
\end{tabular}
\end{table}

In Fig.~\ref{fig:lepsearcheslimits}, the analyses were performed under the assumption of an SM-like production mechanism with a reduced production rate. The Higgs boson decay modes were assumed to be either SM-like or 100\% into a pair of $\tau$ leptons or $b$ quarks. The important feature of these limits is that they can be interpreted directly as constraints on the $HZZ$ coupling, rather than on the Higgs boson mass alone. The sensitivity therefore depends on the strength of the coupling to the $Z$ boson, illustrating that even at fixed energy, exclusion power is controlled by coupling structure rather than purely by kinematic reach.

For the limits shown in Fig.~\ref{fig:lepsearcheslimits2}, the production of the heavier scalar Higgs boson in association with a $Z$ boson was analysed. The heavier Higgs boson decays into a lighter Higgs boson, whose subsequent decays lead to various final states.

Figures~\ref{fig:lepsearchespair} and~\ref{fig:lepsearchespair2} show limits on Higgs boson pair production, resulting in final states containing $\tau$-lepton or $b$-quark pairs. In addition, limits are shown for the $h$-max scenario of the Minimal Supersymmetric extension of the Standard Model (MSSM). Together with Higgsstrahlung, these channels possess a powerful complementarity: if the coupling governing one production mode is suppressed, the other is typically enhanced. This interplay is the underlying reason why large regions of MSSM parameter space could be excluded at LEP despite the limited centre-of-mass energy.

\begin{figure}[h!]
\vspace{-5mm}
\centering
\includegraphics[width=0.51\columnwidth]{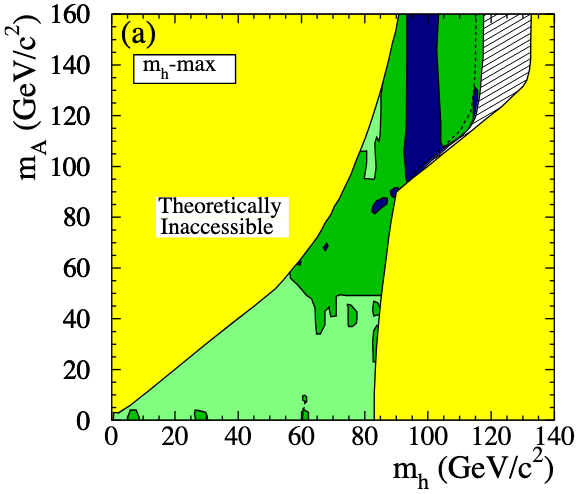}
\includegraphics[width=0.47\columnwidth]{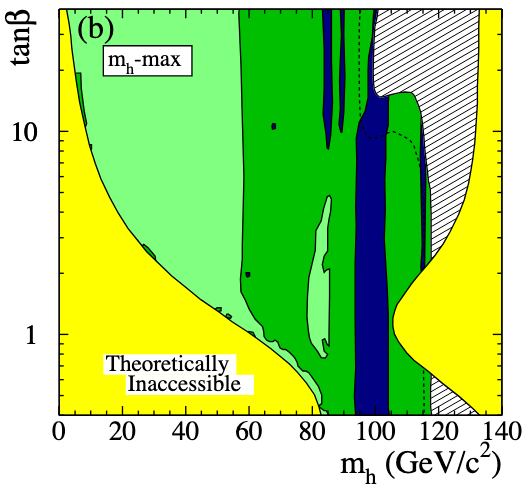}
\caption{LEP results on the CP-conserving MSSM $h$-max benchmark scenario. 
Contour regions of $1-\mathrm{CL}_b$; light-green: $1\sigma$,
dark-green: $1$--$2\sigma$, blue: $>2\sigma$.
In the hatched region an apparent excess would not be significant.
Left:  $(m_h, m_A)$ plane
($m_h = 0$--$140$\,GeV, $m_A = 0$--$160$\,GeV).
Right:  $(m_h, \tan\beta)$ plane ($m_h = 0$--$140$\,GeV, $\tan\beta = 0.4$--$40$).
From~\cite{sopczak2006mssmhiggsbosonsearches}.}
\label{fig:lepmssm}
\end{figure}

Interpretations are given in the MSSM, as shown in Fig.~\ref{fig:lepmssm}, for the $h$-max scenario in the usual parameter spaces $(m_h,m_A)$ and $(m_A,\tan\beta)$. The importance of radiative corrections was emphasised in Refs.~\cite{Sopczak1992PhD,Rosiek:269364}. 
An interpretation including the $>2\sigma$ excess near $m_h = 100$\,GeV as a Three-Higgs-Boson hypothesis was given~\cite{Sopczak:2000pp}.

\begin{figure}[hbtp]
\centering
\includegraphics[width=0.52\columnwidth]{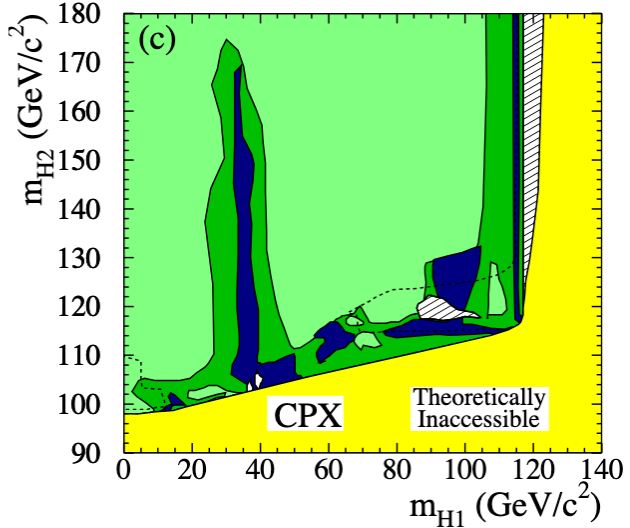}
\includegraphics[width=0.46\columnwidth]{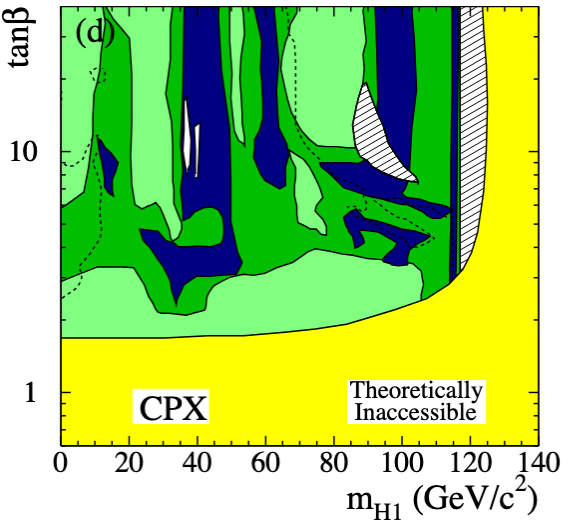}
\caption{LEP results on the CP-violating CPX benchmark scenario.
Contour regions of $1-\mathrm{CL}_b$; light-green: $1\sigma$,
dark-green: $1$--$2\sigma$, blue: $>2\sigma$.
In the hatched region an apparent excess would not be significant.
Left: 
$(m_{H_1}, m_{H_2})$ plane
($m_{H_1} = 0$--$140$\,GeV, $m_{H_2} = 90$--$180$\,GeV);
note the different $m_{H_2}$ axis range compared to $m_A = 0$--$160$\,GeV in Fig.~\ref{fig:lepmssm}.
Right:  $(m_{H_1},\tan\beta)$ plane ($m_{H_1} = 0$--$140$\,GeV, $\tan\beta = 0.6$--$40$);
note the different $\tan\beta$ axis range compared to $\tan\beta = 0.4$--$40$ in Fig.~\ref{fig:lepmssm}.
From~\cite{sopczak2006mssmhiggsbosonsearches}.}
\label{fig:cpxlimits}
\end{figure}

CP-violating results are shown in Fig.~\ref{fig:cpxlimits}.
Table~\ref{tab:benchmarks} summarises benchmark limits at 95\% CL.
In this model, $(m_{H1},m_{H2})$ and $(m_{H2},\tan\beta)$ correspond to $(m_h,m_A)$ and $(m_A,\tan\beta)$ in the CP-conserving model.

\begin{table}[tp]
\centering
\caption{\mbox{Benchmark scenarios for MSSM Higgs searches.}}
\label{tab:benchmarks}

\begin{tabular}{lcccc}
\toprule
\textbf{Benchmark scenario} & 
\textbf{$m_t$ (GeV)} & 
\textbf{$m_h$ (GeV)} & 
\textbf{$m_A$ (GeV)} & 
\textbf{Exclusion of $\tan\beta$} \\
\midrule
$m_h$-max     & 169.3 & 92.9 (94.8) & 93.4 (95.1) & 0.6--2.6 (0.6--2.7) \\
               & 174.3 & 92.8 (94.9) & 93.4 (95.2) & 0.7--2.0 (0.7--2.1) \\
\addlinespace
no-mixing     & 169.3 & excl. (excl.) & excl. (excl.) & excl. \\
               & 174.3 & 93.6 (96.0) & 93.6 (96.4) & 0.4--10.2 (0.4--19.4) \\
\addlinespace
large-$\mu$   & 169.3 & excl. (excl.) & excl. (excl.) & excl. \\
               & 174.3 & excl. (excl.) & excl. (excl.) & excl. \\
\addlinespace
gluophobic    & 169.3 & 90.6 (93.2) & 95.7 (98.2) & 0.4--10.3 (0.4--21.5) \\
               & 174.3 & 90.5 (92.3) & 96.3 (98.0) & 0.4--5.4 (0.4--6.4) \\
\addlinespace
small-$\alpha_{\text{eff}}$ & 169.3 & 88.2 (90.0) & 98.2 (99.6) & 0.4--6.1 (0.4--7.4) \\
               & 174.3 & 87.3 (89.0) & 98.8 (100.0) & 0.4--4.2 (0.4--4.5) \\
\bottomrule
\end{tabular}
\end{table}

Figure~\ref{fig:cpxlimits} shows the results of the LEP searches interpreted in the CP-violating CPX benchmark scenario of the MSSM~\cite{sopczak2006mssmhiggsbosonsearches}, which introduces maximal CP-violating phases $\arg(A_t) = \arg(m_{\tilde{g}}) = 90^\circ$ in the soft-breaking trilinear coupling $A_t$ and the gluino mass $m_{\tilde{g}}$. Unlike the CP-conserving case (Fig.~\ref{fig:lepmssm}), all three neutral Higgs bosons $H_1$, $H_2$, $H_3$ mix: they are no longer CP eigenstates. The parameter scanned in the CPX scenario is the charged Higgs mass $m_{H^\pm}$, from which the neutral masses $m_{H_1}$, $m_{H_2}$, $m_{H_3}$ are derived by diagonalising the one-loop radiatively corrected neutral Higgs mass matrix~\cite{sopczak2006mssmhiggsbosonsearches}.

In the CPX benchmark scenario, the MSSM parameters are fixed to
\[
M_{\mathrm{SUSY}} = 500~\mathrm{GeV}, \qquad
\mu = 2000~\mathrm{GeV}, \qquad
|A_t| = |A_b| = 1000~\mathrm{GeV},
\]
with large CP-violating phases~\cite{Carena:2000yi}. The lightest neutral Higgs mass,
$m_{H_1}$, is therefore not an independent input parameter but is obtained by
diagonalising the radiatively corrected neutral Higgs mass matrix at each point
in the $(m_{H^\pm},\tan\beta)$ parameter space. The loop corrections induce
scalar--pseudoscalar mixing, modifying the neutral Higgs mass eigenvalues and
couplings~\cite{Carena:2001fw}. As a consequence, only a restricted range of
$m_{H_1}$ values is predicted for a given choice of $(m_{H^\pm},\tan\beta)$.
Regions requiring larger values of $m_{H_1}$ than those obtainable from the
corrected mass matrix do not correspond to self-consistent solutions of the
benchmark and are therefore labelled \emph{theoretically inaccessible}. These
regions reflect limitations of the benchmark parameter space rather than the
experimental sensitivity~\cite{LEPHiggs:2006sj}.

The coloured contours in both panels of Fig.~\ref{fig:cpxlimits} represent three levels of statistical compatibility between data and the background-only hypothesis, expressed via $1 - \mathrm{CL}_b$: \emph{light green} = $1\sigma$ region ($1 - \mathrm{CL}_b > 0.16$); \emph{dark green} = $1$--$2\sigma$ region; \emph{blue} = $>2\sigma$ region. 
The left panel presents these contours in the $(m_{H_1}, m_{H_2})$ plane; the right panel in the $(m_{H_1}, \tan\beta)$ plane.
These are not experimental exclusions but indicators of where the data exhibit the most signal-like behaviour. 
All coloured regions are derived from the \emph{same} statistical combination of all search channels simultaneously: Higgsstrahlung ($e^+e^- \to H_1 Z$) in the four-jet, missing-energy, leptonic, and tau-lepton final states; the cascade $H_2 Z \to (H_1 H_1) Z$; pair production ($e^+e^- \to H_1 H_2$) in the $b\bar{b}b\bar{b}$, $b\bar{b}\tau^+\tau^-$, and $\tau^+\tau^-\tau^+\tau^-$ modes; and the triple-Higgs cascade $H_2 H_1 \to (H_1 H_1)H_1$. Additional constraints from the $Z$-width ($\Delta\Gamma_Z < 2$\,MeV) and Yukawa-production ($b\bar{b}H_1$) searches are also included.

In the CP-conserving MSSM (Fig.~\ref{fig:lepmssm}), the sum rule $\sin^2(\beta-\alpha)+\cos^2(\beta-\alpha)=1$ ensures that if the $hZZ$ coupling is suppressed, the pair-production cross-section $e^+e^-\to hA$ is correspondingly enhanced, closing the parameter space without gaps. In the CPX scenario this complementarity breaks down: $H_1$ can be nearly CP-odd in composition, simultaneously suppressing both the Higgsstrahlung and pair-production rates. This creates a prominent unexcluded ``hole'' near $\tan\beta \approx 4$--$5$ and $m_{H_1} \approx 40$--$70$\,GeV with no analogue in the CP-conserving case. Taken together, Figs.~\ref{fig:lepmssm} and~\ref{fig:cpxlimits} illustrate a general principle: CP-violating MSSM Higgs sectors are harder to exclude at LEP than their CP-conserving counterparts because kinematic complementarity between Higgsstrahlung and pair production is no longer guaranteed. The key lesson of Fig.~\ref{fig:cpxlimits} is therefore that once this complementarity is broken, sizeable regions of parameter space can evade experimental exclusion even at accessible masses---demonstrating explicitly that the reach of Higgs boson searches is ultimately limited by coupling configurations rather than by energy alone.

The discussion above clarifies how experimental sensitivity, statistical interpretation, and model structure interplay in this benchmark.

\subsection{LEP Charged Higgs Boson Results}

\begin{figure}[h!]
\vspace{-5mm}
\centering
\includegraphics[width=0.49\columnwidth]{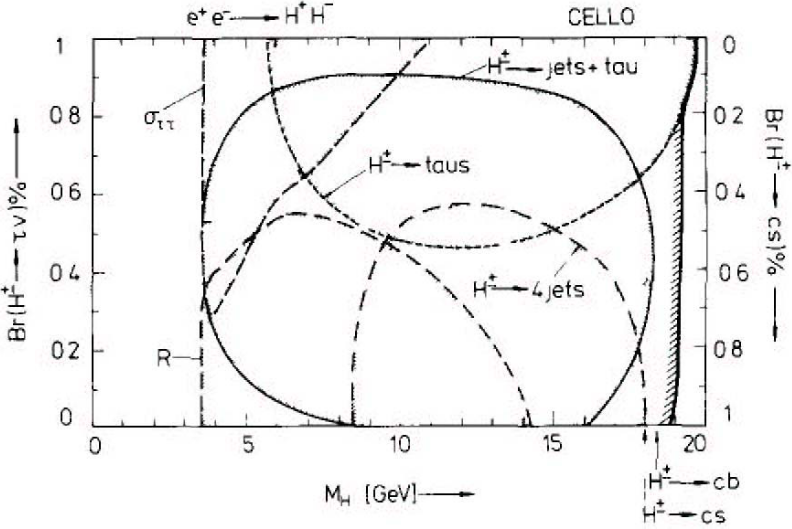}
\includegraphics[width=0.49\columnwidth]{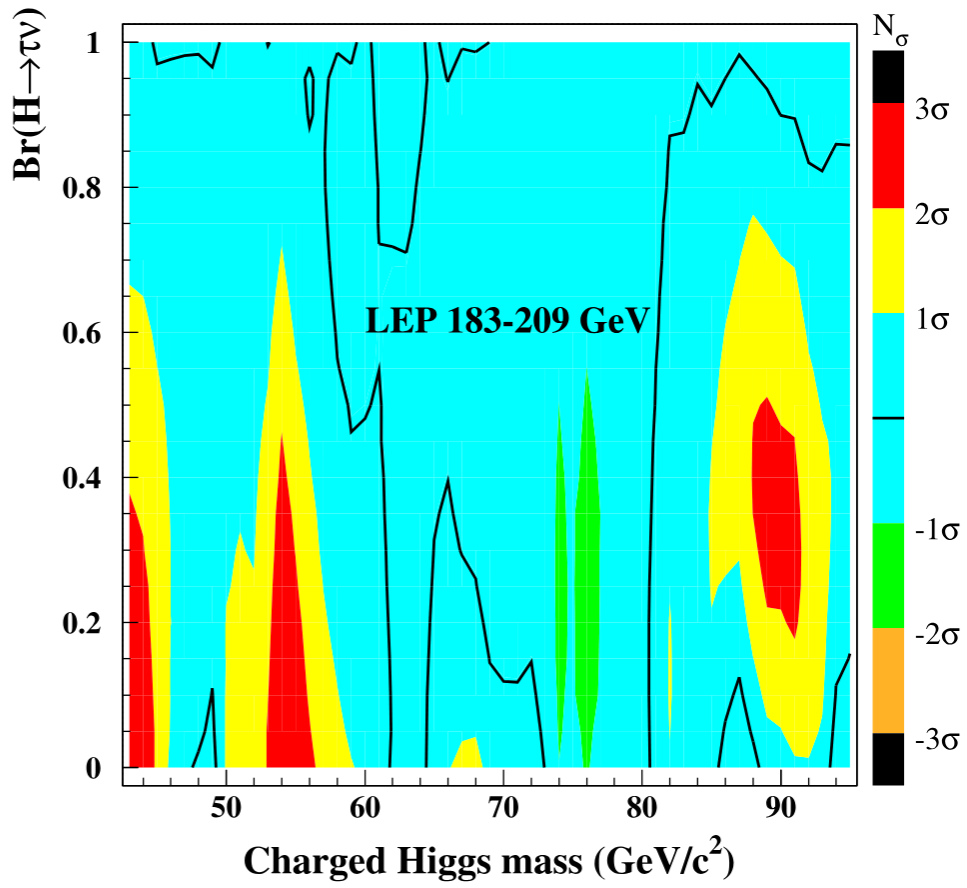}
\caption{CELLO and LEP limits on charged Higgs bosons in the $(m_{H^+},\tan\beta)$ plane.
Left: CELLO results covering $m_{H^+}$ up to $\approx20$\,GeV; right: LEP combined results extending the mass reach to $\approx95$\,GeV. Note the different $m_{H^+}$ axis ranges: 0--20\,GeV (left) and 0--95\,GeV (right).
From~\cite{cello} and~\cite{sopczak2006mssmhiggsbosonsearches}.}
\label{fig:cello}
\end{figure}

The charged Higgs boson searches at LEP focused on the decay modes $c\bar{s}$ and $\tau\nu$. For charged Higgs boson pair production, these modes lead to the signatures $cscs$, $cs\tau\nu$, and $\tau\nu\tau\nu$. These decay channels had already been investigated at the CELLO experiment~\cite{cello}. The mass range up to about 20\,GeV was covered, as shown in Fig.~\ref{fig:cello}. 
The usual parameter space is $(m_{H^+},\tan\beta)$, which determines the $H^+H^-$ production rate and the charged Higgs boson decay branching ratios. 
The figure also illustrates that LEP extended the accessible mass range to about 95\,GeV. 

\begin{figure}[h!]
\vspace{-7mm}
\centering
\includegraphics[width=0.49\columnwidth]{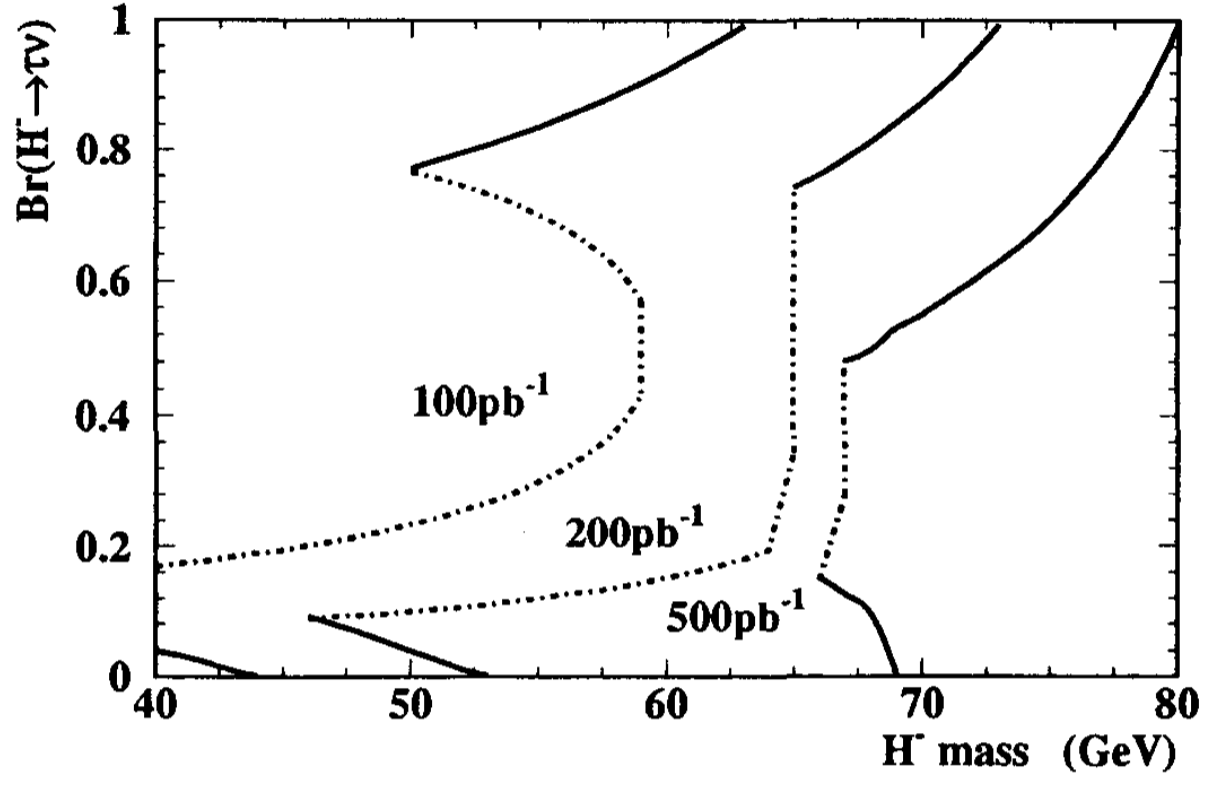}
\includegraphics[width=0.49\columnwidth]{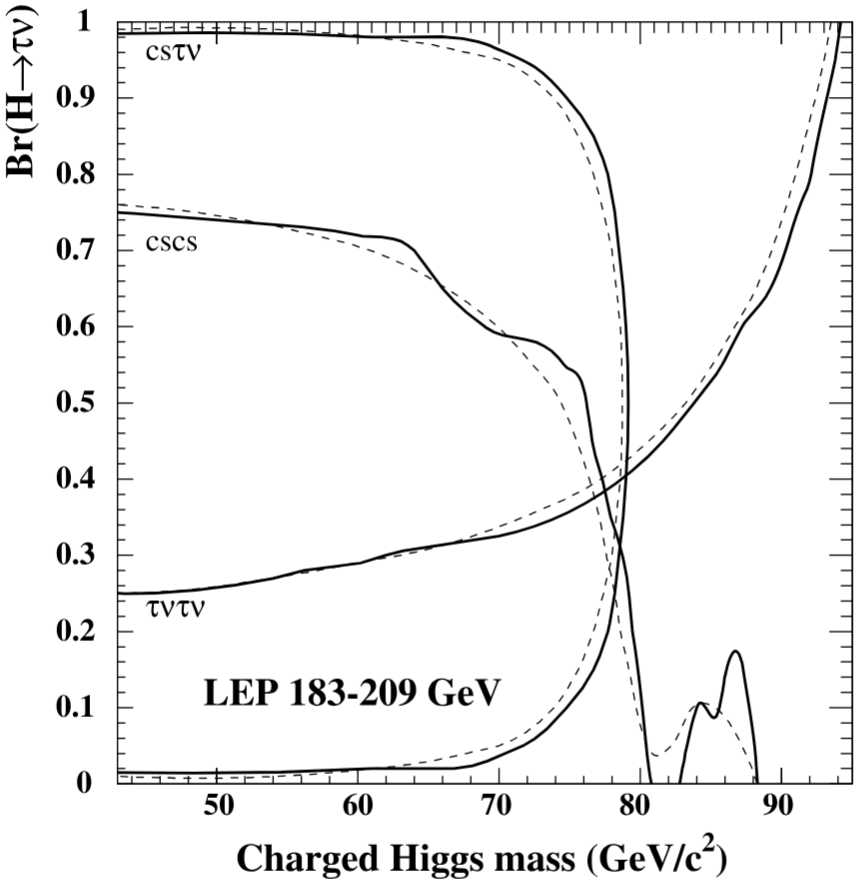}
\caption{Simulated expectations per LEP experiment and observed LEP limits on charged Higgs bosons. 
The LEP  limits are combined results from the four LEP experiments with a total luminosity of about 2600\,pb$^{-1}$.
Note the different $m_{H^+}$ axis ranges: 40--80\,GeV (left) and 40--95\,GeV (right).
From~\cite{Sopczak:1993jt,ALEPH:2013htx}.}
\label{fig:lephplus}
\vspace{-7mm}
\end{figure}

Figure~\ref{fig:lephplus} shows the expected charged Higgs boson search sensitivity from 1993 and the final LEP results from 2013.
The achieved experimental sensitivity is consistent with the per-experiment expectations, taking into account that the final result corresponds to the combination of all four experiments. These results illustrate that the interpretation of charged Higgs searches depends crucially on the assumed branching ratios, and hence on the underlying model parameters such as $\tan\beta$. The experimental sensitivity is therefore not universal but tied to specific coupling patterns, reinforcing the need for complementary searches in multiple decay modes.

\clearpage
\subsection{Discussion of LEP Results}

The focus of the SM Higgs boson research at LEP was on extending the discovery reach to higher Higgs boson masses. Before the start of the LEP era, there were no solid experimental or theoretical lower bounds on the Higgs boson mass. Considerable excitement arose within the LEP experimental collaborations when the LEP collider began operation in the summer of 1989, shortly after the first lower bounds on the Higgs boson mass were published, for example excluding the mass range from 3.0\,GeV to 19.3\,GeV~\cite{Akrawy1990224}. 

Approximately thirteen years later, the combined results from the four LEP experiments pushed the lower bound on the SM Higgs boson mass to 114.4\,GeV~\cite{200361}. It is remarkable that the combined efforts in LEP accelerator operation, detector performance, and data analysis achieved a sensitivity exceeding the expectations of 80--100\,GeV that had been projected in 1989~\cite{Gunion:1989we}.

While these limits assumed SM Higgs boson production cross-sections and decay branching fractions, the LEP experiments also extended their searches to scenarios with reduced production cross-sections relative to the SM expectation, while still assuming SM decay branching fractions. In BSM scenarios, limits were set in the Higgs boson mass range from 10\,GeV to 100\,GeV on the production rate relative to the SM prediction, typically between 2\% and 20\%. Since the Higgs boson production rate is proportional to the square of the HZZ coupling, these results directly constrained modifications of this coupling.

Specific Higgs boson decay modes, such as $H\rightarrow b\bar b$ and ${\rm H}\rightarrow \tau\tau$, were investigated
 in dedicated analyses, and corresponding limits were derived. Already during the LEP era, the presentation formats for Higgs boson search results that are still used today were established. Typical exclusion plots display the expected limit as a dashed line, the observed limit as a solid line, and green and yellow bands representing the one- and two-standard-deviation uncertainty ranges, respectively.

In addition, the LEP collaborations investigated the pair production of neutral and charged Higgs bosons. The dominant decay modes of neutral Higgs bosons were decays into bottom-quark pairs and tau-lepton pairs. Searches for charged Higgs bosons targeted both hadronic decay modes and decays involving tau leptons. The results were interpreted either within the MSSM or within general two-Higgs-doublet models (2HDM).
 
 In the MSSM, the Higgs boson mass is subject to both lower and upper bounds.
 In this model, the search for the SM Higgs boson excluded a large parameter region.
 However, 
 radiative corrections play a crucial role. Detailed studies revealed regions of parameter space that remained unexcluded~\cite{Rosiek:269364}. 
For charged Higgs boson searches, the theoretical lower bound on the charged Higgs boson mass in the MSSM corresponds approximately to the W-boson mass, placing it largely beyond the kinematic reach of LEP; consequently, interpretations were provided primarily within the general 2HDM framework.

Finally, it should be noted that searches for invisibly decaying Higgs bosons were already initiated at LEP. These searches were subsequently continued at the Tevatron and later at the LHC.

\subsubsection*{Is there room below 125\,GeV for a lighter Higgs boson with non-standard decays?}

A subtle but important caveat to the LEP exclusion deserves explicit discussion. The lower bound of 114.4\,GeV~\cite{200361} applies specifically to a Higgs boson with SM-like production cross-section \emph{and} SM-like decay branching ratios. As analysed in detail by Chang, Dermisek, Gunion and Weiner~\cite{Chang:2008cw}, if the Higgs boson decays non-standardly---for instance via $h \to 2a \to 4x$, where $a$ is a light pseudoscalar and $x$ is a SM final state---the LEP searches lose sensitivity and the mass exclusion no longer applies. The simplest such cascade has $x = \tau$, which is experimentally unconstrained for $m_h > 86$\,GeV, while no LEP analysis explicitly addressed the case of a light unflavoured jet final state for pseudoscalar masses above 10\,GeV; the LEP collaborations were not in a position to make explicit statements covering all possible non-standard decay topologies~\cite{Chang:2008cw}. This opens the possibility, theoretically well motivated within the Next-to-Minimal Supersymmetric Standard Model (NMSSM), of a lighter CP-even or CP-odd Higgs boson coexisting alongside the observed 125\,GeV state.

Intriguingly, the LEP data themselves contain a hint pointing in this direction: the combined results from the four LEP experiments (ALEPH, DELPHI, L3 and OPAL)~\cite{200361} showed a $2.3\sigma$ local excess in $e^+e^- \to Zh$, $h \to b\bar{b}$ events at approximately 98\,GeV, where $h$ denotes a light neutral Higgs boson, already visible in Fig.~7 of Ref.~\cite{200361} as a deviation of the observed confidence level from the background expectation, consistent with an NMSSM singlet-like scalar with suppressed coupling to the $Z$ boson, as later analysed in Ref.~\cite{Belanger:2012tt}. This excess was never sufficient for a claim of observation, but has motivated further searches.

The LHC experiments have taken up the challenge of searching for such a light scalar below 125\,GeV. CMS reported a $2.9\sigma$ local excess in the diphoton final state at approximately 95.4\,GeV using the full Run-2 dataset at 13\,TeV, while ATLAS observed a compatible excess of $1.7\sigma$ at the same mass; combining the two experiments yields a joint significance of approximately $3.1\sigma$~\cite{Biekoetter:2023jld}. Simultaneous excesses in the $b\bar{b}$ and $\tau\tau$ final states at comparable masses further motivate the interpretation as a singlet-like scalar in NMSSM-type models. No confirmed observation has been made: the excesses remain statistically suggestive but fall short of the $5\sigma$ discovery threshold, and analyses of Run-3 data will be decisive. Whether the ${\sim}95$\,GeV region harbours a new scalar, and whether it is connected to the long-standing LEP excess at ${\sim}98$\,GeV first seen in Ref.~\cite{200361}, is one of the most concrete open questions in BSM Higgs physics, directly bridging the LEP legacy and the ongoing LHC programme.

\section{Tevatron Experimental Searches}
\subsection{Tevatron Neutral Higgs Boson Results}
\begin{figure}[h!]
\vspace{-1mm}
\centering
\includegraphics[width=0.8\columnwidth]{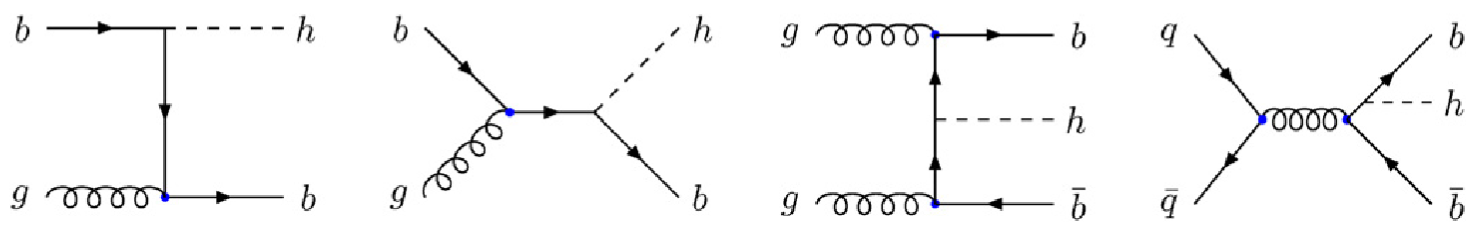}\\
\includegraphics[width=0.65\columnwidth]{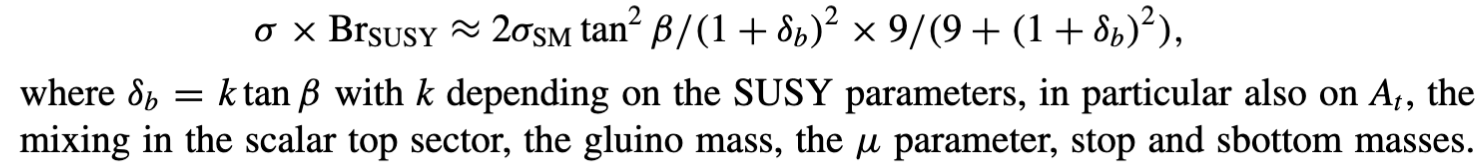}
\includegraphics[width=0.32\columnwidth]{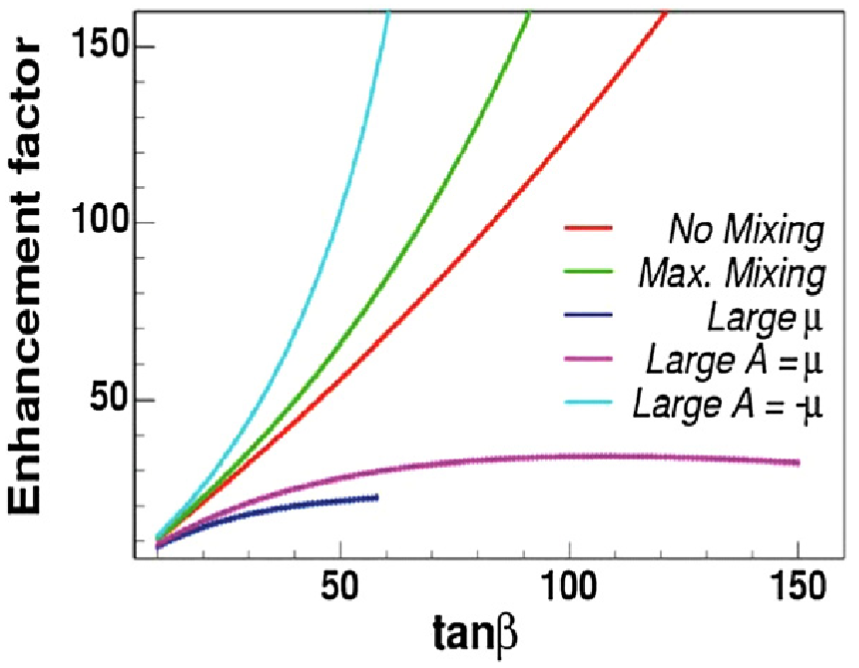}
\caption{Tevatron Feynman diagram of associated Higgs boson with b-quark production, and enhancement factor of the production as a function of $\tan\beta$. From~\cite{sopczak2009statushiggsbosonsearches}.}
\label{fig:tevatronbbh}
\end{figure}

\begin{figure}[h!]
\vspace{-4mm}
\centering
\includegraphics[width=0.49\columnwidth]{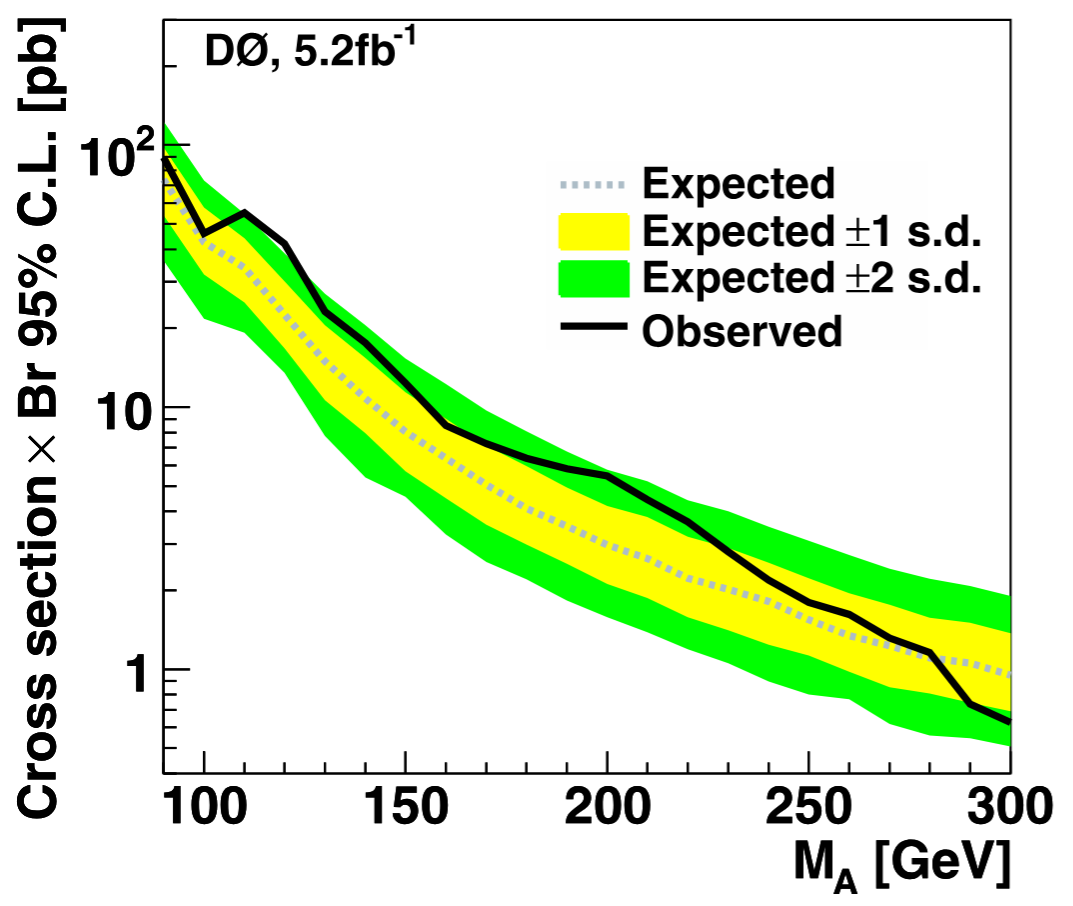}
\includegraphics[width=0.49\columnwidth]{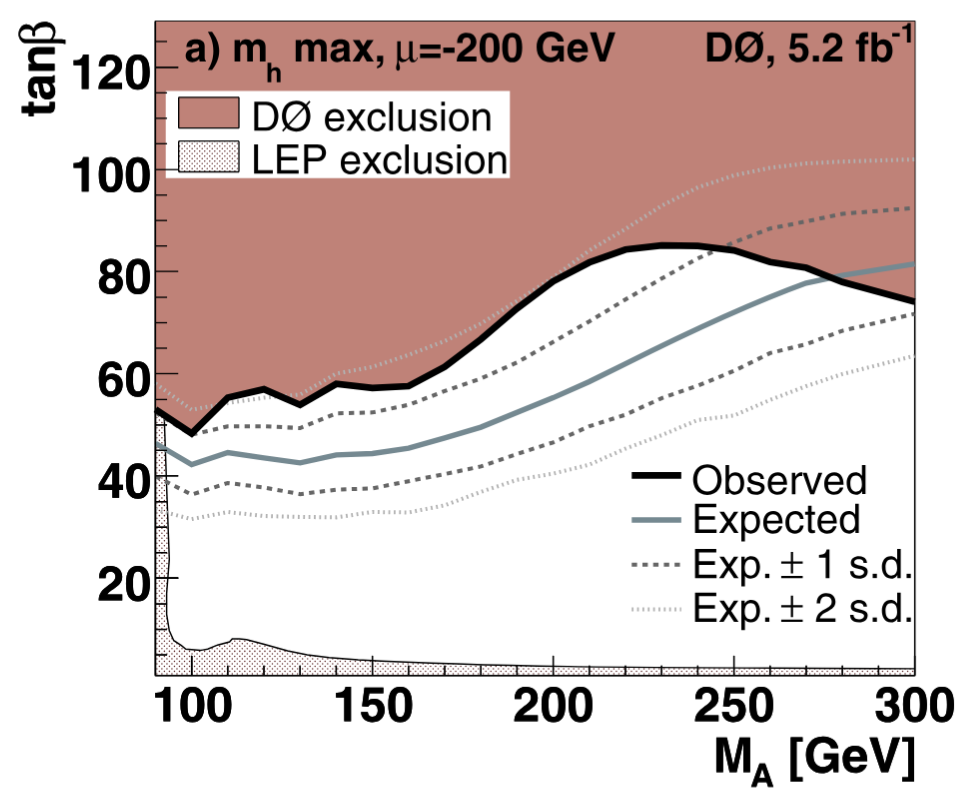}
\caption{Tevatron 95\% CL upper limits on $\sigma(p\bar p\to b\phi+X)\times{\rm BR}(\phi\to b\bar b)$ as a function of the pseudoscalar Higgs boson mass $m_{\rm A}$ (left), and excluded region in the $(m_{\rm A},\tan\beta)$ plane of the MSSM $m_h^{\rm max}$ scenario with $\mu=-200$\,GeV (right). From~\cite{abazov2011search}.}
\label{fig:tevatronbbhlimits}
\vspace{-2mm}
\end{figure}

\begin{figure}[h!]
\centering
\includegraphics[width=0.49\columnwidth]{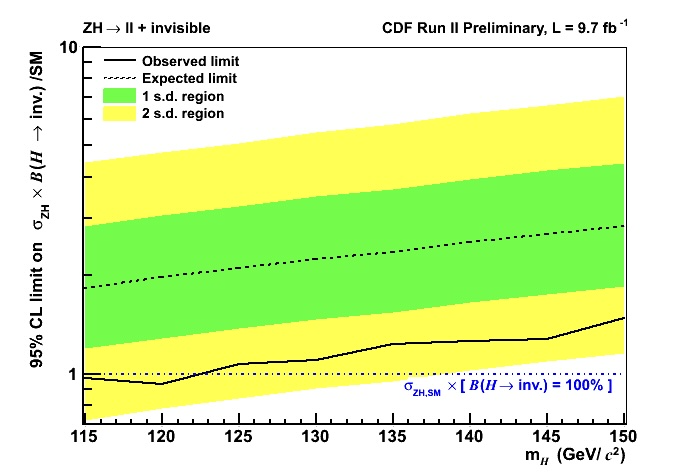}
\caption{Tevatron 95\% CL upper limits on $\sigma_{ZH}\times{\rm BR}(H\to{\rm inv})$, normalised to the SM $ZH$ cross-section, as a function of the Higgs boson mass $m_{\rm H}$.
From~\cite{Principato:2016qmv}.}
\label{fig:CDFinv}
\vspace{-2mm}
\end{figure}

At the Tevatron, experimental searches for BSM Higgs bosons were performed in association with a $b$-quark. Figure~\ref{fig:tevatronbbh} shows the corresponding Feynman diagrams, as well as the enhancement factor of the production as a function of $\tan\beta$.
The results of the searches are expressed as 95\% CL upper limits on the product of the production cross-section and the branching ratio, $\sigma(p\bar p\to b\phi+X)\times{\rm BR}(\phi\to b\bar b)$ with $\phi = h, H, A$, as a function of the pseudoscalar Higgs boson mass $m_{\rm A}$ in the range 90--300\,GeV (Fig.~\ref{fig:tevatronbbhlimits}, left). Interpreted in the MSSM $m_h^{\rm max}$ benchmark scenario with $\mu=-200$\,GeV, these limits exclude values of $\tan\beta$ above about 50 for $m_{\rm A}=100$\,GeV and above about 85 for $m_{\rm A}\approx 250$\,GeV (Fig.~\ref{fig:tevatronbbhlimits}, right), extending the region excluded by LEP at low $\tan\beta$. Presenting the limits on $\sigma\times{\rm BR}$ as a function of the Higgs boson mass has the advantage that they are largely model-independent: they can be compared directly with the predictions of any model, e.g.\ the MSSM, general two-Higgs-doublet models or singlet extensions, whereas an excluded region in the $(m_{\rm A},\tan\beta)$ plane is specific to the chosen benchmark scenario. The same applies to the limits shown in Figs.~\ref{fig:CDFinv} and~\ref{fig:tevatronhpluslimits}. The enhancement shown in Fig.~\ref{fig:tevatronbbh} makes clear that the Tevatron sensitivity was driven not by extending the accessible mass range, but by amplifying specific couplings at large $\tan\beta$. This represents a qualitatively different strategy from LEP, where sensitivity relied primarily on kinematic accessibility combined with coupling complementarity.

At the Tevatron, the search for a neutral Higgs boson decaying into invisible particles was performed. The result from CDF~\cite{Principato:2016qmv} is shown in Fig.~\ref{fig:CDFinv}. The limits extend the LEP mass range from 115\,GeV to 150\,GeV. The limits are set on $\sigma_{ZH}\times{\rm BR}(H\to{\rm inv})$ relative to the SM $ZH$ cross-section. The observed limit is close to unity between 115 and 125\,GeV and rises to about 1.5 at 150\,GeV. Thus, a Higgs boson produced at the SM $ZH$ rate and decaying 100\% invisibly is at the edge of the sensitivity for the lower masses, while smaller invisible branching ratios could not be probed at the Tevatron.

\subsection{Tevatron Charged Higgs Boson Results}
\begin{figure}[h!]
\vspace{-2mm}
\centering
\vspace{-3mm}
\includegraphics[width=0.49\columnwidth]{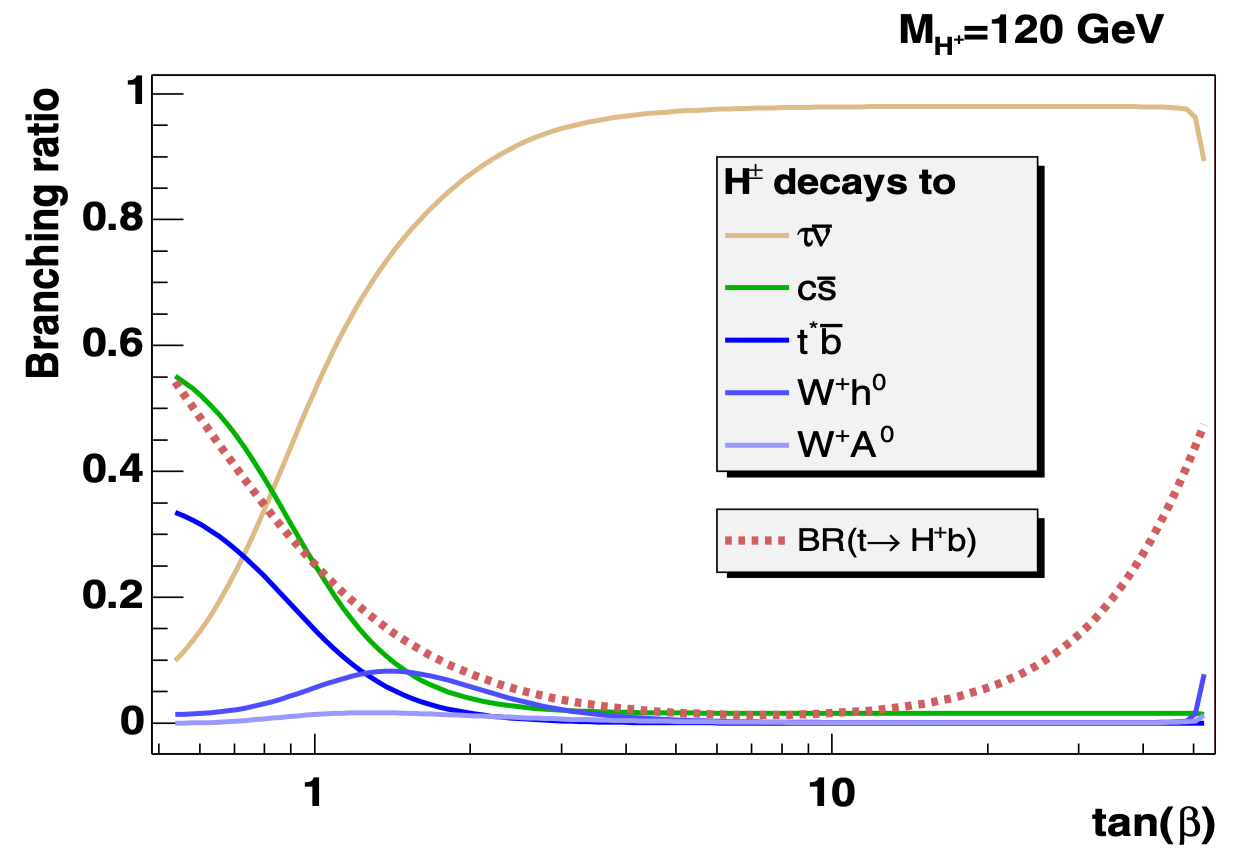}
\caption{Branching ratios of $t$-quark and $H^\pm$ as a function of $\tan\beta$. From~\cite{Sopczak:2824511}.}
\label{fig:tevatronhplus}
\end{figure}

\begin{figure}[h!]
\vspace{-10mm}
\centering
\includegraphics[width=0.49\columnwidth]{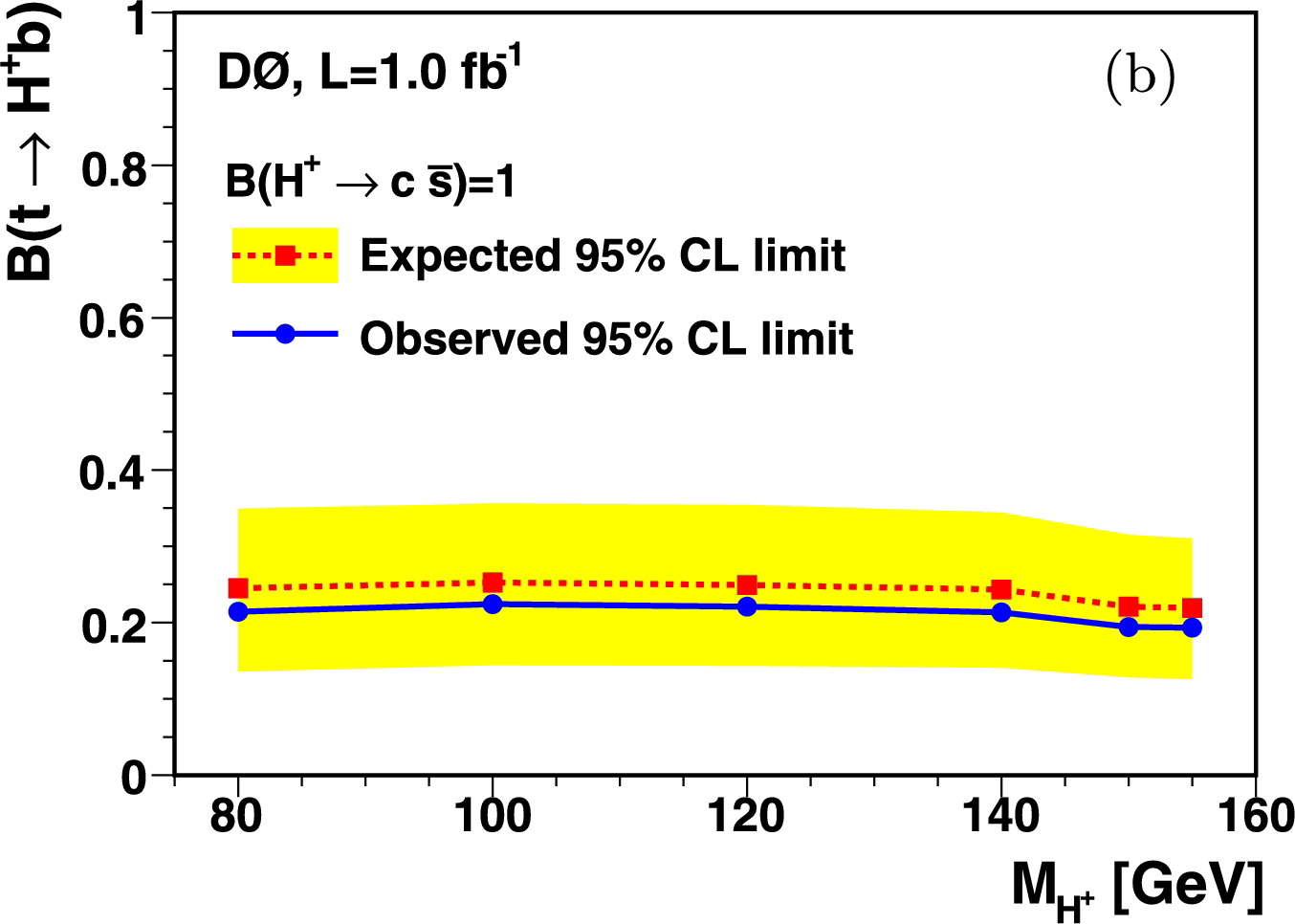}
\includegraphics[width=0.49\columnwidth]{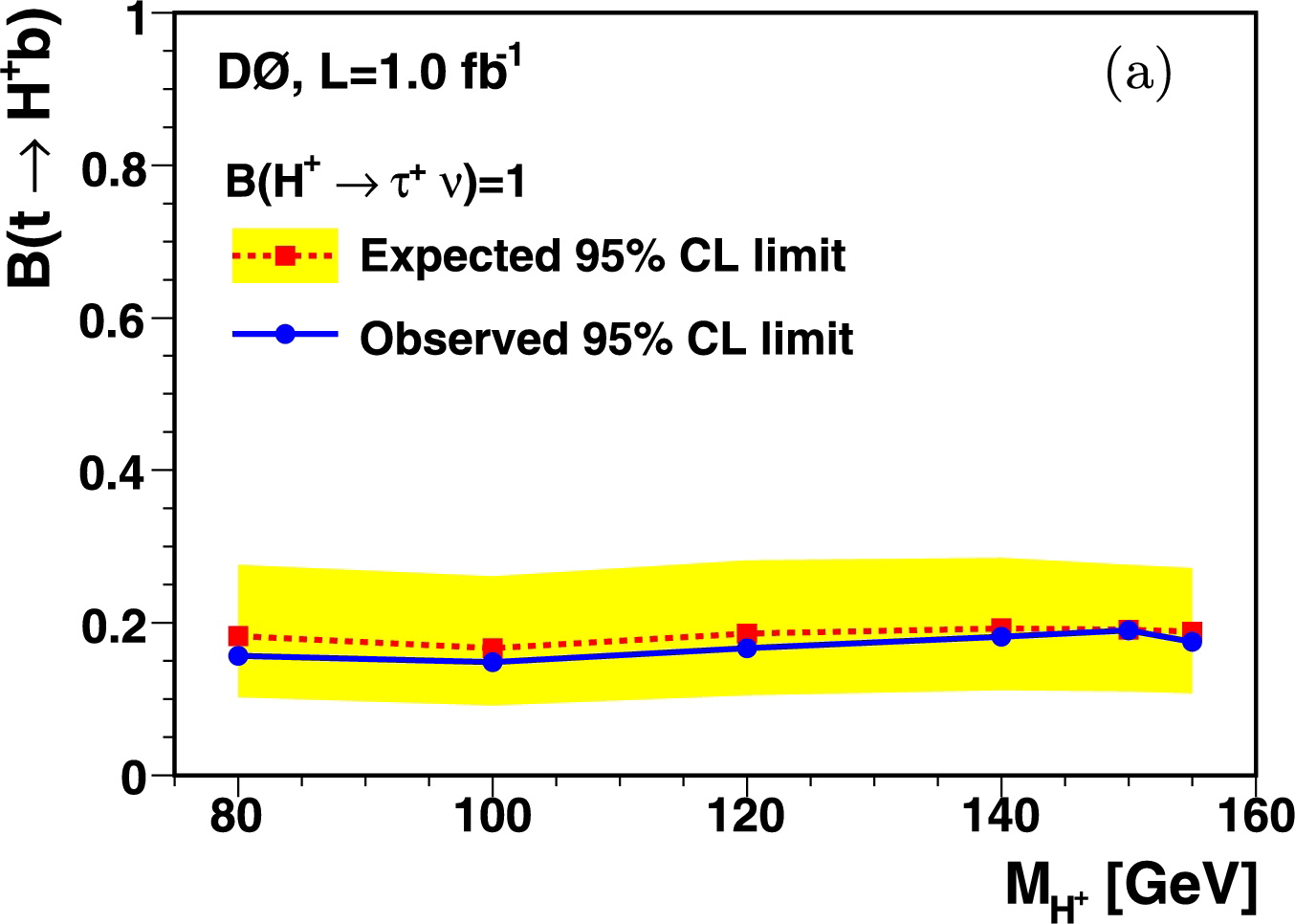}
\caption{Tevatron 95\% CL upper limits on $\mathcal{B}(t\to H^+b)$ as a function of $m_{H^\pm}$.
Left: $\mathcal{B}(H^+ \to c\bar{s}) = 1$.
Right: $\mathcal{B}(H^+ \to \tau\nu) = 1$.
From~\cite{abazov2009search}.}
\label{fig:tevatronhpluslimits}
\end{figure}

\begin{figure}[tp]
\centering
\includegraphics[width=0.49\columnwidth]{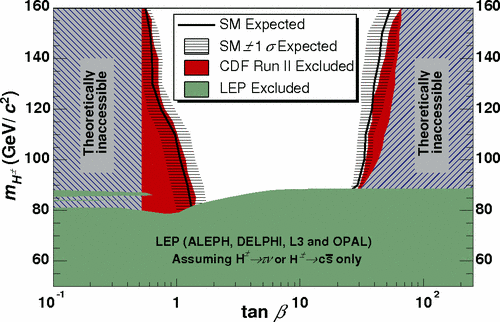}
\vspace{-2mm}
\caption{Tevatron limits in the $(\tan\beta, m_{\rm H^\pm})$ plane at 95\% CL.
From~\cite{PhysRevLett.96.042003}.}
\label{fig:tevatronhpluslimits2}
\end{figure}

\begin{figure}[tp]
\centering
\includegraphics[width=0.49\columnwidth]{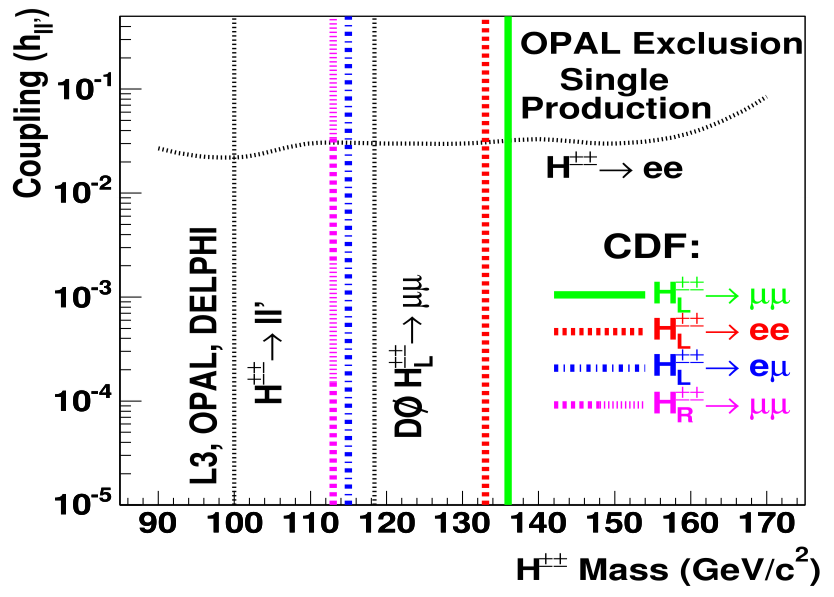}
\includegraphics[width=0.49\columnwidth]{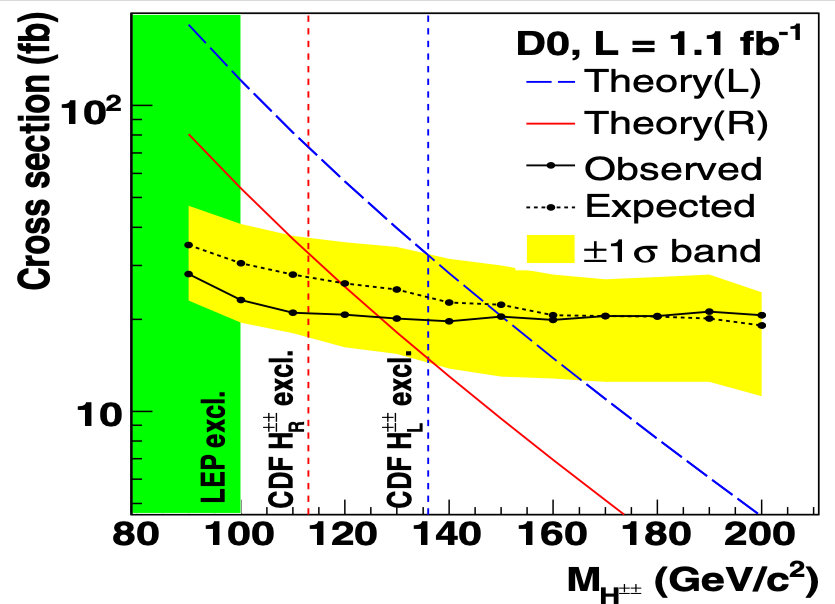}
\vspace{-2mm}
\caption{Tevatron limits on doubly charged Higgs bosons as a function of the $m_{\rm H^{\pm\pm}}$ mass at 95\% CL. From~\cite{Acosta_2005} and~\cite{Abazov_2008}.}
\label{fig:tevatronhplusplus}
\vspace{-2mm}
\end{figure}

At the Tevatron, the search for charged Higgs bosons was performed in the decay of $t$-quarks. For small and large $\tan\beta$ values an enhanced decay branching ratio is expected, as shown in Fig.~\ref{fig:tevatronhplus}.
The figure also shows the different $H^{\pm}$ decay modes as a function of $\tan\beta$. These lead to different final-state compositions compared with the SM decay $t\rightarrow Wb$. Measurements were consistent with the SM expectation and limits were set on the $t$-quark decay branching ratio and in the $(m_{\rm H^\pm},\tan\beta)$ plane at 95\% CL (Figs.~\ref{fig:tevatronhpluslimits} and~\ref{fig:tevatronhpluslimits2}). Assuming $\mathcal{B}(H^+\to c\bar{s})=1$, the D\O\ data exclude $\mathcal{B}(t\to H^+b)$ above about 0.2, and assuming $\mathcal{B}(H^+\to\tau\nu)=1$ above about 0.15--0.2, for charged Higgs boson masses between 80 and 155\,GeV (Fig.~\ref{fig:tevatronhpluslimits}). Interpreted in the MSSM (Fig.~\ref{fig:tevatronhpluslimits2}), these limits exclude the regions of small ($\tan\beta\lesssim 1$) and large ($\tan\beta\gtrsim 40$--60) values of $\tan\beta$, where $\mathcal{B}(t\to H^+b)$ is enhanced (Fig.~\ref{fig:tevatronhplus}), for $m_{H^\pm}$ up to about 160\,GeV. The intermediate $\tan\beta$ region, where $\mathcal{B}(t\to H^+b)$ is small, remains unconstrained by these searches, while LEP excludes $m_{H^\pm}\lesssim 80$\,GeV independently of $\tan\beta$. 
Doubly-charged Higgs bosons lead to a clear signature of lepton pairs with the same electric charge. The resulting limits from the searches are shown in Fig.~\ref{fig:tevatronhplusplus}. Depending on the lepton flavours and on the chirality of the coupling ($H_L^{\pm\pm}$ or $H_R^{\pm\pm}$), doubly-charged Higgs bosons are excluded for masses below about 110--150\,GeV; for example, D\O\ excludes $H_L^{\pm\pm}$ ($H_R^{\pm\pm}$) decaying exclusively into muon pairs below about 150 (127)\,GeV, well beyond the LEP limit of about 100\,GeV.

\subsection{Discussion of Tevatron Results}

The Tevatron served as a bridge between the LEP era and the start of the LHC. For a SM Higgs boson with a mass of 125\,GeV, the highest sensitivity of the Tevatron experiments, CDF and D\O, was achieved in the
$VH\rightarrow Vb\bar b$ channel. 
This channel directly probes the Higgs boson coupling to bottom quarks. While the expected SM production rates were not sufficient for a Higgs boson observation at the Tevatron~\cite{PhysRevD.88.052014}, a variety of searches for BSM  Higgs bosons were performed.

These searches were motivated by the possibility that the $Hb\bar b$ coupling could be enhanced in BSM scenarios. In particular, large enhancement factors were expected for large values of $\tan\beta$. 
Higgs boson mass ranges between approximately 90\,GeV and 300\,GeV were probed. In addition, the Tevatron experiments conducted searches for invisibly decaying Higgs bosons. These analyses focused on leptonic decays of the Z boson and extended the mass range explored previously at LEP.

The top quark was studied in great detail at the Tevatron due to its high production rate. In the SM, the top quark decays almost exclusively into a W boson and a bottom quark, with the subsequent W-boson decays leading to well-known final states. Any deviation from the SM branching fractions would therefore provide a clear indication of BSM physics in top-quark decays. In particular, the W boson could be replaced by a charged Higgs boson, modifying the top-quark decay branching ratios. Searches for such effects represented a novel aspect of the Tevatron physics programme, as the LEP collider did not have sufficient centre-of-mass energy to produce top quarks.

\subsubsection*{Could the Tevatron have discovered the 125\,GeV Higgs boson?}

The question deserves a direct answer. As reviewed at the beginning of the LHC era~\cite{Sopczak:2824511}, the Tevatron experiments CDF and D\O\ achieved their highest sensitivity at $m_H = 125$\,GeV in associated production $q\bar{q} \to VH \to Vb\bar{b}$, which directly probes the Higgs coupling to bottom quarks. With approximately 10\,fb$^{-1}$ per experiment collected before the Tevatron shutdown in September 2011, the expected combined sensitivity at $m_H = 125$\,GeV was approximately $1.9\sigma$~\cite{PhysRevD.88.052014}. In their final combined result, CDF and D\O\ reported in 2012, around the time of the LHC Higgs boson discovery, a local excess of $3.0\sigma$ at $m_H = 125$\,GeV~\cite{PhysRevD.88.052014}. This falls short of the $5\sigma$ threshold required for a discovery claim, but represents the $3\sigma$ evidence level. To reach $5\sigma$ at this mass in the $VH \to Vb\bar{b}$ channel would have required roughly four times the collected luminosity based on the projected sensitivity~\cite{Haber:2010bq}---approximately 40\,fb$^{-1}$ per experiment---which was not achievable within the operational timeline of the Tevatron. The conclusion is therefore clear: the Tevatron could not have made a five-sigma discovery, but it did achieve about the three-sigma evidence level that the available luminosity permitted.

This is itself a remarkable result. The $VH \to Vb\bar{b}$ channel probed by the Tevatron was inaccessible to the LHC at early luminosities due to the much larger QCD multi-jet background at higher centre-of-mass energies, and the Tevatron evidence in the fermionic decay mode complemented the LHC discovery, which was established primarily through bosonic channels ($H \to \gamma\gamma$, $H \to ZZ^* \to 4\ell$). Although the Tevatron reached the evidence level, it would not have been possible to establish a discovery for a 125\,GeV SM Higgs boson.

Overall, the Tevatron analyses provided valuable experience and methodological groundwork that proved essential for the subsequent Higgs boson searches and measurements at the LHC~\cite{aglietti2007tevatronforlhcreporthiggs}.

\section{LHC and HL-LHC Experimental \\ Searches}
The LHC experiments at Run-1 collected data in 2010 and 2012. 
At the beginning of Run-1, the status of Higgs boson searches in the SM and BSM was summarised in a review~\cite{Sopczak:2824511}.

\subsection{LHC and HL-LHC Neutral Higgs Boson Results}
\begin{figure}[h!]
\vspace{-5mm}
\centering
\begin{minipage}[c]{0.63\columnwidth}\includegraphics[width=\linewidth]{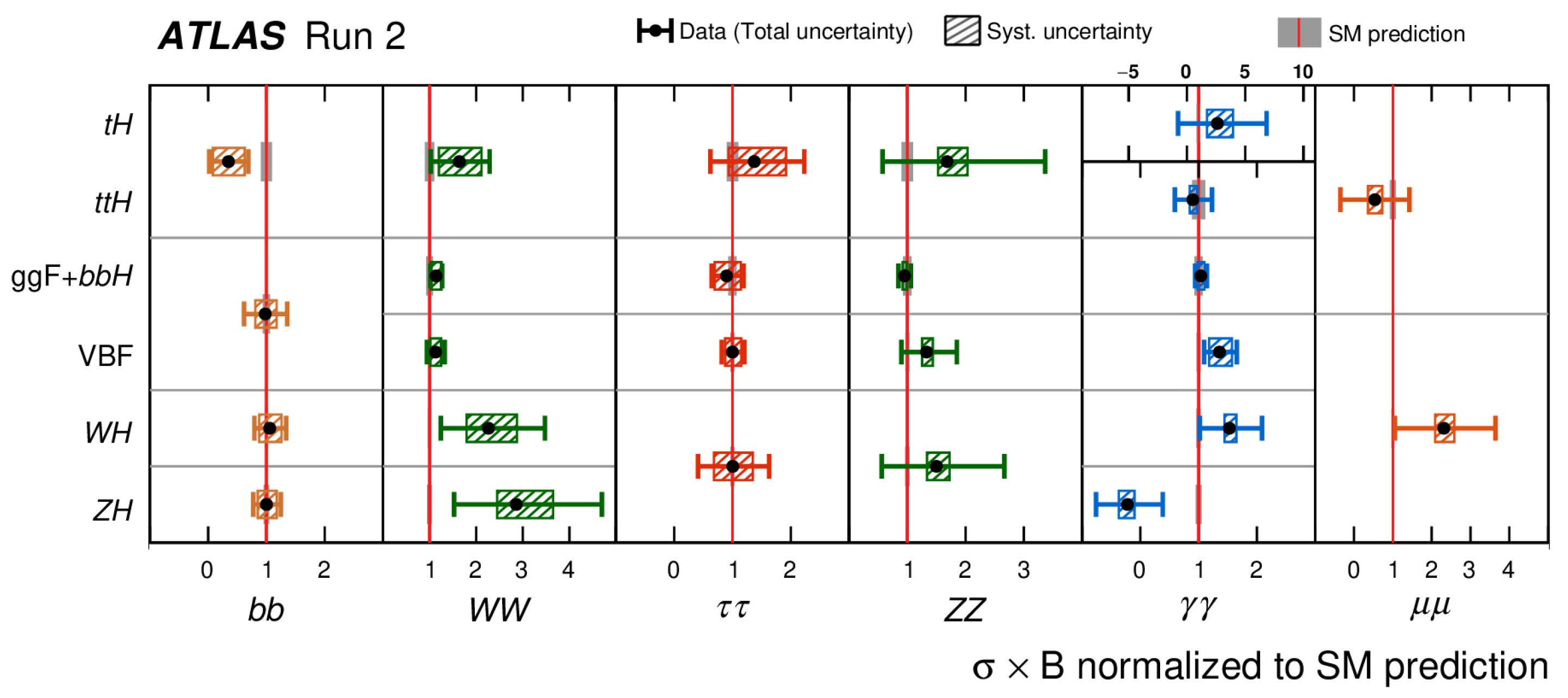}\end{minipage}\hfill
\begin{minipage}[c]{0.35\columnwidth}\includegraphics[width=\linewidth]{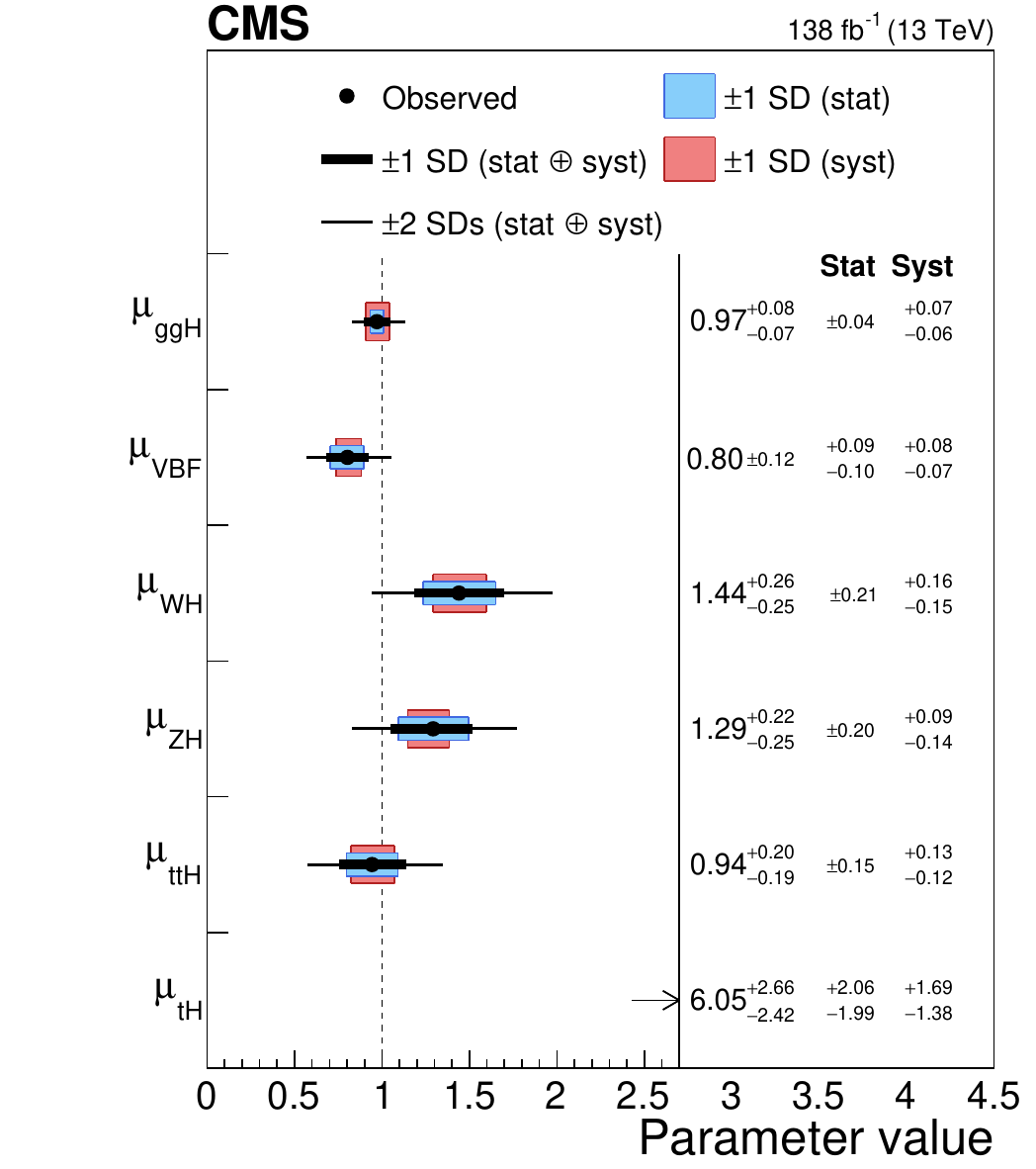}\end{minipage}
\caption{Left: ATLAS measurements of the production of the Higgs boson in various modes and their decays compared to the SM expectation. Right: CMS signal strengths for the individual production modes, assuming SM decay branching fractions. From~\cite{ATLAS:2022vkf} and~\cite{Tumasyan:2814513}.}
\label{fig:atlascouplings}
\end{figure}

Ten years after the discovery of the Higgs boson, the \mbox{ATLAS} and CMS collaborations reviewed the knowledge about the Higgs boson properties~\cite{ATLAS:2022vkf,Tumasyan:2814513}. These reviews gave detailed information about the production of the Higgs boson in the various modes and their decays, compared to the SM expectation. There is no indication for BSM deviations, as shown in Fig.~\ref{fig:atlascouplings}. Both experiments reach the same conclusion: ATLAS measures a global signal strength of $\mu = 1.05 \pm 0.06$ and CMS $\mu = 1.002 \pm 0.057$, and the signal strengths for the individual production modes (ggH, VBF, $WH$, $ZH$, $t\bar tH$ and $tH$) and decay channels are all consistent with the SM expectation. The important feature of Fig.~\ref{fig:atlascouplings} is that it combines multiple production and decay channels into a global test of Higgs couplings. The agreement with the Standard Model therefore reflects the consistency of a wide network of independent measurements, rather than the precision of any single observable.

\begin{figure}[h!]
\centering
\vspace{-12mm}
\includegraphics[width=0.49\columnwidth]{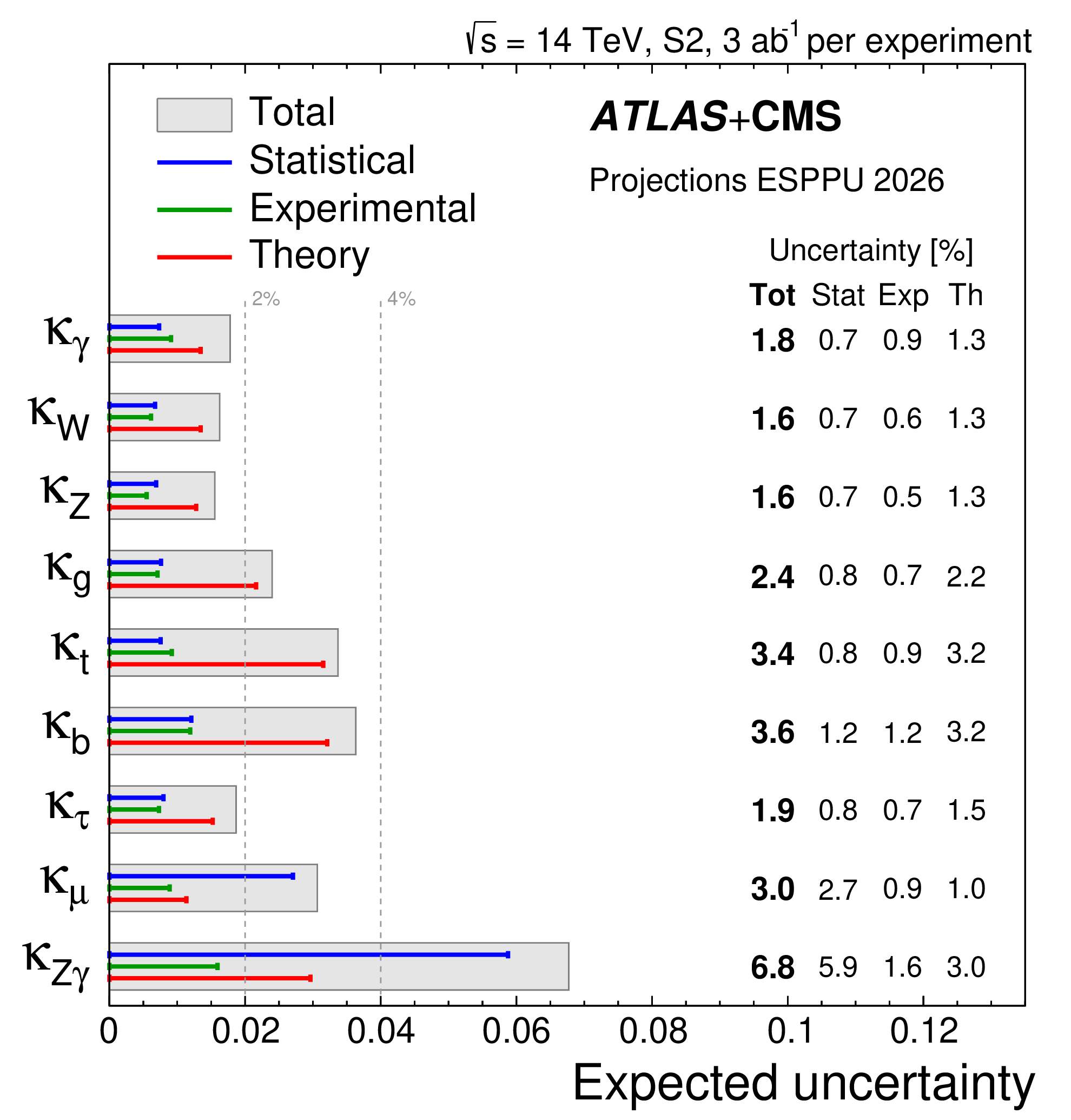}
\includegraphics[width=0.49\columnwidth]{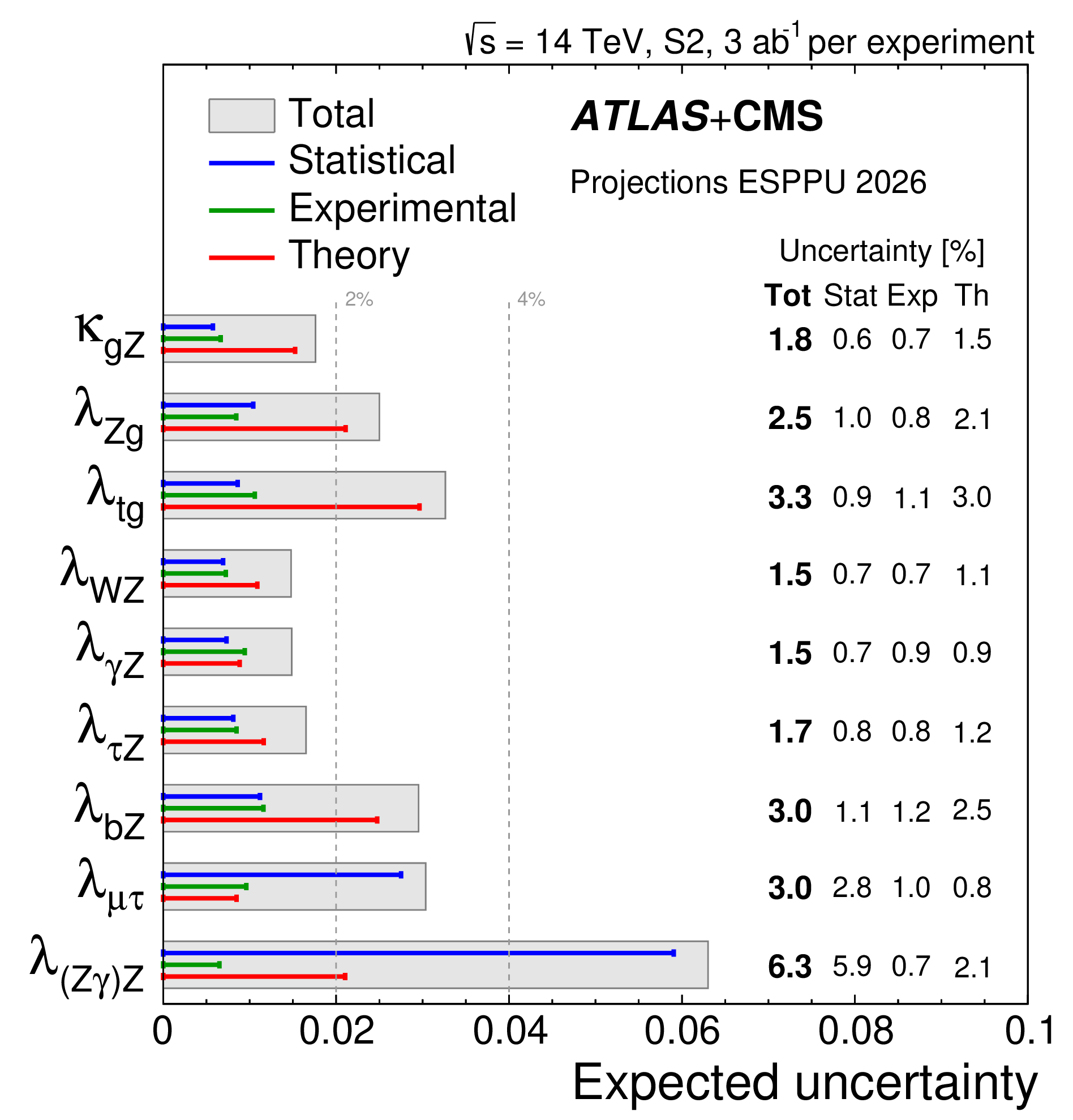}
\caption{HL-LHC predictions of the Higgs boson coupling to fermions and bosons in the $\kappa$ and $\lambda$ framework. 
From~\cite{atlas2025highlightshllhcphysicsprojections}.
}
\label{fig:atlaskappalambda}
\vspace{-2mm}
\end{figure}

A summary of the 
main measured Higgs boson couplings to fermions
and bosons is given in Fig.~\ref{fig:atlascouplings}. Using the coupling modifier $\kappa$ and self-coupling $\lambda$ framework~\cite{atlas2025highlightshllhcphysicsprojections}, Fig.~\ref{fig:atlaskappalambda} shows the projections for the HL-LHC. With 3\,ab$^{-1}$ per experiment, the coupling modifiers to $W$ and $Z$ bosons, photons and $\tau$ leptons are expected to be determined with a total uncertainty of 1.6--1.9\%, those to gluons, muons, top and bottom quarks with 2.4--3.6\%, and the rare $Z\gamma$ coupling with about 7\%. These and the other HL-LHC projections shown in this section (Figs.~\ref{fig:HHdevelopment}, \ref{fig:HHkappa}, \ref{fig:selfcoupling} and~\ref{fig:stability}) are combined ATLAS and CMS results prepared for the 2026 update of the European Strategy for Particle Physics~\cite{atlas2025highlightshllhcphysicsprojections}. Note that the theory uncertainty is mostly dominant~\cite{Sopczak:2025ptc}.

It is important to be precise about what these ``theory uncertainties'' are, since they are not, in general, a consequence of our inability to evaluate higher-order Feynman diagrams. The dominant sources are of a different character. First, the uncertainties on parton distribution functions (PDFs) and on $\alpha_s$ contribute at the few-percent level to Higgs production cross-sections at the LHC; these are parametric uncertainties constrained by global QCD analyses of experimental data~\cite{xsec}, and reducing them requires better measurements of hard-scattering processes, not more perturbative calculations. Second, the $\kappa$-framework itself introduces a source of theory uncertainty through its simplifying assumptions~\cite{xsec}: it presupposes that the observed Higgs boson has exactly the Lorentz structure of the SM, that no new particles contribute in production or decay loops, and that off-shell effects are negligible. In reality, BSM contributions to loop-induced couplings ($\kappa_g$, $\kappa_\gamma$) can mimic or partially cancel SM contributions in ways not resolved within the $\kappa$ parametrisation. Third, for the top-quark Yukawa coupling, the choice of renormalisation scheme for the top-quark mass introduces a non-negligible theory uncertainty that is not simply reduced by computing additional loops. These considerations suggest that progress on the theory frontier at the HL-LHC requires advances in PDF determinations, development of Standard Model Effective Field Theory (SMEFT) frameworks~\cite{Brivio:2017vri} beyond the $\kappa$ approach, and improved treatment of scheme and matching uncertainties. Future lepton colliders such as FCC-ee, operating in a cleaner environment with precisely known initial-state kinematics, would circumvent several of these hadronic uncertainties entirely, providing one of the primary physics motivations for precision Higgs studies at such facilities~\cite{CERN-2025-005}. These uncertainties therefore reflect limitations of the experimental and theoretical framework, not calculational precision alone.

\begin{figure}[tbp]
\centering
\includegraphics[width=0.40\columnwidth]{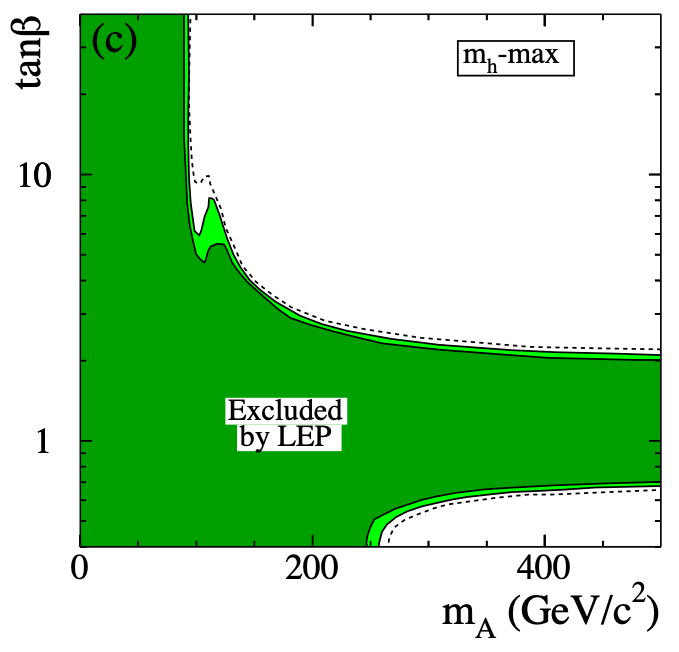}\\[1mm]
\includegraphics[width=0.72\columnwidth]{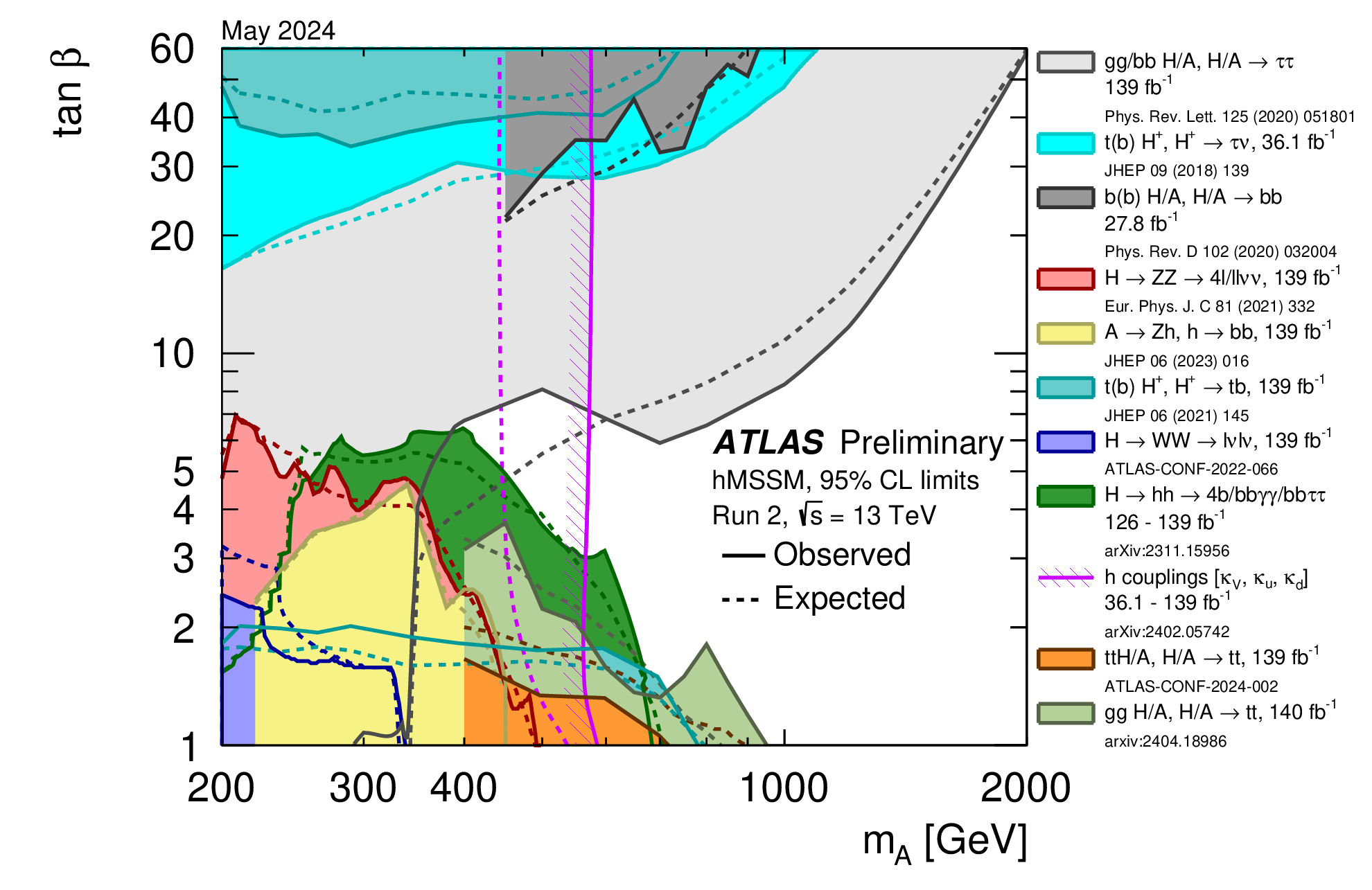}\\[1mm]
\includegraphics[width=0.42\columnwidth]{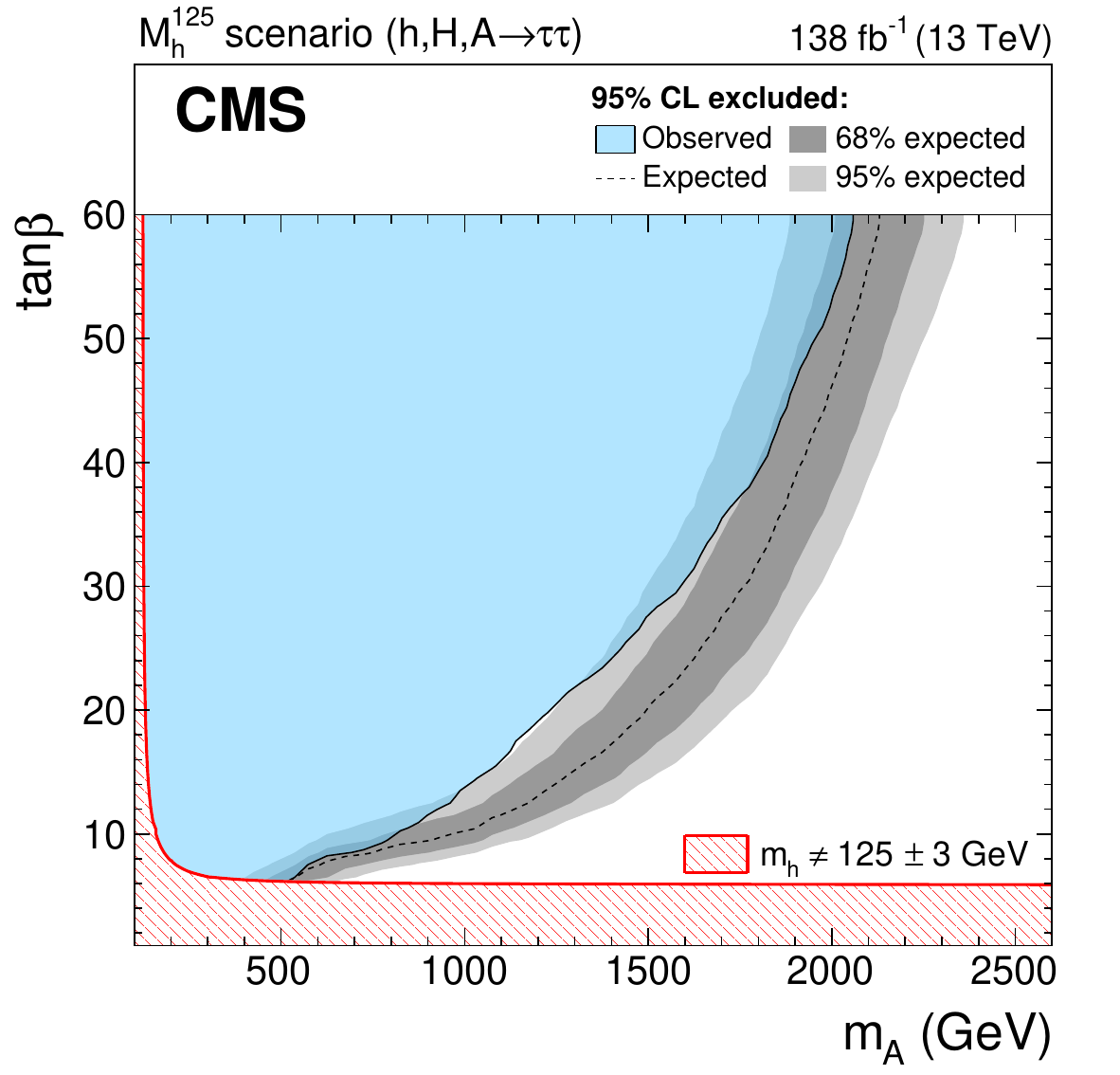}
\vspace{-1mm}
\caption{Progress from LEP to the LHC in BSM Higgs boson searches, shown in the MSSM interpretation.
Upper: LEP exclusion in the $(m_A, \tan\beta)$ plane of the $m_h^{\rm max}$ scenario ($m_A = 0$--$500$\,GeV);
middle: ATLAS exclusion in the $(m_A, \tan\beta)$ plane of the hMSSM scenario ($m_A = 200$--$2000$\,GeV, a qualitatively different mass range);
lower: CMS exclusion from the $h,H,A\to\tau\tau$ searches in the $(m_A, \tan\beta)$ plane of the $M_h^{125}$ scenario ($m_A$ up to about 2.6\,TeV); the hatched region is incompatible with $m_h = 125 \pm 3$\,GeV.
Note the different $\tan\beta$ ranges: $0.4$--$40$ (upper) and up to 60 (middle and lower).
The $x$-axis scales differ substantially, reflecting the very different mass regimes probed by the two colliders.
From~\cite{sopczak2006mssmhiggsbosonsearches,ATL-PHYS-PUB-2024-008,CMS:2022goy}.}
\label{fig:lepatlas}
\end{figure}

The progress from LEP to the LHC in BSM Higgs boson searches is shown in Fig.~\ref{fig:lepatlas} in the MSSM interpretation. The upper panel shows the LEP result in the $m_h^{\rm max}$ benchmark scenario: the region $m_{\rm A}\lesssim 90$\,GeV is excluded for all values of $\tan\beta$, and in addition a band of low $\tan\beta$ values (approximately $0.7<\tan\beta<2$) is excluded for all $m_{\rm A}$ shown, since there the predicted mass of the light CP-even Higgs boson $h$ lies below the LEP limit. The middle panel shows the LHC Run-2 exclusion obtained by ATLAS in the hMSSM scenario~\cite{ATL-PHYS-PUB-2024-008}, in which the mass of the light CP-even Higgs boson is fixed to the measured value of 125\,GeV. Different channels dominate in different regions: at large $\tan\beta$, the enhanced couplings to down-type fermions make $gg/b\bar b\to H/A\to\tau\tau$ the most sensitive channel, excluding $m_{\rm A}$ up to about 1.5--2\,TeV for $\tan\beta=60$, while at low $\tan\beta$ the searches for $A\to Zh$, $H\to hh$, $H\to WW$ and $H/A\to t\bar t$ provide the sensitivity. In addition, the measurements of the couplings of the 125\,GeV Higgs boson exclude $m_{\rm A}$ below about 500\,GeV for all $\tan\beta$ values shown, since the couplings of $h$ would deviate from the SM values for a light $A$. Comparable results were obtained by CMS (Fig.~\ref{fig:lepatlas}, lower): in the $M_h^{125}$ MSSM benchmark scenario, the $h,H,A\to\tau\tau$ searches alone exclude $\tan\beta$ values above about 8 for $m_{\rm A}\approx 500$\,GeV and $m_{\rm A}$ up to about 2\,TeV for $\tan\beta=60$, while low values of $\tan\beta$ are incompatible with the observed Higgs boson mass in this scenario; additional Higgs bosons with masses below 350\,GeV are excluded~\cite{CMS:2022goy}. The comparison of the LEP and LHC panels shows that the LHC extended the probed $m_{\rm A}$ range by about an order of magnitude and covers both the large- and small-$\tan\beta$ regions, leaving mainly a wedge of intermediate $\tan\beta$ values (about 5--10) at $m_{\rm A}$ above about 500\,GeV unexcluded. The use of different benchmark scenarios in the LEP and LHC panels reflects the evolution of the theoretical interpretation after the Higgs boson discovery. The main comparative conclusion is a change in the role of the searches: at LEP, the MSSM was constrained mainly through the non-observation of the light CP-even Higgs boson $h$, which excluded low $m_{\rm A}$ and low $\tan\beta$; at the LHC, with $h$ identified with the observed 125\,GeV state, the constraints arise from direct searches for the heavy states $H$, $A$ and $H^\pm$ and from the precision measurements of the $h$ couplings, which together push the allowed heavy Higgs boson masses to the TeV scale.

\begin{figure}[h!]
\centering
\begin{minipage}[c]{0.56\columnwidth}\includegraphics[width=\linewidth]{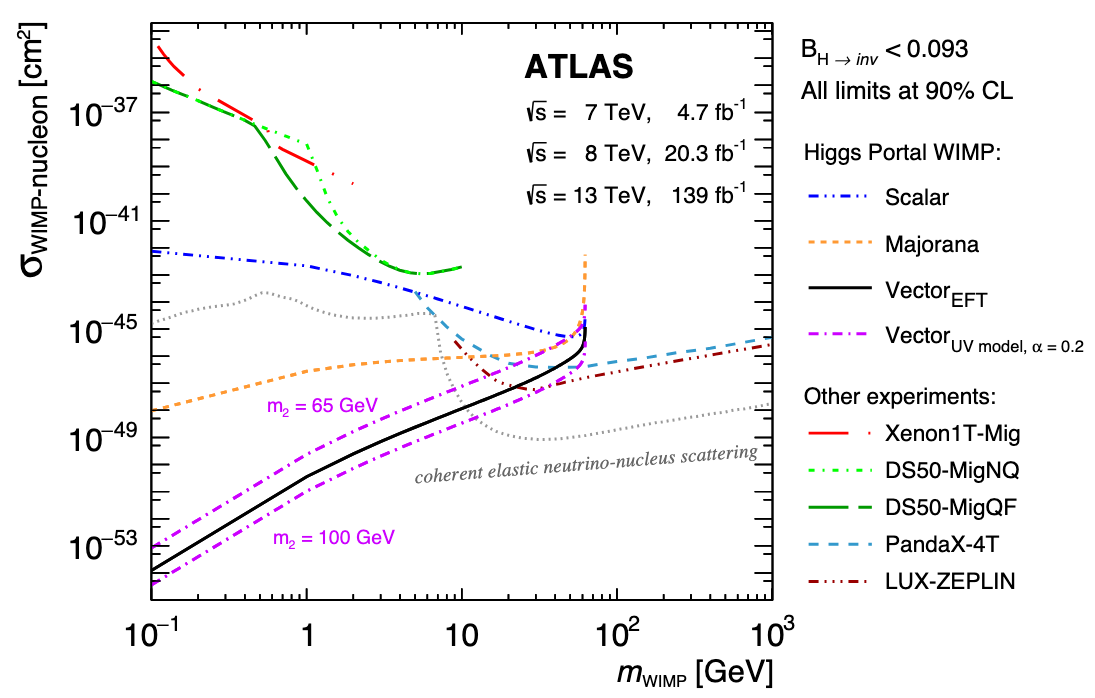}\end{minipage}\hfill
\begin{minipage}[c]{0.42\columnwidth}\includegraphics[width=\linewidth]{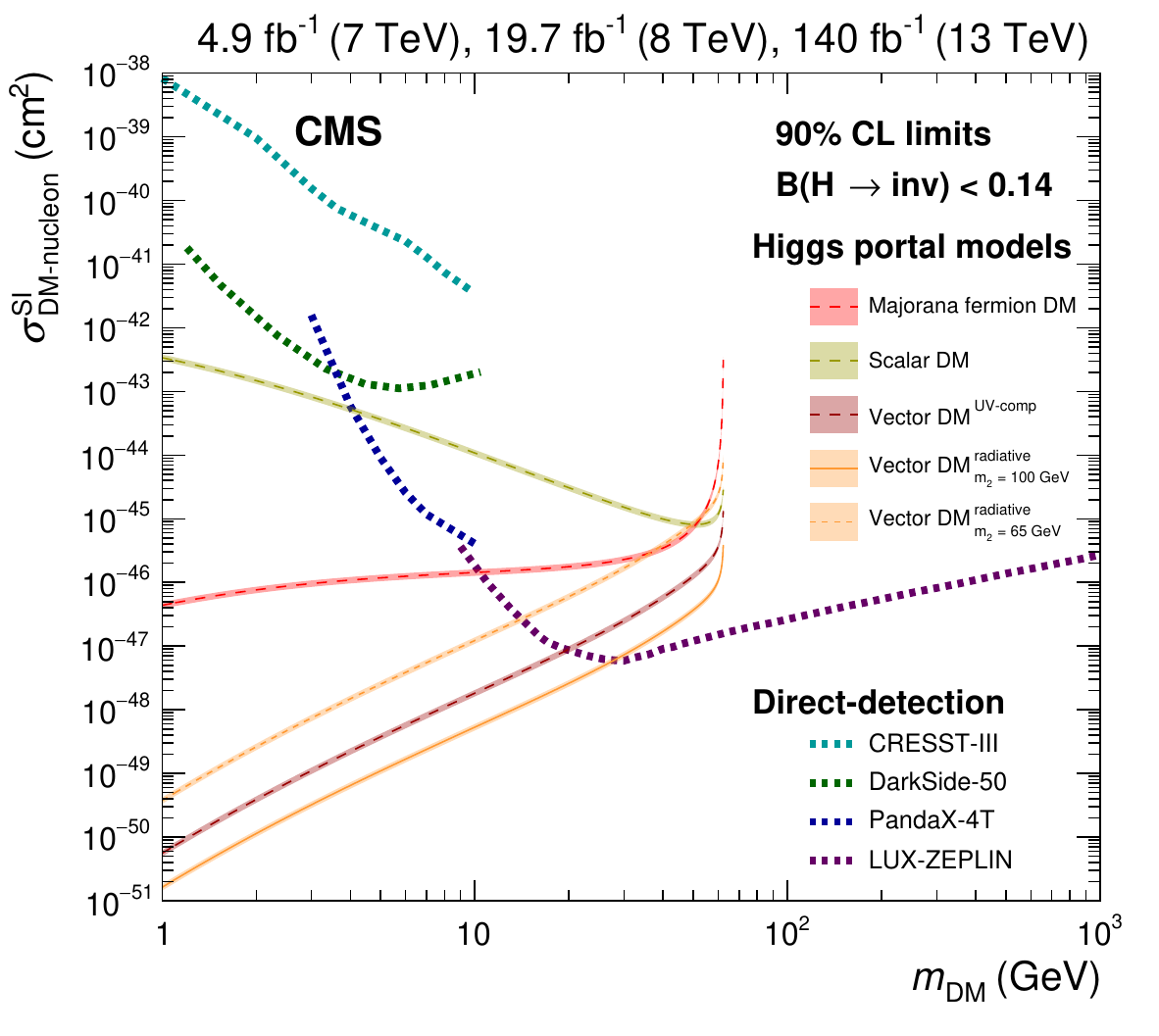}\end{minipage}
\caption{Upper limits on the spin-independent WIMP--nucleon scattering cross-section in Higgs-portal dark-matter models, derived from the searches for invisibly decaying Higgs bosons, compared with results from direct-detection experiments, as a function of the dark-matter mass. Left: ATLAS (90\% CL) for scalar, Majorana-fermion and vector dark matter; right: CMS for fermion, scalar and vector dark matter.
From~\cite{ATLAS:2023tkt,Tumasyan_2023}.
}
\label{fig:DM}
\end{figure}

The possibility that a Higgs boson decays invisibly was searched for both at LEP and the LHC. A summary is given in Table~\ref{tab:invleplhc}. 
The prospects for the HL-LHC are
$BR(H\rightarrow inv) < 0.038$ with 3\,ab$^{-1}$, as expected by CMS from vector-boson-fusion production~\cite{CMS-PAS-FTR-18-016} (see also Ref.~\cite{cepeda2019higgsphysicshllhchelhc}).
The invisible particle could be a candidate for a Weakly Interacting Massive Particle (WIMP), thus a candidate for Dark Matter (DM). Figure~\ref{fig:DM} shows limits at 90\% CL from the search for invisibly decaying Higgs bosons. In Higgs-portal models, in which the Higgs boson couples to a dark-matter particle with a mass below $m_H/2$, the ATLAS upper limit on $BR(H\to{\rm inv})$ (0.093 at 90\% CL) is translated into upper limits on the spin-independent WIMP--nucleon scattering cross-section, shown as a function of the WIMP mass for scalar, Majorana-fermion and vector dark matter~\cite{ATLAS:2023tkt}. For WIMP masses below a few GeV, where direct-detection experiments lose sensitivity, these collider limits are by far the most stringent, whereas for masses approaching $m_H/2$ the limits from direct-detection experiments become comparable or stronger. The two approaches are therefore complementary. CMS obtains similar constraints (Fig.~\ref{fig:DM}, right) from its combination of searches for invisible Higgs boson decays, with an observed (expected) upper limit of $BR(H\to{\rm inv}) < 0.15$ (0.08) at 95\% CL~\cite{Tumasyan_2023}, and also finds that the collider limits are the most stringent for light dark-matter candidates. Searches for invisibly decaying Higgs bosons probe a qualitatively different aspect of the Higgs sector, namely the possibility of couplings to hidden or dark sectors. The sensitivity of these searches is determined not by visible decay signatures but by recoil techniques and missing energy observables, adding an essential complementary probe of BSM physics.
\begin{table*}[h!]
\centering
\caption{Limits on the production of invisibly decaying Higgs bosons from past and current colliders.}
\label{tab:invleplhc}
\small
\begin{tabular}{p{2.2cm}p{2.5cm}p{2.7cm}p{2.6cm}p{1.2cm}}
\toprule
Collider &
Energy / dataset &
Method &
95\% CL limit on $BR(H\rightarrow \mathrm{inv})$ &
$H$ mass  \\
\midrule
LEP (ALEPH, DELPHI, L3, OPAL) &
$\sqrt{s}$ up to 209\,GeV ($\approx 2.5$\,fb$^{-1}$ total) &
$e^+e^- \rightarrow ZH$, with $Z \rightarrow \ell\ell$, $q\bar q$ and recoil against MET &
Excluded\newline $m_H < 114.4$\,GeV,\newline $BR = 100\%$~\cite{LEPHiggsWG:2001inv} &
$m_H \lesssim 114$\,GeV \\
LHC Run-1+2 (ATLAS) &
7, 8, 13\,TeV, up to 139\,fb$^{-1}$ &
VBF, ZH, WH, $t\bar tH$ with MET &
$BR(H\rightarrow \mathrm{inv}) < 0.107$~\cite{ATLAS:2023tkt} &
$m_H = 125$\,GeV \\
LHC Run-1+2 (CMS) &
7, 8, 13\,TeV, up to 138\,fb$^{-1}$ &
VBF, ZH, WH, $t\bar tH$ with MET &
$BR(H\rightarrow \mathrm{inv}) < 0.15$~\cite{Tumasyan_2023} &
$m_H = 125$\,GeV \\
\bottomrule
\end{tabular}
\end{table*}

\begin{figure}[h!]
\centering
\includegraphics[width=0.29\columnwidth]{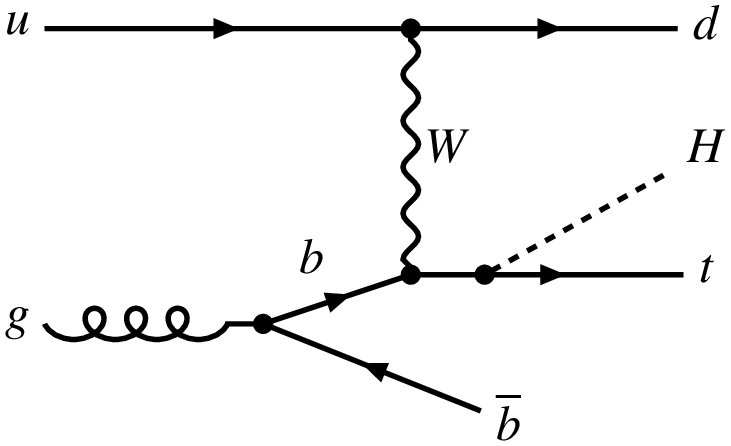}
\caption{LHC single top-Higgs production Feynman diagram.}
\label{fig:tbh}
\end{figure}

\begin{figure}[h!]
\centering
\includegraphics[width=0.49\columnwidth]{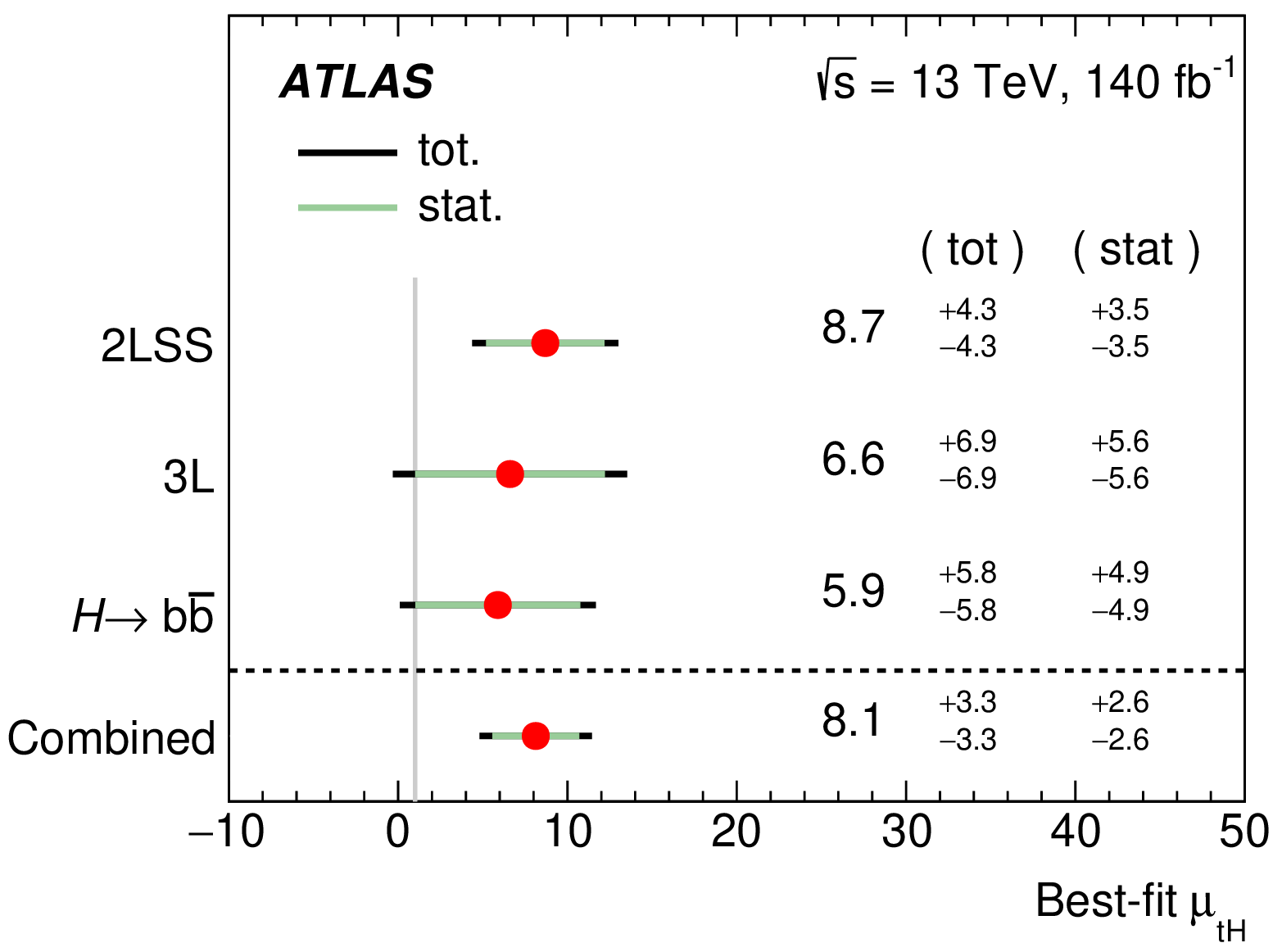}
\caption{LHC best fit on $\mu$(tH) for $\kappa_t = 1$.
From~\cite{ATLAS:2025irr}.
}
\label{fig:tbhlimitSM}
\end{figure}

\begin{figure}[h!]
\centering
\vspace{-3mm}
\includegraphics[width=0.49\columnwidth]{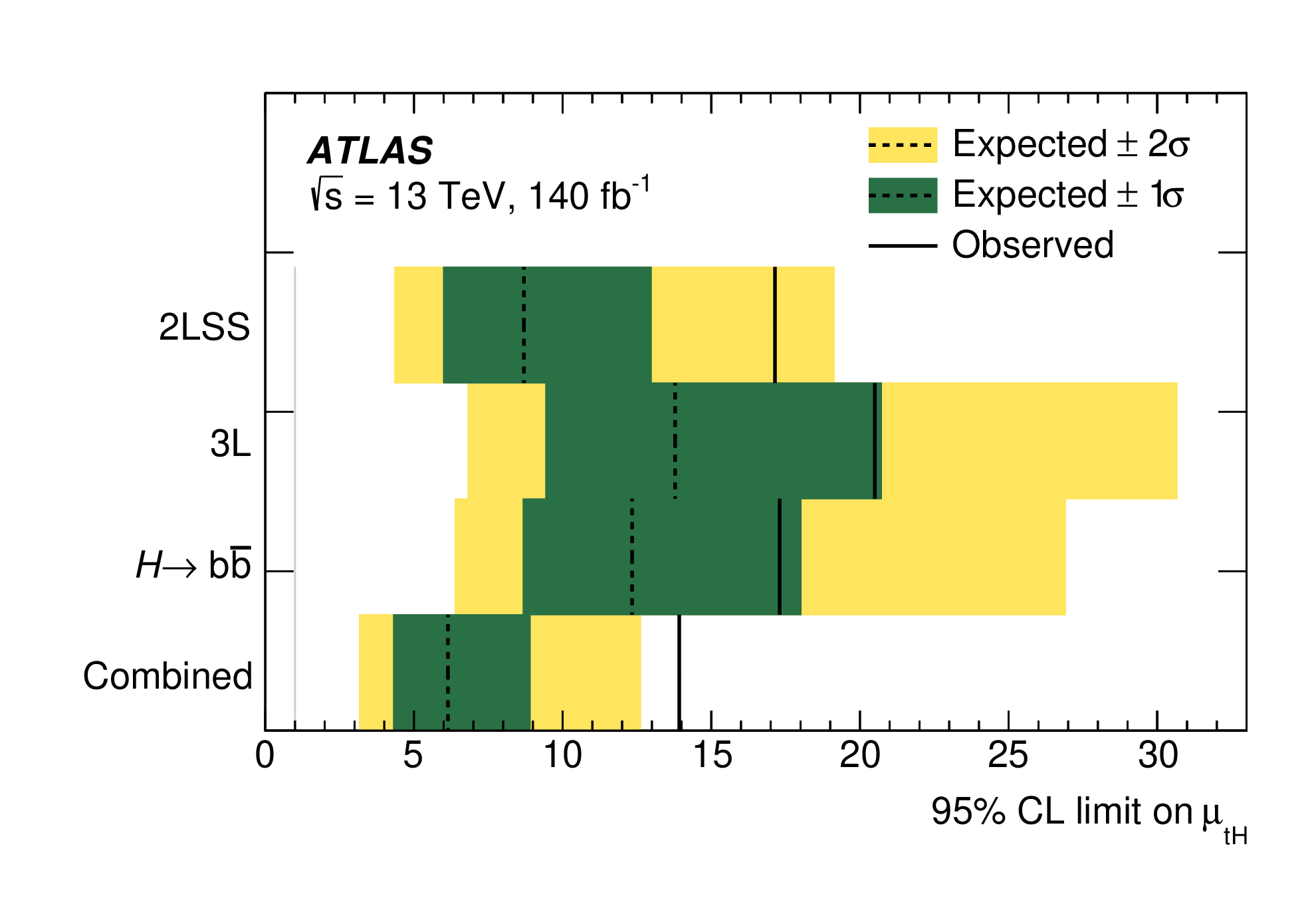}
\vspace{-3mm}
\caption{LHC limit on $\mu$(tH) for $\kappa_t = 1$.
From~\cite{ATLAS:2025irr}.
}
\label{fig:tbhlimit2SM}
\end{figure}

\begin{figure}[h!]
\centering
\includegraphics[width=0.49\columnwidth]
{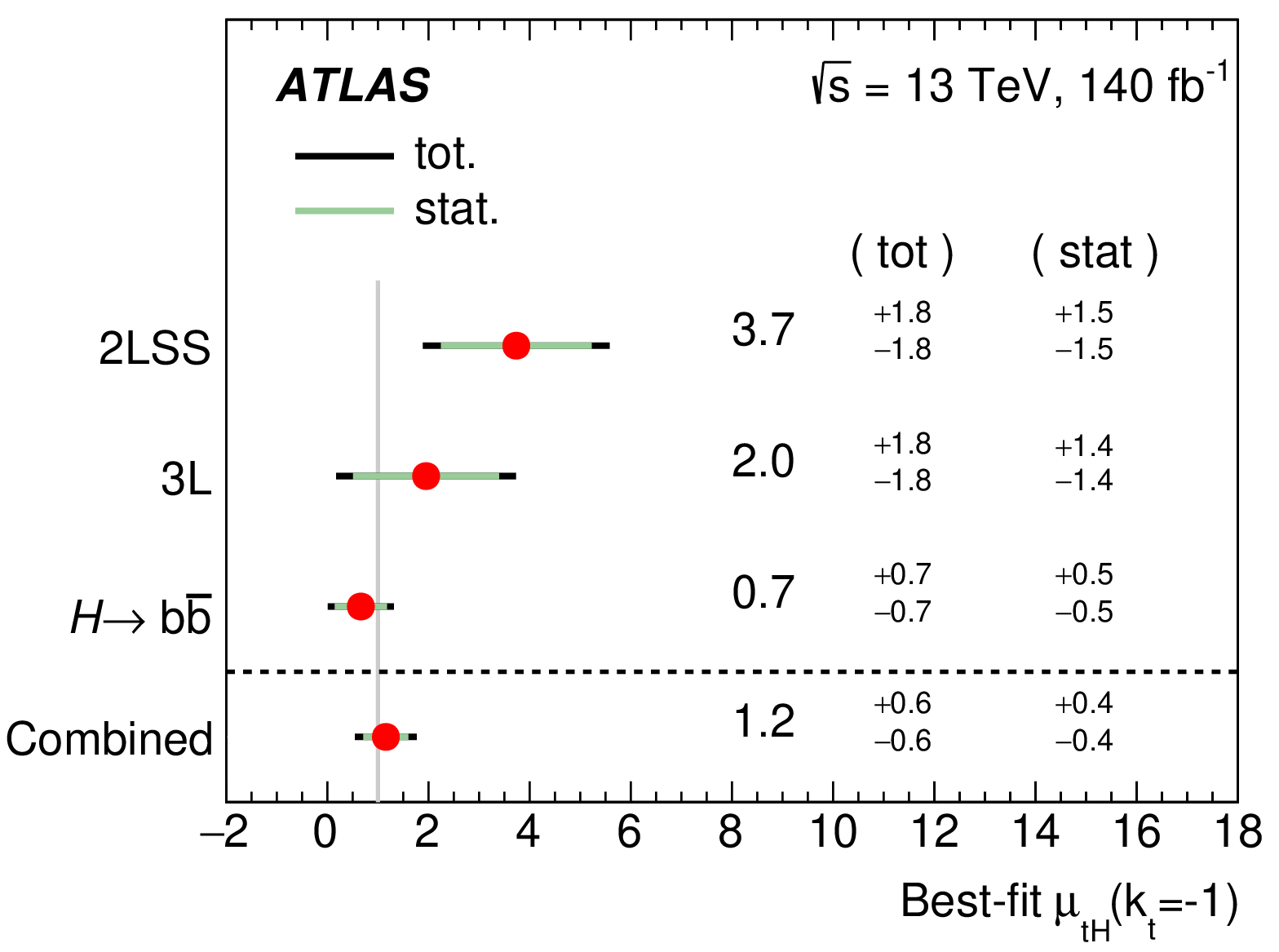}
\caption{LHC best fit on $\mu$(tH) for $\kappa_t=-1$.
From~\cite{ATLAS:2025irr}.
}
\label{fig:tbhlimit}
\end{figure}

\begin{figure}[h!]
\centering
\vspace{-3mm}
\includegraphics[width=0.49\columnwidth]{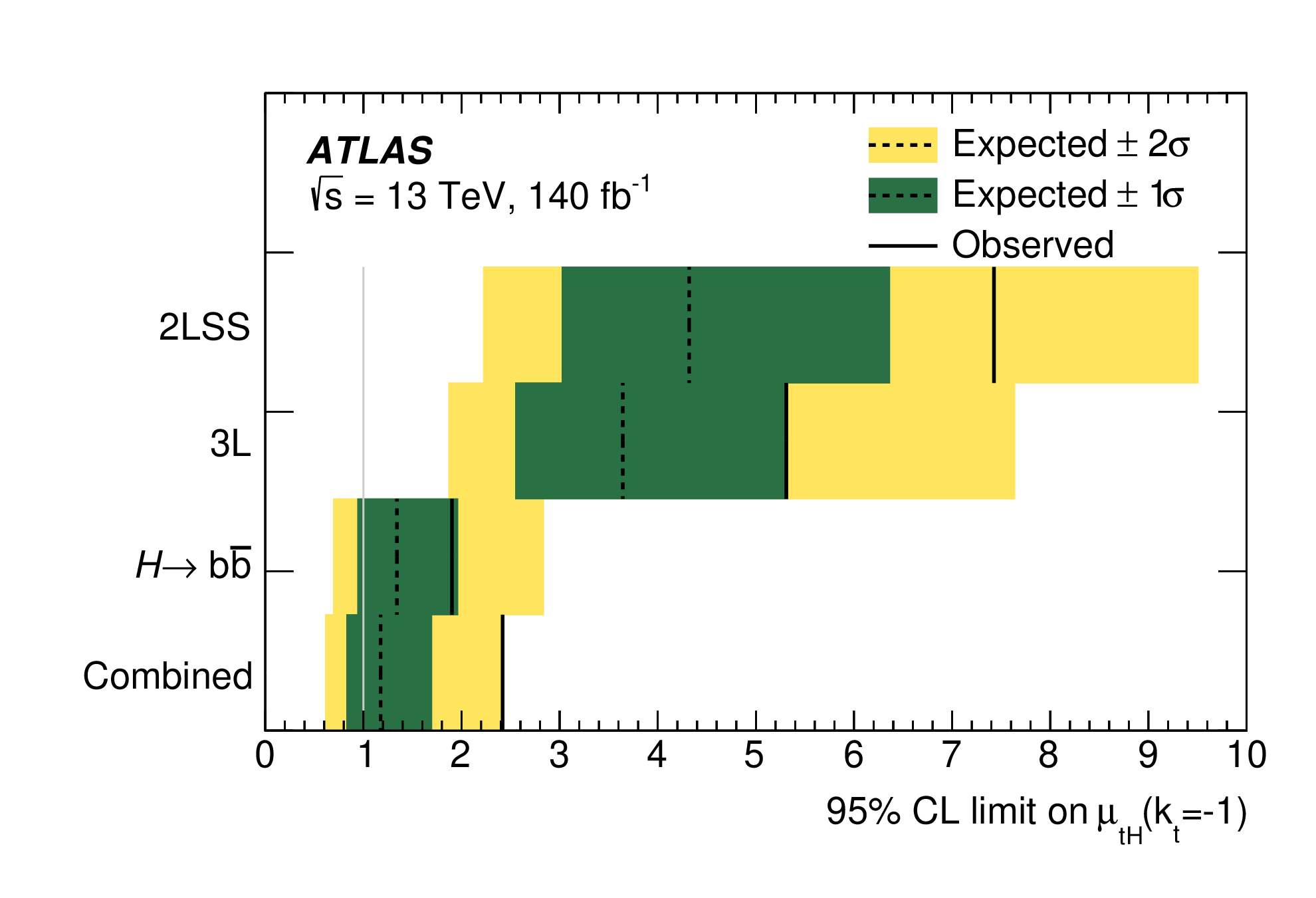}
\vspace{-3mm}
\caption{LHC limit on $\mu$(tH) for $\kappa_t=-1$.
From~\cite{ATLAS:2025irr}.
}
\label{fig:tbhlimit2}
\end{figure}

At the LHC searches for single top-Higgs production, $tH$, were performed. A corresponding Feynman diagram is shown in Fig.~\ref{fig:tbh}.
The observed signal strength, $\mu({ tH})$, defined as the ratio between the measured cross-section and the predicted SM value, is $\mu({tH})= 8.1 \pm 2.6 ({\rm stat.}) \pm 2.0 ({\rm syst.})$~\cite{ATLAS:2025irr}. 
The corresponding significance of the  observed (expected) signal above the background-only 
expectation is 2.8 (0.4) standard deviations. 
Figure~\ref{fig:tbhlimitSM} and~\ref{fig:tbhlimit2SM} show the best fit value of $\mu$(tH) and the corresponding upper limit at 95\% CL for the SM value $\kappa = 1$. Figure~\ref{fig:tbhlimit} and~\ref{fig:tbhlimit2} show the corresponding limits for $\kappa = -1$.
CMS measured $\mu({tH})= 5.7 \pm 2.7 ({\rm stat.}) \pm 3.0 ({\rm syst.})$~\cite{Sirunyan_2021}. 
The corresponding significance of the  observed (expected) signal above the background-only 
expectation is 1.4 (0.3) standard deviations based on Run-2 data (0.14\,ab$^{-1}$, 13\,TeV). 
Both ATLAS and CMS observe $\mu({tH}) > 1$, indicating a mild excess above the SM expectation in both experiments. Although not yet statistically significant, this consistent pattern makes $tH$ analyses a high-priority target for current research.
Extending the analysis with LHC Run-3 data (0.33\,ab$^{-1}$, 13.6\,TeV) and later at the HL-LHC (3\,ab$^{-1}$, 14\,TeV),
ATLAS and CMS will largely reduce both statistical and systematic uncertainties, improving the sensitivity to the sign of 
$\kappa_t$ and reducing the uncertainty on the $\mu({\rm tH})$ measurement~\cite{Sopczak:2025ptc}.

Before turning to Higgs boson pair production, it is instructive to place the individual Higgs boson production mechanisms in context, since together they form a complete and complementary experimental programme for measuring all major Higgs boson couplings (Fig.~\ref{fig:HHxsec}). The dominant mechanism, gluon--gluon fusion (ggF, $pp \to H$), proceeds through a virtual top-quark triangle loop and accounts for approximately 87\% of the total Higgs production cross-section at 13\,TeV. Because the loop is sensitive to any new heavy coloured particles that couple to the Higgs boson, deviations in the effective $Hgg$ coupling provide an indirect window onto BSM physics at scales beyond the direct reach of the LHC. Vector boson fusion (VBF, $qq \to qqH$), producing approximately 7\% of total events, proceeds via $t$-channel $W$ or $Z$ exchange and probes the $HWW$ and $HZZ$ couplings at tree level; its experimental signature---two forward jets with a large rapidity gap---provides a clean environment for $H \to \tau^+\tau^-$, $H \to WW^*$, and searches for invisible Higgs decays. Associated production with a $W$ or $Z$ boson ($WH$ and $ZH$) probes the same $HWW$ and $HZZ$ couplings in a complementary kinematic regime; crucially, the leptonic $W$ or $Z$ decay provides a trigger handle that renders the otherwise overwhelming $H \to b\bar{b}$ decay mode accessible, making $VH$, $H \to b\bar{b}$ the primary channel for directly measuring the Higgs coupling to the bottom quark. It was precisely this channel that drove the Tevatron's sensitivity to the SM Higgs boson, and the $3\sigma$ evidence established there~\cite{PhysRevD.88.052014} was the first experimental indication of the Higgs coupling to fermions. In addition, $ZH$ production with $Z \to \ell^+\ell^-$ and $H$ decaying invisibly provides the most sensitive collider probe of invisible Higgs decays and hence of potential dark matter portal couplings (Table~\ref{tab:invleplhc}).

Associated production with a top-quark pair ($t\bar{t}H$) offers the only direct, leading-order measurement of the top Yukawa coupling $y_t$, which is the largest coupling of the Higgs boson and therefore both the most important to measure precisely and the most sensitive to BSM modifications; the observation of $t\bar{t}H$ in 2018~\cite{ATLAS:2018ttH,CMS:2018ttH} was a landmark result confirming the predicted Yukawa structure of the SM. Bottom-quark associated production ($bbH$) is highly suppressed in the SM but becomes dramatically enhanced in Two-Higgs-Doublet Models and the MSSM at large $\tan\beta$, where the bottom Yukawa coupling is enhanced by a factor of $\tan\beta$; consequently, $bbH$ is one of the primary search channels for neutral heavy Higgs bosons in the MSSM, directly connecting the Tevatron searches at high $\tan\beta$ to the ongoing LHC programme. Finally, single top-quark associated production ($tH$) has the smallest cross-section but carries unique physical information: in the SM, the $tH$ amplitude receives large destructive interference between the top Yukawa and the $HWW$ vertex diagrams. If the sign of the top Yukawa coupling is inverted ($\kappa_t = -1$), this interference becomes constructive, enhancing the $tH$ cross-section by roughly an order of magnitude. The $tH$ channel is therefore the only production mode sensitive to the \emph{sign} of $\kappa_t$ rather than merely its magnitude---information entirely complementary to $t\bar{t}H$.

Taken together, the seven production mechanisms shown in Fig.~\ref{fig:HHxsec} probe every major Higgs coupling independently and with different experimental systematics, background compositions, and BSM sensitivities; no single reaction suffices to establish the full picture. The remarkable consistency with SM expectations across all modes and decay channels, as shown in Fig.~\ref{fig:atlascouplings}, already provides some of the most stringent indirect constraints on BSM physics in the Higgs sector. Figure~\ref{fig:HHxsec} therefore encapsulates the central strategy of Higgs physics at the LHC: the full structure of the Higgs sector can only be reconstructed by combining multiple production mechanisms, each probing different couplings and offering sensitivity to different classes of BSM effects.

\begin{figure}[h!]
\centering
\includegraphics[width=0.39\columnwidth]{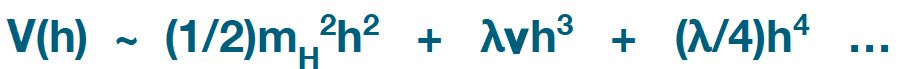}\hfill
\includegraphics[width=0.2\columnwidth]{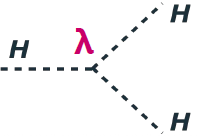}
\includegraphics[width=.39\columnwidth]{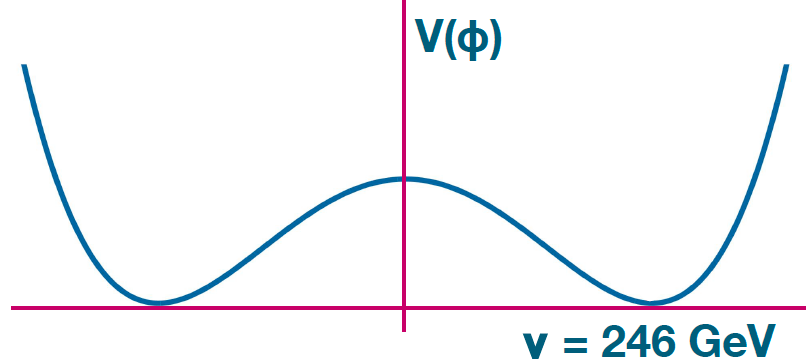}
\caption{Higgs boson potential.}
\label{fig:potential}
\end{figure}

\begin{figure}[h!]
\centering
\vspace{-8mm}
\includegraphics[width=0.49\columnwidth]{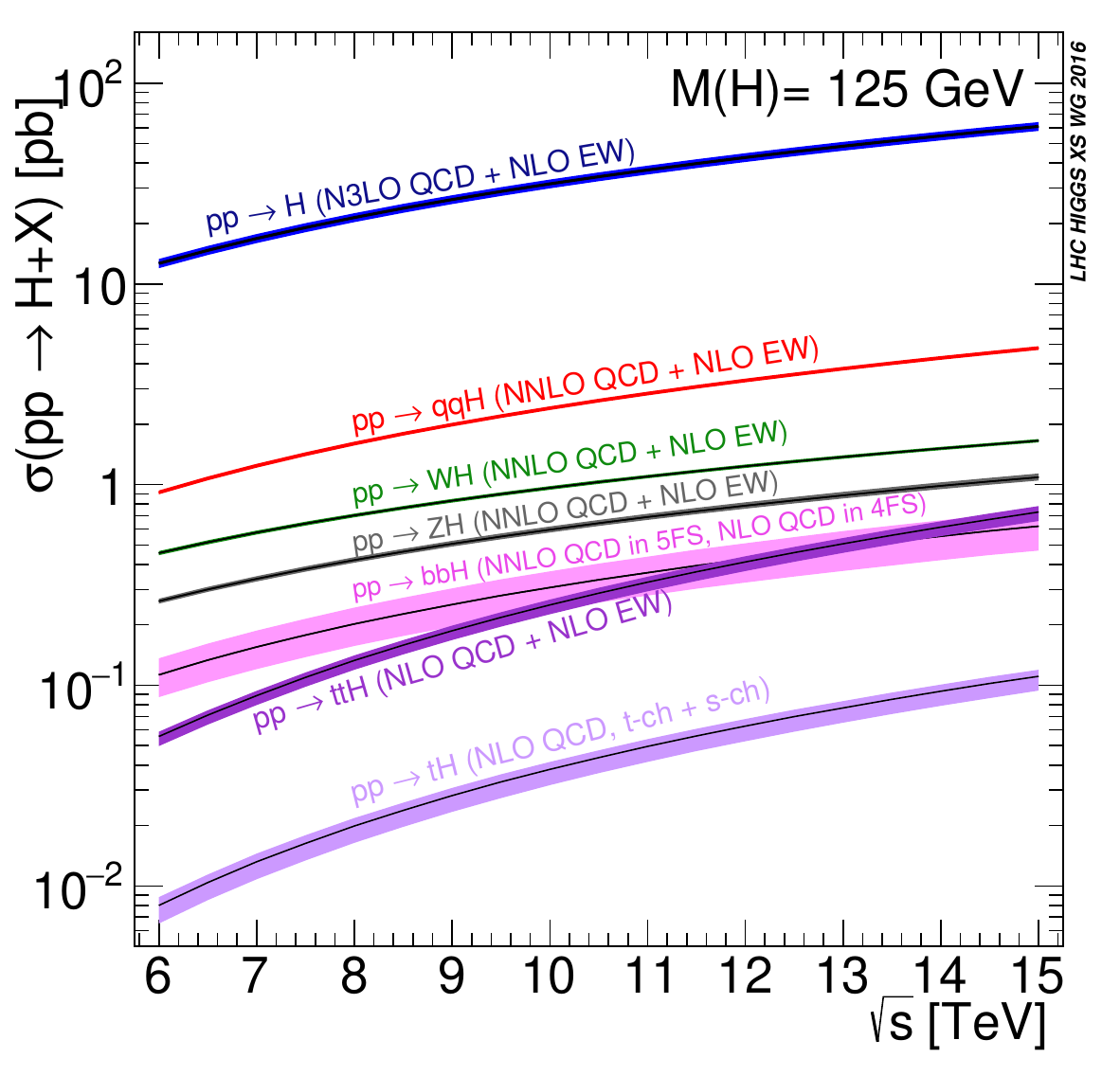}
\vspace{-1mm}
\caption{Higgs boson production cross-sections as a function of the LHC centre-of-mass energy $\sqrt{s}$, for the seven main production mechanisms: gluon--gluon fusion (ggF), vector boson fusion (VBF), associated production with a $W$ or $Z$ boson ($WH$, $ZH$), associated production with a top-quark pair ($t\bar{t}H$), bottom-quark associated production ($bbH$), and single top-quark associated production ($tH$). From~\cite{xsec}.}
\label{fig:HHxsec}
\end{figure}

\begin{figure}[h!]
\centering
\includegraphics[width=.49\columnwidth]
{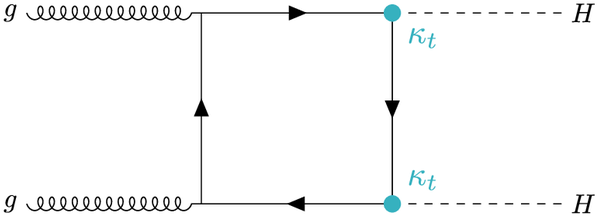}
\includegraphics[width=0.49\columnwidth]{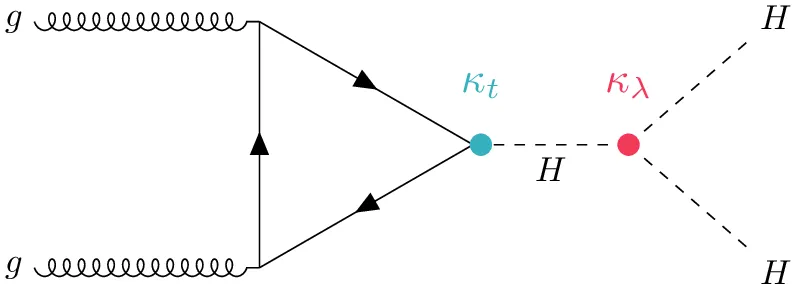}
\caption{Feynman diagrams HH production.}
\label{fig:HH}
\end{figure}

\begin{figure}[H]
\centering
\includegraphics[width=0.49\columnwidth]{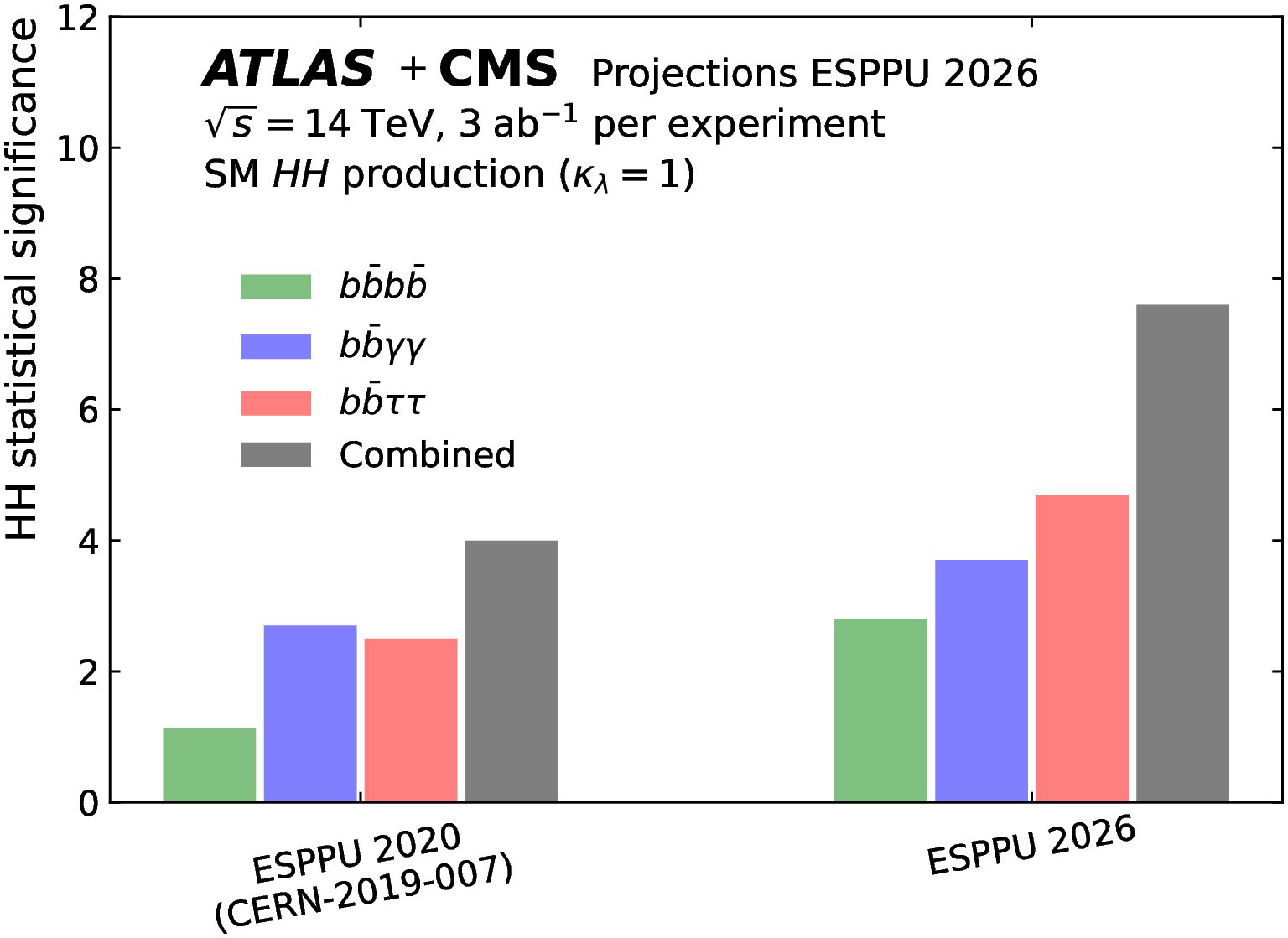}
\caption{Development of HH sensitivity predictions.
From~\cite{atlas2025highlightshllhcphysicsprojections}.
}
\label{fig:HHdevelopment}
\end{figure}

\begin{figure}[h!]
\centering
\includegraphics[width=0.49\columnwidth]{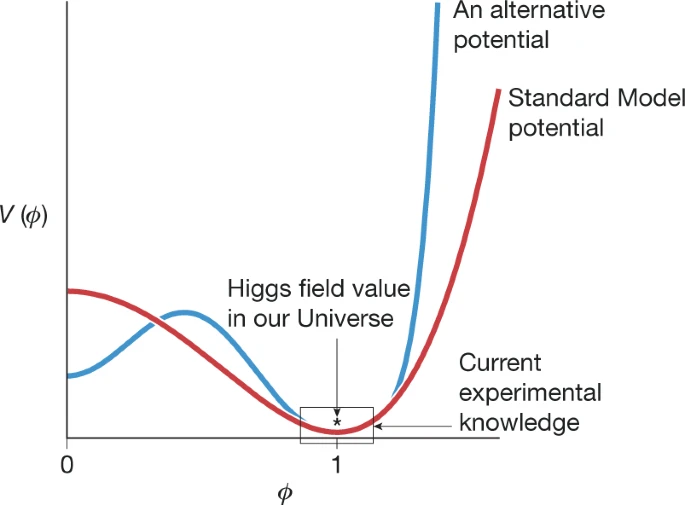}
\caption{BSM Higgs boson potential. From~\cite{Zghiche:2025qig}.}
\label{fig:BSMpotential}
\end{figure}

\begin{figure}[h!]
\centering
\includegraphics[width=0.49\columnwidth]{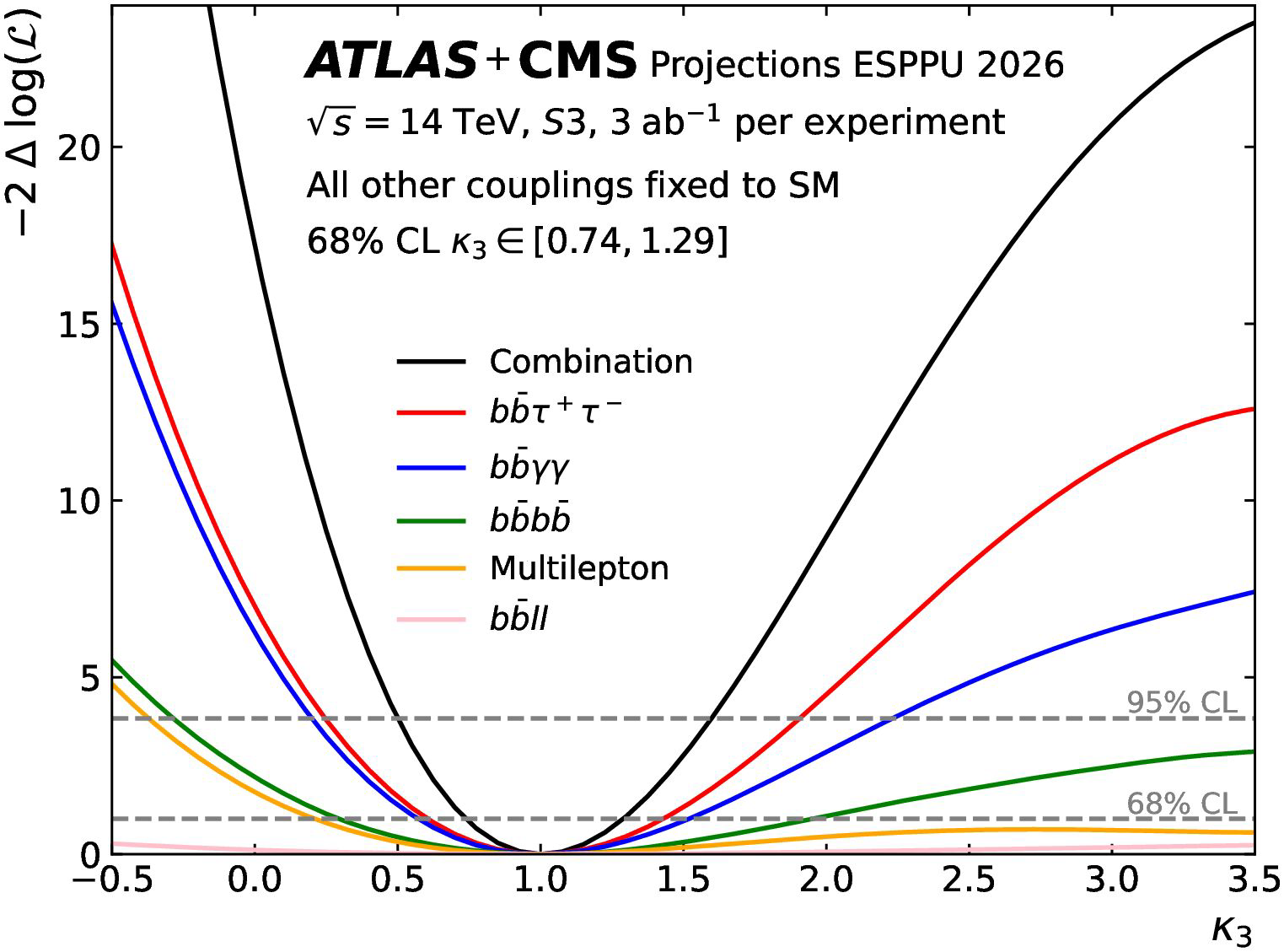}
\caption{Expected sensitivity of the self-coupling measurement with respect to the SM expectation $\kappa_\lambda (=\kappa_3)$.
From~\cite{atlas2025highlightshllhcphysicsprojections}.
}
\label{fig:HHkappa}
\end{figure}

\begin{figure}[h!]
\centering
\vspace{-3mm}
\includegraphics[width=0.49\columnwidth]{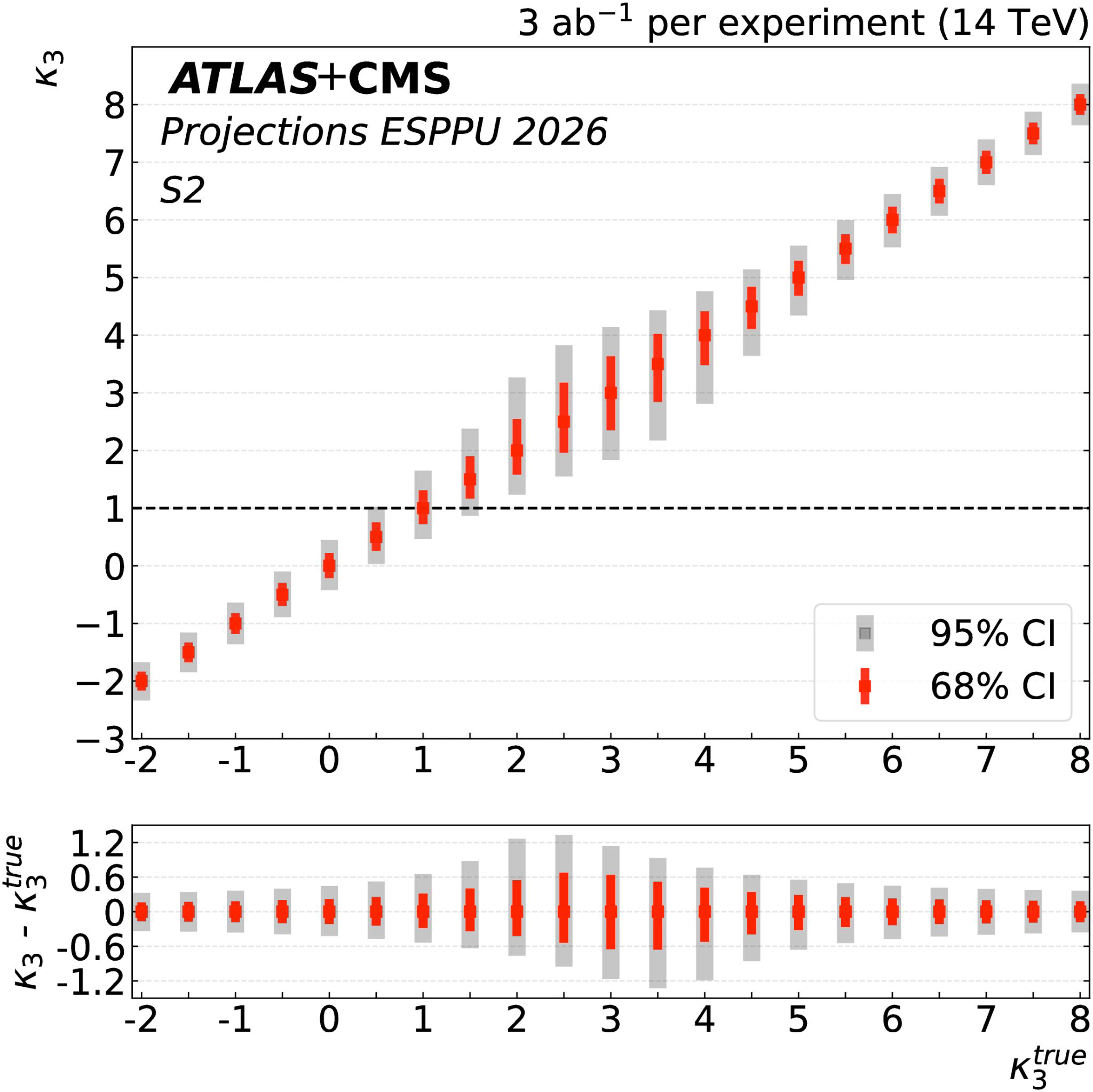}
\caption{HL-LHC expected linearity of the self-coupling measurement.
From~\cite{atlas2025highlightshllhcphysicsprojections}.
}
\label{fig:selfcoupling}
\vspace{-2mm}
\end{figure}

\begin{figure}[H]
\centering
\includegraphics[width=0.49\columnwidth]{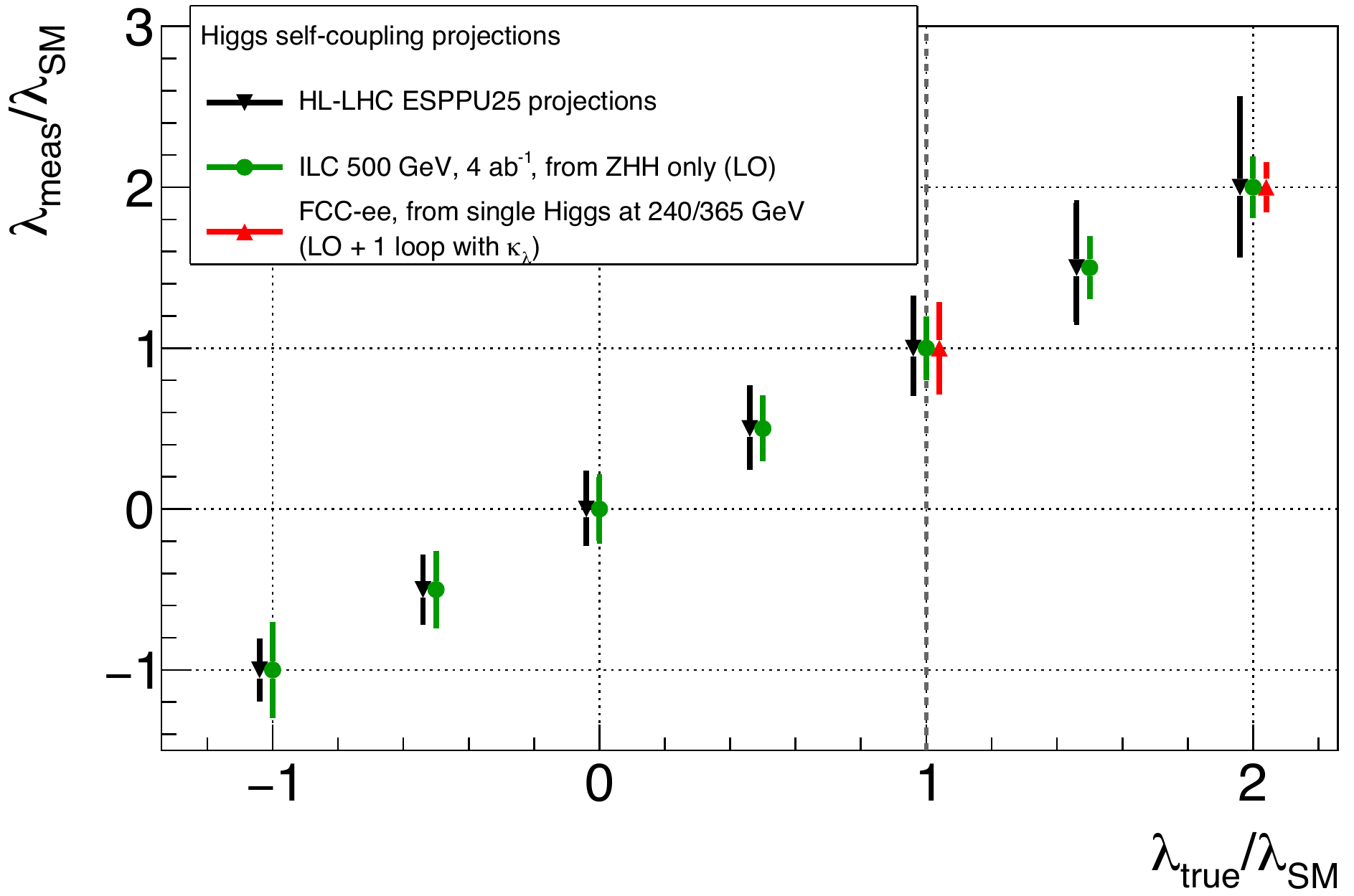}
\caption{Expected linearity on the self-coupling measurement HL-LHC, and comparison with ILC and FCC.
``LS'' stands for long shutdown.
From~\cite{CERN-2025-005}.}
\label{fig:selfcouplingcompare}
\end{figure}

The measurement of the Higgs boson potential will be a high-priority research area in the coming decades.
Figure~\ref{fig:potential} illustrates the Higgs boson potential. The Higgs boson mass has been measured with a precision of 0.09\%~\cite{Aad:2867152}. The measurement of the self-coupling follows and can then be compared with the SM expectation, $\lambda=m_H^2/(2v^2)$.
The Higgs boson pair-production $HH$ is a very rare process and amounts to only about 0.1\% of the total Higgs boson production cross-section (Fig.~\ref{fig:HHxsec}).
The SM box and triangle production diagrams exhibit a large destructive interference (Fig.~\ref{fig:HH}).
As a consequence, even small BSM contributions can significantly enhance the $HH$ production rate. The strong destructive interference highlighted in Fig.~\ref{fig:HH} is precisely why Higgs pair production is particularly sensitive to BSM effects: small modifications to the Higgs self-coupling or to loop contributions can significantly alter the rate. This makes double-Higgs production a uniquely sensitive probe of the Higgs potential, despite its small cross-section.
Detailed studies have been performed to determine the sensitivity of $HH$ production at the HL-LHC.
Three scenarios are investigated:
\begin{itemize}
    \item S1, 2\,ab$^{-1}$: $6\sigma$
    \item S2, 3\,ab$^{-1}$: $7.2\sigma$
    \item S3, 3\,ab$^{-1}$: $7.6\sigma$,
\end{itemize}
where the difference between scenarios S2 and S3 are improved object (electron, muon, tau and jet) identification.
These results from the ESPPU2026~\cite{atlas2025highlightshllhcphysicsprojections} improved significantly the previous ESPPU2020~\cite{EuropeanStrategyGroup:2020pow} studies. The statistical significances are shown in Fig.~\ref{fig:HHdevelopment}.
BSM physics could modify the Higgs boson potential as illustrated in Fig.~\ref{fig:BSMpotential}.
The ratio of the self-coupling to the SM expectation $\kappa_\lambda (=\kappa_3)$ has been studied for several final states from HH decays, as shown in Fig.~\ref{fig:HHkappa}. 
The resulting precisions at 68\% CL are:
\begin{itemize}
    \item S1, 2\,ab$^{-1}$: -32\%/+37\%,
    \item S2, 3\,ab$^{-1}$: 
    -27\%/+31\%,
    \item S3, 3\,ab$^{-1}$: 
    -26\%/+29\%.
\end{itemize}
In addition, the linearity of the measured self-coupling \mbox{versus} the true self-coupling has been studied, as shown in Fig.~\ref{fig:selfcoupling}.
Good linearity is expected within the uncertainties.
Figure~\ref{fig:selfcouplingcompare} compares the expected sensitivities between HL-LHC, ILC and FCC~\cite{CERN-2025-005}.

\begin{figure}[H]
\centering
\vspace{-1mm}
\includegraphics[width=0.49\columnwidth]{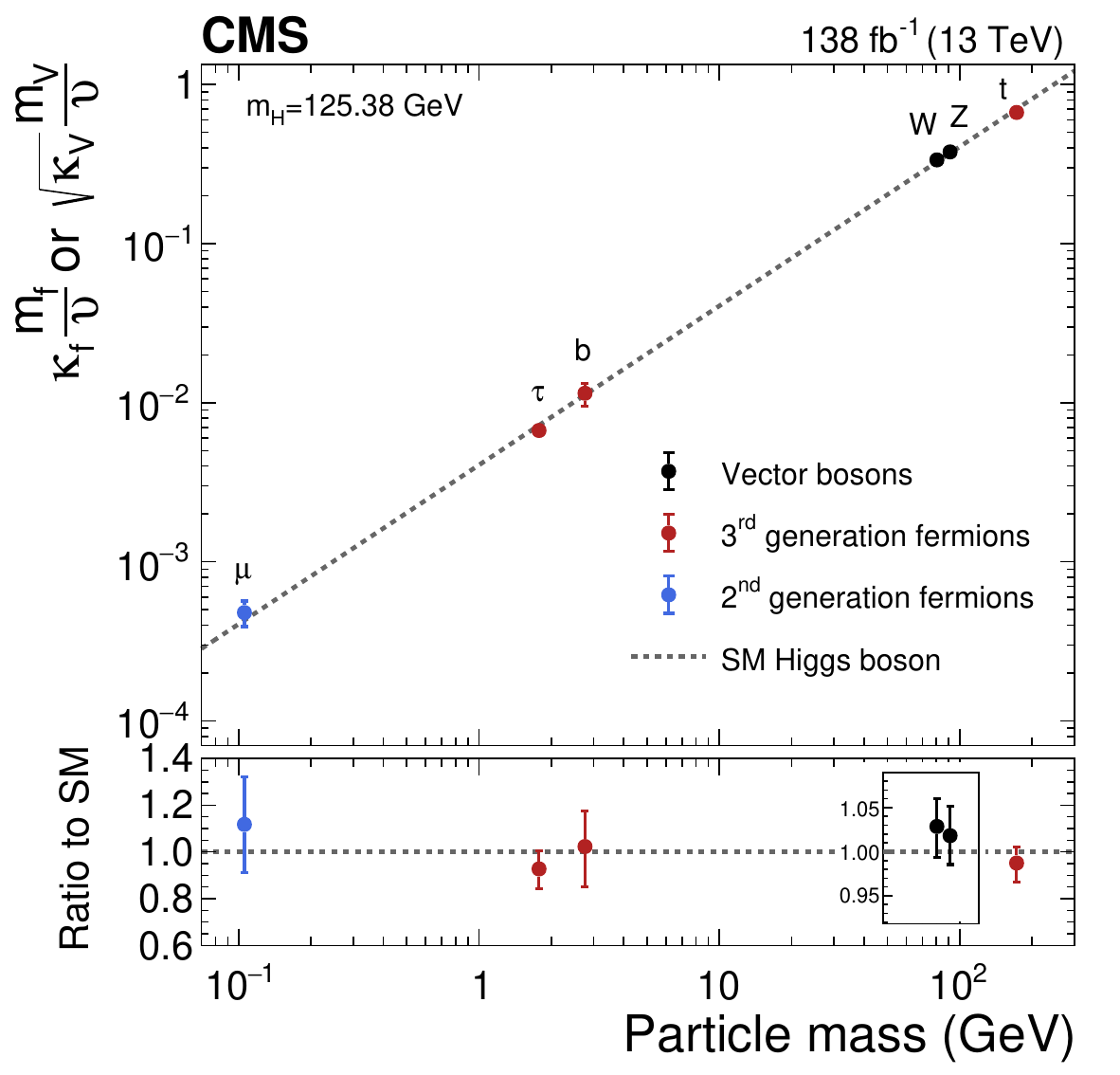}
\caption{LHC relation of coupling and fermion/vector boson mass.
From~\cite{Tumasyan:2814513}.
}
\label{fig:linear}
\vspace{-1mm}
\end{figure}

\begin{figure}[H]
\centering
\vspace{-1mm}
\includegraphics[width=0.49\columnwidth]{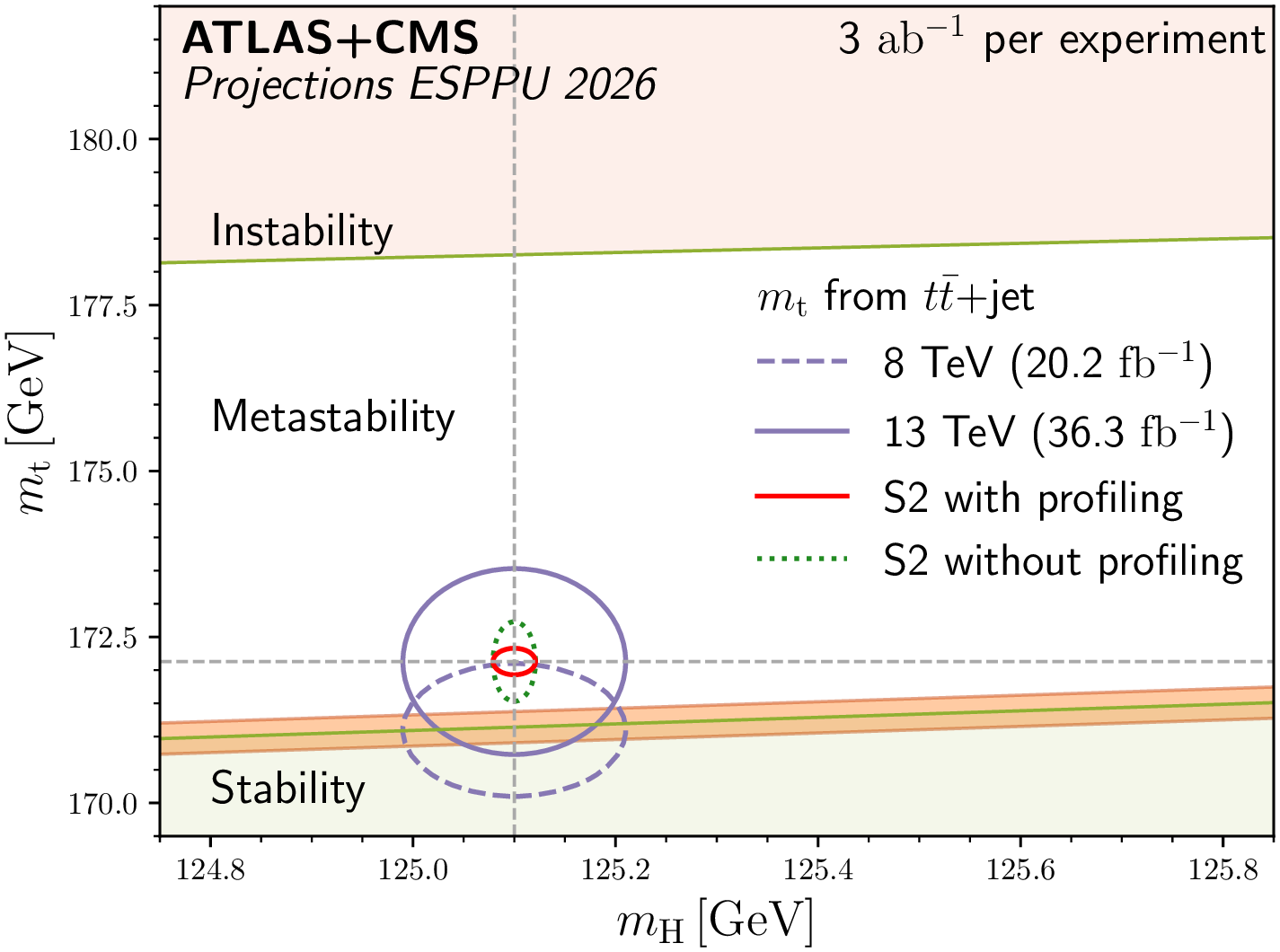}
\caption{Development and prospects of the precision on the Higgs boson mass. Unstable, metastable and stable vacuum region are shown in the mass plane $(m_H,m_t)$.
From~\cite{atlas2025highlightshllhcphysicsprojections}.
}
\label{fig:stability}
\vspace{-2mm}
\end{figure}

The linear relation between the Higgs boson coupling strength and the mass of the fermion or vector boson to which it couples has been measured at the LHC, as shown in Fig.~\ref{fig:linear}, confirming the Yukawa structure of the Standard Model across nearly four orders of magnitude in particle mass~\cite{Tumasyan:2814513}. The excellent agreement with SM expectations at the current level of experimental precision is indeed a striking result---and yet the conclusion that this agreement leaves little room for BSM deviations must be carefully qualified, because it conflates two distinct questions: whether BSM physics is \emph{excluded}, and whether it is \emph{detectable} with current instruments.

In the framework of the Standard Model Effective Field Theory (SMEFT), BSM physics at a scale $\Lambda \gg v$ modifies Higgs couplings through operators of dimension six, generating corrections of order $(v/\Lambda)^2$, where $v = 246$\,GeV. For $\Lambda = 1$\,TeV---entirely consistent with existing LHC direct search limits---this yields coupling deviations  at the level of a few percent, below the sensitivity of the current LHC measurements for which individual $\kappa$ modifiers are constrained at the 5--15\% level~\cite{ATLAS:2022vkf,Tumasyan:2814513}. The data are compatible with SM expectations, but they do not exclude few-percent BSM effects. This distinction matters, because several well-motivated BSM scenarios predict precisely such deviations in the 1--5\% range. In composite Higgs models, where the Higgs is a pseudo-Nambu--Goldstone boson of a new strongly interacting sector with scale $f$, all Higgs couplings are universally suppressed by $\sqrt{1-v^2/f^2}$, yielding a fractional deviation of order $v^2/(2f^2) \simeq 3$\% for $f = 1$\,TeV. In the decoupling limit of the MSSM, the $HZZ$ coupling deviates from its SM value by a term of order $m_Z^2\cos^2 2\beta/m_A^2$, which for $m_A \sim 500$\,GeV is at the percent level and entirely consistent with current data. Similarly, any mixing of the SM Higgs with a real singlet scalar induces a universal suppression of all couplings, with a mixing angle of a few degrees generating deviations below 1\%. None of these scenarios are excluded by the LHC.

For the strongest coupling of the top-quark, the global (individual) fit uncertainty is currently 12\% (10\%); at the HL-LHC, these uncertainties could be reduced to 5.1\% (3.7\%)~\cite{CERN-2025-005}. This is important progress, yet hadronic production uncertainties from PDF and QCD scale variations are irreducible in a hadron collider environment and will prevent the HL-LHC from reaching the sub-percent regime. It is precisely the FCC-ee physics programme that is designed to exploit this gap. Operating at $\sqrt{s} = 240$--$250$\,GeV at the $ZH$ production threshold, FCC-ee can measure the total $e^+e^- \to ZH$ cross-section in a fully model-independent way via the $Z$-boson recoil technique, determining the $HZZ$ coupling to a precision of approximately 0.2\%~\cite{CERN-2025-005}. This would probe compositeness scales up to several TeV and definitively test the composite Higgs, singlet mixing, and MSSM decoupling scenarios described above. The conclusion from Fig.~\ref{fig:linear} is therefore not that little room remains for BSM physics in the Higgs sector, but rather that the SM is confirmed at the level of sensitivity accessible to the LHC, and that the next generation of precision lepton colliders will probe the \emph{natural} BSM range predicted by weakly coupled extensions of the SM at the TeV scale. Far from closing the door on BSM Higgs physics, the LHC results define exactly where and at what precision the crucial measurements must be made---providing a strong and well-motivated physics case for a future $e^+e^-$ Higgs factory. The linear trend in Fig.~\ref{fig:linear} should therefore be interpreted as a global consistency check of the Yukawa structure at the current level of precision, rather than as a definitive exclusion of percent-level deviations expected in many BSM scenarios.

\begin{figure}[H]
\centering
\includegraphics[width=0.49\columnwidth]{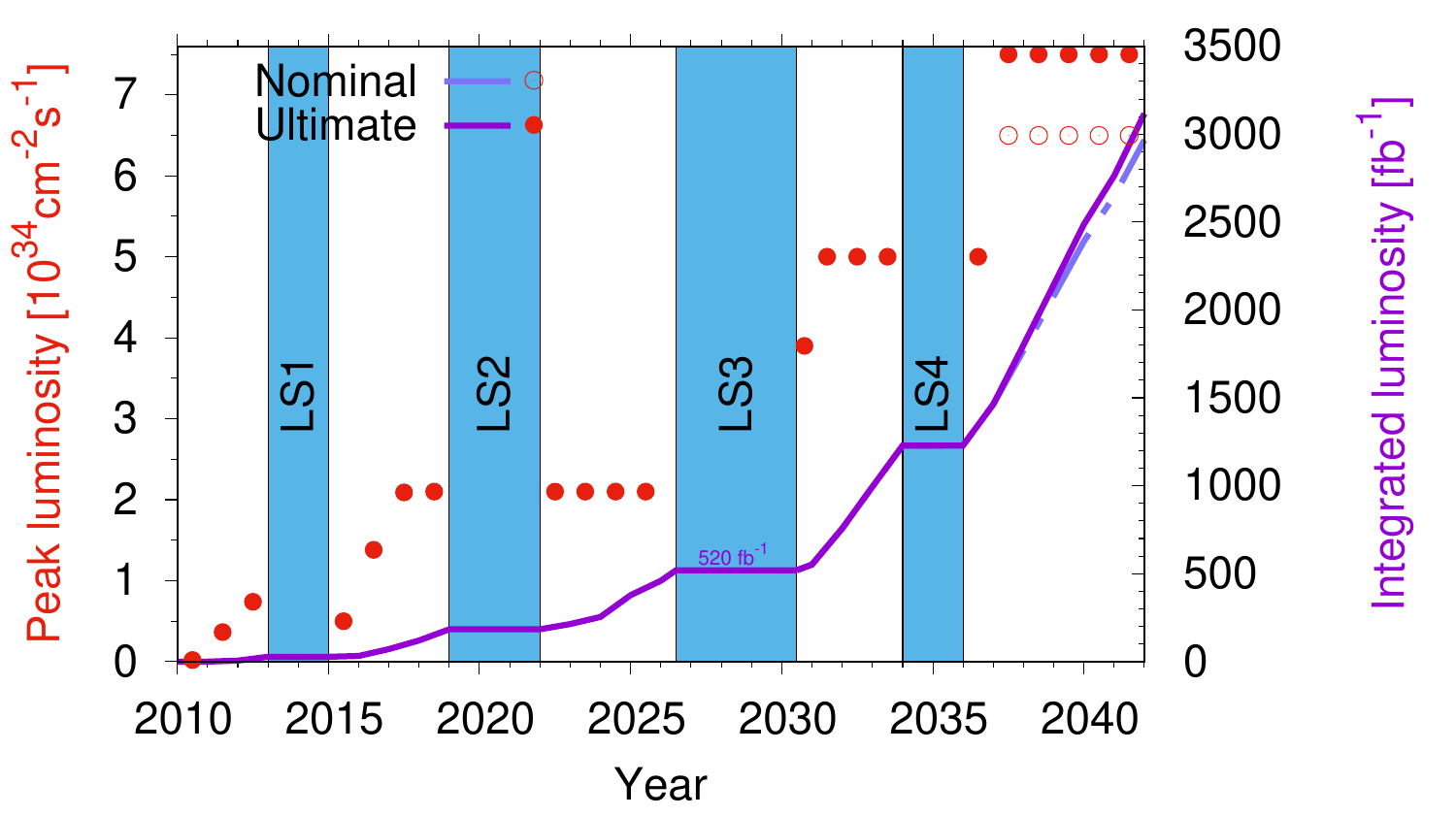}
\caption{Time scale of LHC and HL-LHC. ``LS'' stands for long shutdown. From~\cite{ATL-UPGRADE-PUB-2025-001}.}
\label{fig:HL-LHC}
\vspace{-3mm}
\end{figure}

The Higgs boson and top-quark masses are directly related to the question of vacuum stability. 
From the current uncertainty 110\,MeV~\cite{Aad:2867152},
these projections result in a total uncertainty on 
$m_H$ from 
$H \rightarrow ZZ^* \rightarrow 4\ell 
(\ell = e$ or $\mu)$ 
events of 38\,MeV for \mbox{ATLAS} and 26\,MeV for CMS, 
leading to a combined uncertainty of 21\,MeV~\cite{atlas2025highlightshllhcphysicsprojections}.
The development and the prospects of the precision on the Higgs boson mass are shown in Fig.~\ref{fig:stability}.

\subsection{LHC Charged Higgs Boson Results}
\begin{figure}[H]
\centering
\vspace{-3mm}
\includegraphics[width=0.49\columnwidth]{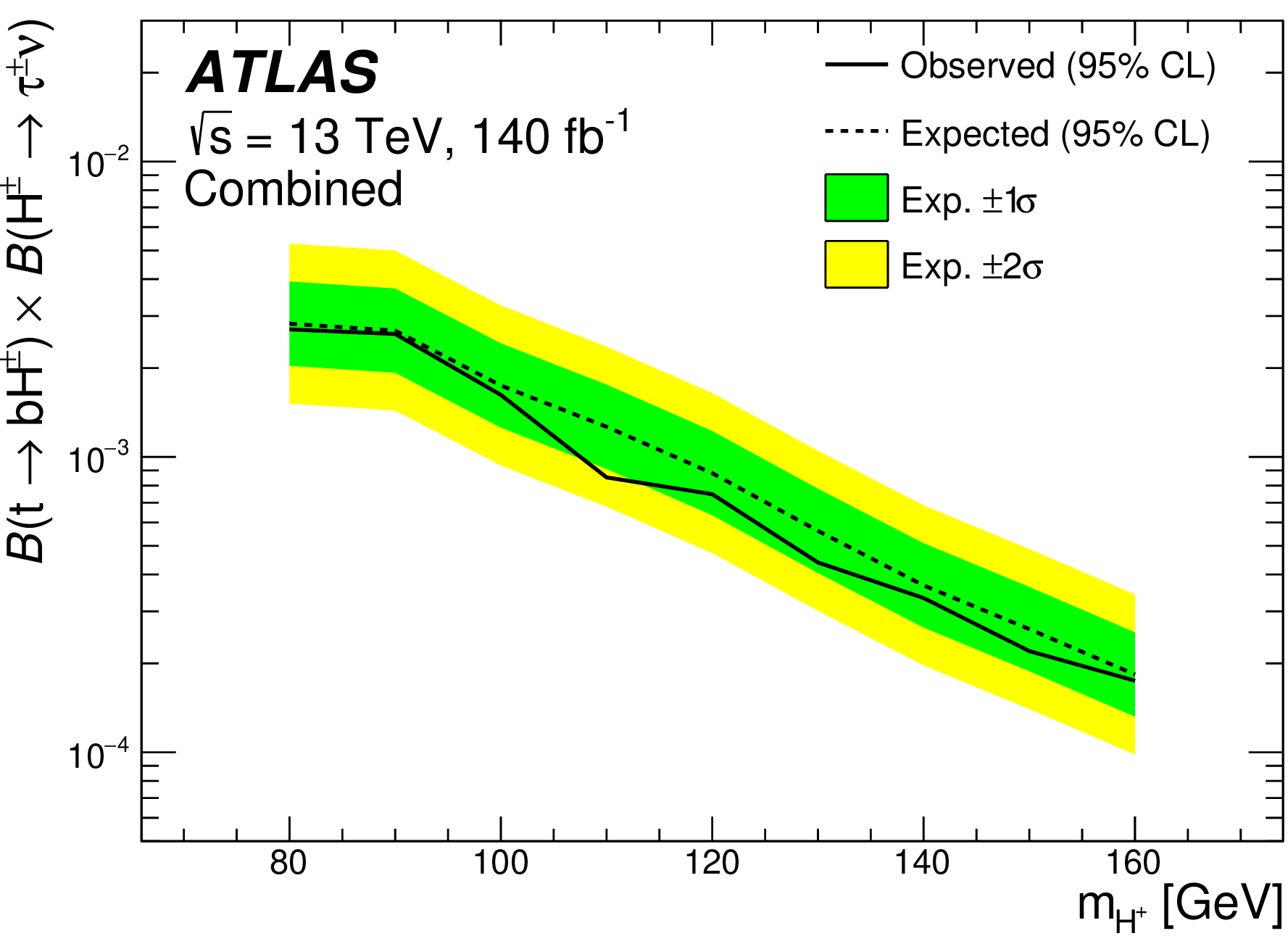}
\caption{ATLAS 95\% CL upper limits on $BR(t\rightarrow H^+b)\times BR(H^+ \rightarrow \tau\nu)$ as a function of $m_{H^\pm}$ (80--160\,GeV).
From~\cite{ATLAS:2024hya}.
}
\label{fig:LHChplus}
\vspace{-8mm}
\end{figure}

\begin{figure}[H]
\centering
\includegraphics[width=0.49\columnwidth]{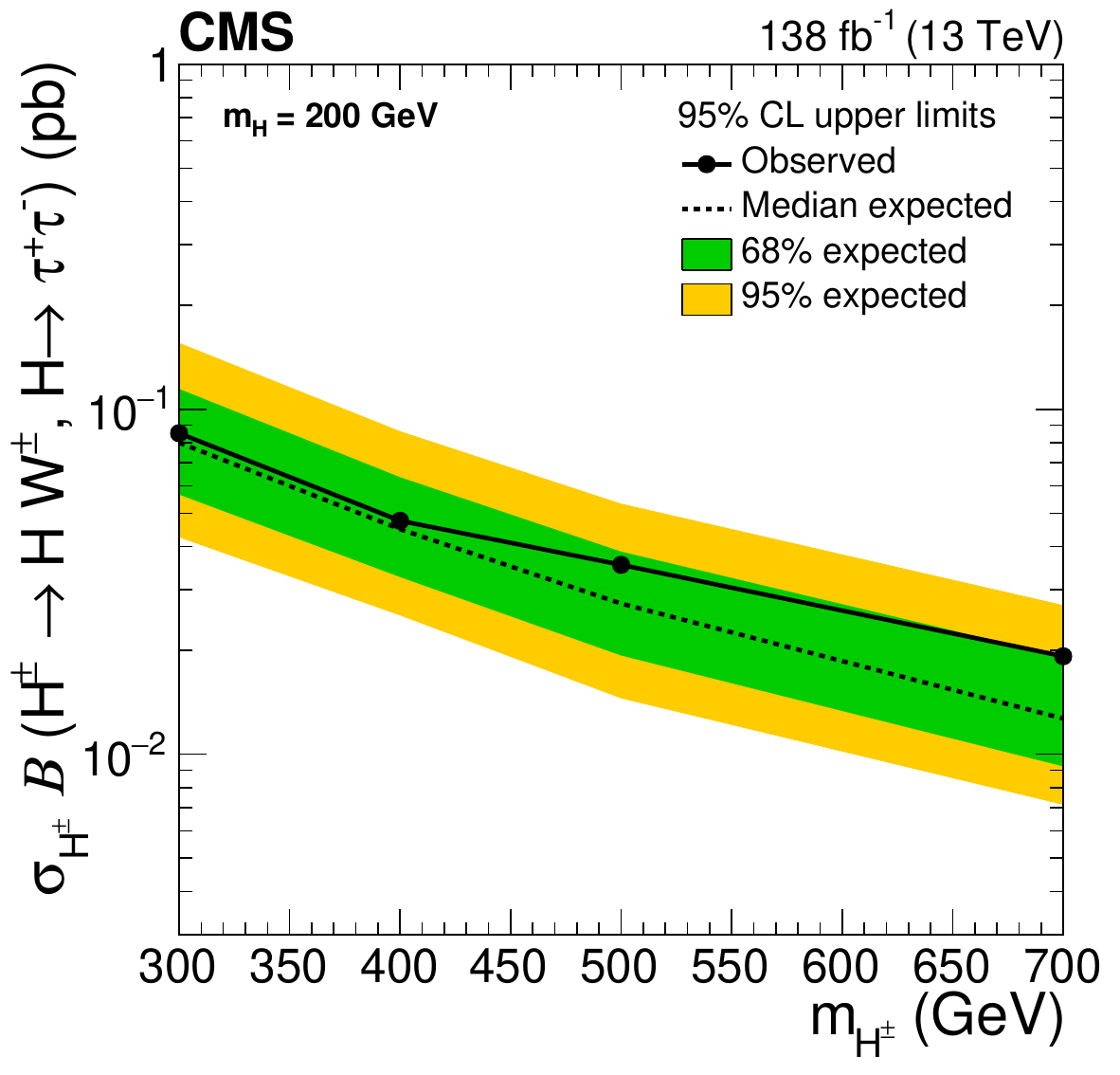}
\caption{CMS 95\% CL upper limits on $\sigma_{H^\pm}\,\mathcal{B}(H^\pm\to H W^\pm, H\to\tau^+\tau^-)$ as a function of $m_{H^\pm}$ (300--700\,GeV) for $m_H = 200$\,GeV.
From~\cite{CMS2023}.
}
\label{fig:LHChplusCMS}
\end{figure}

The sensitivity for charged Higgs bosons has much increased at the LHC compared to the LEP and the Tevatron colliders. At the LHC, ATLAS sets upper limits between 0.27\% and 0.02\% on $BR(t\rightarrow H^+b)\times BR(H^+ \rightarrow \tau\nu)$ for charged Higgs boson masses between 80 and 160\,GeV, using 140\,fb$^{-1}$ at 13\,TeV~\cite{ATLAS:2024hya}, as shown in Fig.~\ref{fig:LHChplus}. CMS obtained upper limits between 0.36\% (at 80\,GeV) and 0.08\% (at 160\,GeV) in the same mass range with 35.9\,fb$^{-1}$ of 2016 data~\cite{CMS:2019bfg}. These limits are two to three orders of magnitude more stringent than the Tevatron limits of about 15--20\% (Fig.~\ref{fig:tevatronhpluslimits}). 
Important systematic uncertainties are present in the $t\bar t$ background modelling and the determination of fake-taus. These uncertainties will be reduced with higher statistics data. In addition, CMS searched for the decay of a heavy charged Higgs boson into a $W$ boson and a heavy neutral Higgs boson, $H^\pm\to W^\pm H$, with $H\to\tau^+\tau^-$ and $m_H = 200$\,GeV, using 138\,fb$^{-1}$ at 13\,TeV~\cite{CMS2023}. The data agree with the SM background expectation, and upper limits on $\sigma_{H^\pm}\,\mathcal{B}(H^\pm\to H W^\pm, H\to\tau^+\tau^-)$ are set, ranging from 0.085\,pb at $m_{H^\pm} = 300$\,GeV to 0.019\,pb at 700\,GeV (Fig.~\ref{fig:LHChplusCMS}). These are the first LHC limits in this decay channel, which becomes important in models where the charged Higgs boson is heavier than the additional neutral Higgs boson. 


\subsection{Discussion of LHC Results}

The LHC had a spectacular start with the discovery of the Standard Model Higgs boson at a mass of 125\,GeV in 2012~\cite{ATLAS:2012Higgs,CMS:2012Higgs}. Just one year later, the Nobel Prize in Physics was awarded to Fran\c{c}ois Englert and Peter W.~Higgs
``for the theoretical discovery of a mechanism that contributes to our understanding of the origin of mass of subatomic particles, and which was recently confirmed through the discovery of the predicted fundamental particle by the ATLAS and CMS experiments at CERN's Large Hadron Collider''~\cite{NobelPrizePhysics2013}.

Increasingly precise measurements of the properties of the newly discovered particle in the following years have led to stringent constraints on BSM Higgs boson physics. So far, no observation has been made that indicates the presence of BSM effects in the Higgs sector.

A highlight of the Higgs boson searches was the observation of the $t\bar tH$ coupling~\cite{ATLAS:2018ttH,CMS:2018ttH}, which was found to be consistent with the Standard Model expectation. This measurement is particularly sensitive to BSM physics, as the top-quark Yukawa coupling is the strongest coupling of the Higgs boson. The result confirmed the predicted linear relationship between fermion masses and Yukawa couplings over several orders of magnitude.

Searches for neutral Higgs boson pair production have led to significantly extended limits in the MSSM compared to the results obtained at LEP. Another highly sensitive probe of BSM physics is Higgs boson production in association with a single top quark. Further analyses with increased data sets are required to establish an observation of the tH process.

Neutral Higgs boson pair production, which directly probes the Higgs boson self-coupling, has only recently become experimentally accessible due to improvements in analysis techniques. This has opened the possibility of observing this process with the full LHC and HL-LHC data sets. The main challenge in this measurement arises from the small Higgs boson pair production rate and the need to combine multiple final-state signatures resulting from the decays of the two Higgs bosons.

Compared to the first Tevatron results on charged Higgs bosons in top-quark decays, the LHC experiments have reduced the associated uncertainties by approximately an order of magnitude, leading to significantly stronger constraints on charged Higgs bosons. In addition to the charged Higgs boson decay modes into \(\tau\nu\) and \(c\bar{s}\), which were studied previously, recent searches have also considered decays into WH.

A significant increase in sensitivity to BSM physics in the Higgs sector is expected in the coming years. Experience from previous analyses indicates that improvements in sensitivity often exceed what would be expected from the increase in data statistics alone.

Taken together, the LHC results reinforce the central theme of this review: sensitivity to BSM Higgs physics is ultimately determined by coupling structures and the complementarity of multiple probes, not simply by energy reach.

\vspace{-2mm}
\subsection{LHC and HL-LHC Future Planning}
The future planning of the LHC and HL-LHC is well defined for the next decades.
LHC Run-3 operation ended in June 2026, leading to a total recorded integrated luminosity of about 500\,fb$^{-1}$ per experiment. In 2030, the HL-LHC will start operating with expected total luminosities of 3\,ab$^{-1}$. The time scale is shown in Fig.~\ref{fig:HL-LHC}.
At the HL-LHC, the centre-of-mass energy will increase to 14\,TeV, the instantaneous luminosity will increase to $5$--$7 \times 10^{34}$\,cm$^{-2}$s$^{-1}$, and the pileup will increase to 140--200.

\section{Conclusions and Perspective}

The search for Higgs bosons beyond the Standard Model has evolved over more than three decades, from the first systematic exclusions at LEP to increasingly precise measurements at the LHC. While this historical progression is often described in terms of rising collision energy and luminosity, the results reviewed in this article point to a deeper and more instructive conclusion.

The sensitivity of Higgs boson searches is not determined solely by kinematic reach, but by the structure of Higgs couplings to Standard Model particles and by the interplay of production and decay mechanisms. This principle is already evident in the LEP programme: the complementarity between Higgsstrahlung and pair production ensured robust exclusions in CP-conserving scenarios, while its breakdown in CP-violating models led to persistent unexcluded regions. At the Tevatron, sensitivity was driven not by mass reach but by enhanced couplings in specific channels, most notably $VH \to Vb\bar{b}$ at large bottom Yukawa coupling. At the LHC, the richness of Higgs production modes---gluon fusion, vector boson fusion, associated production, and top-associated processes---has enabled independent and complementary probes of all major Higgs couplings.

These observations explain why substantial regions of parameter space in extended Higgs sectors remain unexcluded despite decades of experimental and theoretical effort. Models with suppressed couplings, non-standard decay topologies, or modified loop contributions can evade otherwise powerful searches. The continued interest in light Higgs bosons below 125\,GeV, motivated by small excesses observed at LEP and in LHC Run-2 data, illustrates this point clearly: experimental sensitivity depends as much on what is looked for as on how much energy is available.

At present, measurements of Higgs boson properties at the LHC are in remarkable agreement with Standard Model expectations. However, this agreement must be interpreted with care. In the framework of effective field theory, new physics at the TeV scale generically induces deviations in Higgs couplings at the percent level---well below the current experimental precision. The absence of observed deviations therefore reflects the present sensitivity of hadron-collider measurements rather than the absence of BSM effects.

This leads to a central conclusion: the LHC has established the Standard Model as a consistent and predictive description of the Higgs sector at the current level of precision, but it has not exhausted the discovery potential for extended Higgs dynamics. Instead, it has defined the scale and nature of the next experimental challenges. 
The analysis of the LHC Run-3 data in the following years, followed by the HL-LHC programme collecting 3\,ab$^{-1}$ at 14\,TeV from 2030 onwards, will represent a major step forward: Higgs boson coupling modifiers will be determined with roughly a factor of two to three improvement in precision, observation of Higgs boson pair production will yield the first direct measurement of the self-coupling $\lambda$, and BSM searches across all production and decay channels will be substantially extended.
In particular, percent-level measurements of Higgs couplings---sensitive to weakly coupled new physics at multi-TeV scales---require the clean experimental environment and model-independent techniques available at future lepton colliders such as ILC and FCC-ee.

The overall picture that emerges is therefore one of complementarity, not closure. Each experimental stage has contributed a distinct piece to the understanding of the Higgs sector: LEP established the framework of precision exclusion, the Tevatron demonstrated sensitivity to fermionic couplings, and the LHC has mapped the structure of Higgs interactions across multiple production and decay channels. Future facilities will extend this programme into the precision regime, where subtle deviations from the Standard Model---if they exist---are expected to appear.

The key open question remains whether the observed Higgs boson is the manifestation of a minimal scalar sector or part of a richer structure. The evidence accumulated so far neither confirms nor excludes the latter possibility. Instead, it delineates a well-defined region of parameter space---in coupling strength, mass scale, and decay structure---where new physics may still reside. Exploring this region is the central task of the next generation of Higgs boson experiments.

\section*{Acknowledgements}
The research is supported by the Ministry of
Education, Youth and Sports of the Czech
Republic under the project number LM 2023040.

\bibliography{sn-bibliography}%

\end{document}